\documentstyle[12pt]{article}

\newtheorem{theorem}{Theorem}
\newtheorem{prop}{Proposition}
\newtheorem{lemma}{Lemma}
\newtheorem{cor}{Corollary}

\newcommand{\eqa}{\begin{eqnarray}}
\newcommand{\eeqa}{\end{eqnarray}}
\newcommand{\beq}{\begin{equation}}
\newcommand{\eeq}{\end{equation}}
\newcommand{\nn}{\nonumber}
\newcommand{\pal}{\partial}

\newcommand{\F}{{\cal F}}
\newcommand{\al}{\alpha}
\newcommand{\ga}{\gamma}
\newcommand{\de}{\delta}
\newcommand{\tn}{\widetilde \nabla}
\newcommand{\lm}{\lambda}
\newcommand{\del}{\delta}
\newcommand{\ve}{\varepsilon}

\newcommand{\f}{F^{(1)}}
\newcommand{\tx}{t_{X}}
\newcommand{\txx}{t_{XX}}
\newcommand{\txxx}{t_{XXX}}

\newcommand{\pf}{\noindent{\it Proof \ }}
\newcommand{\epl}{Lemma is proved.$\quad \Box$}
\newcommand{\ept}{Theorem is proved.$\quad \Box$}
\newcommand{\epp}{Proposition is proved.$\quad \Box$}
\begin{document}


%
\begin{center}
{\LARGE Bihamiltonian Hierarchies in 2D Topological Field}\\
\vskip 0.3 cm
{\LARGE Theory At One-Loop Approximation}\\
\vskip 0.6cm
{Boris Dubrovin${}^*$ \ \ Youjin Zhang${}^{**}$}\\
\vskip 0.4cm
{\small {\it ${}^{*}$  SISSA, Via Beirut 2--4, 34014 Trieste, Italy}}\\
{\small {\it email: dubrovin@sissa.it}}\\
{\small {\it ${}^{**}$  Division of Mathematics, Graduate School of Science}}\\
{\small {\it Kyoto University, Kyoto 606-8502, Japan}}\\
{\small {\it email: youjin@kusm.kyoto-u.ac.jp}}
\end{center}

\vskip 1cm
{\bf Abstract.} We compute the genus one correction to the integrable
hierarchy describing coupling to gravity of a 2D topological field theory.
The bihamiltonian structure of the hierarchy is given by a classical 
$W$-algebra; we compute the central charge of this algebra. We also express the
generating function of elliptic Gromov - Witten invariants via
tau-function of the isomonodromy deformation problem arising in the theory
of WDVV equations of associativity.
\newpage

\section{Introduction}
According to \cite{Dij3, Dij5,  Witten1}, 
the primary free energy of the matter
sector of a 2D topological field theory (TFT) with $n$ primaries
as a function $F(t)$ of the coupling constants $t=(t^1,\dots, t^n)$
must satisfy WDVV equations of associativity. The problem
of selection of physical solutions among all the solutions to 
WDVV  equations is still open. Reformulating, the problem is to
understand which part of the building of a 2D TFT can be constructed 
taking an arbitrary solution of WDVV equations as the basement.\par

The first problem to be settled is coupling of a given matter sector 
to topological gravity. In the full theory, besides the primaries
$\ \phi_1=1,\ \phi_2,\dots,\phi_n$\ that we now redenote 
$\ \phi_{1,0},\dots,\phi_{n,0}$,\ there are infinite number of their
gravitational descendents $\ \phi_{1,p},\dots,\phi_{n,p},\ 
p=1,2,\dots.$\ The generating function of their correlators 
is the full free energy of the theory
\beq
\F(T)=\left< e^{\sum T^{\al,p}\,\phi_{\al,p}}\right>,
\eeq
here $\ T^{\al,p}$\ are the coupling constants correspondent to the
fields $\phi_{\al,p}$,
\beq 
\left< ... \right> :=\sum_{g\geq 0} \int_{\Sigma_g} ... e^{-S [ \psi] } 
[ d \psi ]
\eeq
(the sum over the fields $\psi$ living on the surface of genus $g$, $S$ is the classical action).
\ According to the idea of Witten \cite{Witten2} this
procedure of coupling to topological gravity must be described by
an integrable hierarchy of PDEs. The unknown functions of the
hierarchy are the particular two-point correlators
\beq
v_\al :=\left<\phi_{\al,0}\,\phi_{1,0}\,
e^{\sum T^{\beta,q}\,\phi_{\beta,q}}\right>={{\pal^2\,\F(T)}\over
{\pal T^{\al,0}\pal T^{1,0}}},
\eeq
$T^{1,p},\dots,T^{n,p}$ are the times of the $p$-th flow of the
hierarchy, and the cosmological constant $X :=T^{1,0}$ plays the role of
the 
spatial variable of the hierarchy. The partition function of the
full theory is the $\tau$-function of a particular symmetric solution
of the hierarchy. This idea works perfectly well for the case of
pure gravity (the matter sector is trivial, $n=1,\ F(t)=\frac16\,t^3$).\
According to the theory of Witten - Kontsevich \cite{Kon1, Witten3}
the partition function of
2D gravity is a particular $\tau$-function of the KdV hierarchy.\par

For a 2D TFT with a nontrivial matter sector the correspondent integrable
hierarchies are not known, although there are interesting conjectures
about their structure for topological minimal models \cite {Dij2}, for
$CP^1$ topological sigm-model \cite{D3, Egu1, Egu5, Egu6, Egu4}. 
However, the properties of {\it genus expansion} of a 2D TFT provide us with
certain nontrivial assumptions about the structure of the hypothetical
hierarchy. Denote ${\cal F}_g$ the genus $g$ part of the free energy
\beq
\F_g :=\left< e^{\sum T^{\al,p}\,\phi_{\al,p}}\right>_g,
\eeq
\beq
{\cal F} = \sum_{g\geq 0} {\cal F}_g.
\eeq
Particularly, the primary free energy is obtained restricting ${\cal F}_0$ onto
the {\it small phase space} $T^{\alpha, p>0}=0$
\beq
F(t) = \F_0 |_{T^{\al ,0}=t^\al , \quad T^{\al , p>0}=0}.
\eeq
The procedure of genus expansion consists of the following two parts.

1). We introduce slow spatial and time variables rescaling
\beq
X\mapsto \varepsilon X, \quad T^{\alpha ,p} \mapsto \varepsilon T^{\alpha ,p}.
\eeq

2). We change
\beq
\F \mapsto\sum_{g=0}^{\infty} \ve^{2\,g-2}\,\F_g.
\eeq
The indeterminate $\ve$ is called {\it string coupling constant}. As 
$\varepsilon\to 0$ one has a singular limit of the {\it tau-function} (i.e., of the partition function) of the theory
\beq
\tau (T, \varepsilon) : = \exp\left({\varepsilon^{-2}\F_0 + \F_1 + 
\varepsilon^2 \F_2 +...}\right).\label{tau-eps}
\eeq 
\par

Also all of the correlators become series in $\varepsilon$. Particularly,
the series of the two-point correlators (1.2) have the form
\beq
v_\alpha = \sum_{g=0}^\infty \varepsilon^{2g} v_\alpha^g
\eeq
where
\beq
v_\alpha^g := \left< \phi_{\alpha,0} \phi_{1,0} e^{\sum T^{\beta , q} 
\phi_{\beta ,q}} \right>_g = 
\frac {\partial^2 {{\cal F}_g}} {\partial T^{\alpha , 0} \partial X}.
\eeq
\par
The genus zero (i.e., the tree-level) approximation of the theory 
corresponds to the dispersionless approximation of the hierarchy.
The solution $\left( v^0_1, ..., v^0_n\right)$ to  
the hierarchy is given by the genus zero two-point 
correlators $\, v_\al ^0=\left<\phi_{\al,0}\,\phi_{1,0}\,e^{\sum T^{\beta,q}\,
\phi_{\beta,q}}\right>_0$\ as the functions
of the couplings. The solution is specified by the initial data on the small phase space
\beq
v_\al^0 |_{T^{\al , p>0}=0} =\eta_{\al \beta}T^{\beta,0}
\eeq
where the constant ``metric'' $\eta_{\al\beta}$ is specified by the primary correlators of the form
\beq
\eta_{\al\beta} := \left< \phi_{1,0}\phi_{\al ,0}\phi_{\beta ,0}\right>_0 |_{T^{\al ,p>0}=0}.
\eeq\par
The construction of the would-be dispersionless
approximation of the unknown integrable hierarchy for an arbitrary solution
to equations of associativity
and of the needed
$\tau$-function of it was given in \cite{D1} in terms of the geometry
of WDVV equations (see also \cite{D3}). The bihamiltonian structure
of the hierarchy was found in \cite{D2}. We briefly recollect this 
construction in Section 2 below. (We also describe more accurately
the quasihomogeneity property of the hierarchy and of the $\tau$-function
formulated in \cite{D1} only for a generic solution of WDVV equations).\par

One can try to go beyond the tree-level approximation expanding 
the unknown hierarchy in a series w.r.t. $\ve^2$. The string coupling constant
 $\ve$ plays 
the role of the dispersion parameter. The reader can keep in mind 
the dispersion expansion
\beq
u_t=u\,u_{x}+\frac1{12}\,\ve^2\,u_{xxx}
\eeq
of the KdV equation as an example of such a series. Particularly,
for the one-loop (i.e., genus$=1$) approximation of the theory,
it is sufficient to retain the terms of the hierarchy up to the
$\ve^2$ order. Particular solutions of the one-loop approximation
must have the form
\eqa
v_\al &=&v_{\al}^0(T)+\ve^2\,v_{\al}^1(T)+{\cal O}(\ve^4)
\nn\\
&=&\left<\phi_{\al,0}\,\phi_{1,0}\,e^{\sum T^{\beta,q}\,\phi_{\beta,q}}
\right>_0+\ve^2\,
\left<\phi_{\al,0}\,\phi_{1,0}\,e^{\sum T^{\beta,q}\,\phi_{\beta,q}}
\right>_1 +{\cal O}(\ve^4)
\nn\\
&=&\frac{\pal^2}{\pal T^{\al,0}\pal T^{1,0}}\left(
\F_0(T)+\ve^2\,\F_1(T)\right)+{\cal O}(\ve^4).
\eeqa
So for $\ve=0$ the one-loop approximation becomes the already known
tree-level
approximation of the hierarchy. We will call the genus one
approximation of the integrable hierarchy {\it the one-loop
deformation} of the genus zero hierarchy.\par

Our result is that, under the assumption of semisimplicity (see 
below) the one-loop deformation of the hierarchy exists for any
solution of WDVV equations and it is uniquely determined by the
general properties of the genus one correlators proved by Dijkgraaf and 
Witten \cite{Dij1} and by Getzler \cite{Getz}. 
(For the solution of WDVV equations
with one and two primaries the one-loop approximation of the hierarchy was 
constructed in \cite{Dij1, Egu6}). Recall that the genus one part of the
free energy has the form
\beq
\F_1(T)=\left[\frac1{24}\,\log\det M_{\al\beta}(t,\pal_X t)+
G(t)\right]_{t=v^0(T)},
\eeq
where the matrix $M_{\al\beta}$ has the form 
\eqa\label{Def-matrix}
&&M_{\al\beta}(t,\pal_X t)=c_{\al\beta\ga}(t)\,\pal_X t^{\gamma},
\\
&&c_{\al\beta\ga}(t)=\pal_\al\pal_\beta\pal_\ga F(t),
\eeqa
and $G(t)$ is a certain function specified by Getzler's equation \cite{Getz}
(see also Section 6 below).
The first part of the formula becomes trivial on the small phase space
$T^{\al,p}=0$ for $p>0$. The second part describes, in the topological
sigma-models, the genus one Gromov-Witten invariants of the target space.
For this function we derive the following formula
\beq\label{119}
G = \log \frac{\tau_I}{J^{1/24}}
\eeq
(as above, semisimplicity of the solution of WDVV is assumed). Here $J$
is the Jacobian of the transform between canonical and flat coordinates
(see Sect. 2 below). To explain who is $\tau_I$ we recall that, in the
semisimple case, WDVV can be reduced to equations of isomonodromy
deformations of a certain linear differential operator with rational
coefficients \cite{D1}. Our $\tau_I$ is the tau-function of the solution
of these equations of isomonodromy deformations in the sense of
\cite{Sato}. 
According to \cite{Sato}, \cite{Miwa} the tau-function appears as the 
Fredholm determinant of an appropriate Riemann-Hilbert boundary value
problem (see \cite{D4} for reduction of WDVV equations to a boundary 
value problem).
Remarkably, the formula makes sense for an arbitrary
semisimple solution of WDVV equations. 
Using explicit expressions (\ref{ham-i}),
(\ref{tau-i}) for $\tau_I$ one can derive from (\ref{119}) the proof
of main conjectures of the recent paper of Givental \cite{Giv}.\par  

As a byproduct of our computations, we obtained a nice formula 
for the generating function of elliptic Gromov - Witten invariants of
complex projective plane. Namely, the function
\beq
\psi :=\frac{\phi'''-27}{8 (27+2\,\phi'-3\,\phi'')},
\eeq
where 
\eqa\label{Defofphi}
\phi(z)&=&\sum_{k\ge 1} \frac{N_k^{(0)}}{(3\,k-1)!}\,e^{k\,z},\\
N_k^{(0)}&=&\mbox{the number of rational curves of degree}\ k
\nn\\ && 
\mbox{on $CP^2$ passing through generic $3\,k-1$ points}\nn
\eeqa
is the generating function for the numbers $N^{(1)}_k$ of the
elliptic curves of degree $k$ on 
$CP^2$ passing through generic $3\,k$ points:
\beq
\psi(z)=-\frac18+ \sum_{k\ge 1} N^{(1)}_k\,\frac{k}{(3\,k)!}\,
e^{k\,z}.
\eeq
\par
We prove also that the compatible pair of Poisson brackets
describing the tree-level hierarchy admits a unique deformation to
give a bihamiltonian structure, modulo ${\cal O}(\ve^4)$,
of the one-loop hierarchy.
The deformed bihamiltonian structure turns out to be a nonlinear 
extension of the Virasoro algebra (i.e., a classical $W$-algebra) with the
central charge
\beq\label{Cent-char}
c=\frac{12 \varepsilon^2}{(1-d)^2} \left[ \frac{1}{2} n -2 \sum_{\alpha =1}^n 
( q_\alpha -
\frac{1}{2} d)^2 \right].
\eeq
Here $\varepsilon$ is the string coupling constant, $d$ and $q_\alpha$ are 
the ``dimension'' and the ``charges'' of the
theory. In the case of quantum cohomology of $X$  (i.e., the
topological sigma-model with the target space $X$) $d$ coincides
with the complex dimension of the target space $X$ and $q_\alpha$ are
the halfs of the degrees of the basic elements in $H^*(X)$.
Remarkably, this formula works not only in quantum cohomologies.
It gives the correct value for the central charge \cite{FatLuk}
of the classical $W$-algebras for the topological minimal models of A - D - E type
(see below Sect.8)!\par

We can continue this procedure trying to construct higher genera
approximation of the unknown integrable hierarchy. Of course, it 
would be too optimistic to expect that our procedure will go
smoothly for any genus $g$ for an arbitrary solution of equations of associativity. Moreover, from \cite{Egu6} it follows that,
constructing the integrable hierarchy, probably for a generic  solution
of WDVV one cannot go beyond the genus one. However,
our results suggest
that in an arbitrary physical 
2D TFT coupling to gravity is given by an integrable bihamiltonian
hierarchy of $1+1$ PDEs. 
Bihamiltonian structure of the hierarchy is to be described by a classical
$W$-algebra with the prescribed central charge and the conformal
dimensions of the primaries. So, we embed the problem of coupling to
topological gravity into the problem of classification of a certain class
of classical $W$-algebras.

We briefly discuss this project in the final
section, postponing the study of the higher genera corrections
for a subsequent work.\par
  
The paper is organized as follows. In Section 2 we recall some
important points of the theory of WDVV equations of associativity
(equivalently, the theory of Frobenius manifolds) and the construction
of coupling to gravity at tree-level. The main results of the paper
are formulated in Section 3. In Section 4 we derive some useful identities
of the theory of semisimple Frobenius manifolds used in the
proof of the main results. The derivation of the bihamiltonian structure
of the hierarchy in the genus one approximation 
is given in Section 5.
In Section 6 we solve 
Getzler's equations for elliptic Gromov - Witten invariants
for any semisimple Frobenius manifold. The examples of the 
deformed bihamiltonian hierarchies are given in Section 7. In the
last Section 8 we discuss the programme of study of higher
genera corrections in the setting of classical $W$-algebras.

{\it Acknowledgments}.\ The authors thank E. Getzler for fruitful
discussions. We thank G.Falqui for the help with $W$-algebras. The work
of one of the authors (B.D.) was done under partial support of the EC TMR
Programme {\it Integrability, non-perturbative effects and symmetry in 
Quantum Field Theories}, grant FMRX-CT96-0012. The work of Y.Z. was 
supported by the
Japan Society for the Promotion of Science, it was initiated in SISSA 
when he was a post-doc there; he thanks M. Jimbo for valuable discussions.
\setcounter{equation}{0}
\section{WDVV  equations of associativity and the structure of a
2D TFT at genus zero}\par

WDVV equations of associativity is the problem of finding a function
$F(t)=F(t^1,\dots,t^n)$, a constant symmetric nondegenerate matrix
$(\eta^{\al\beta})$,\ numbers $\ q_1,\dots,q_n,\ r_1,\dots,r_n,\ d$\
such that 
\beq\label{wdvv1}
\pal_\al\pal_\beta\pal_\lm F(t)\,\eta^{\lm\mu}\,\pal_\mu\pal_\gamma
\pal_\de F(t)=
\pal_\de\pal_\beta\pal_\lm F(t)\,\eta^{\lm\mu}\,\pal_\mu\pal_\gamma
\pal_\al F(t)
\eeq
for any $\al, \beta, \ga, \de=1,\dots,n$,
\eqa\label{wdvv2}
&&\pal_1\pal_\al\pal_\beta F(t)\equiv \eta_{\al\beta},
\quad\mbox{where}\ (\eta_{\al\beta})=(\eta^{\al\beta})^{-1},
\\
&& \sum \left[(1-q_\al)\,t^\al+r_\al\right]\pal_\al F(t)=
(3-d)\,F(t)+\frac12\,A_{\al\beta}\,t^\al\,t^\beta+B_\al\,t^\al+C
\label{wdvv3}
\eeqa
for some constants $A_{\al\beta}, B_\al, C$. The numbers 
$q_\al, r_\al,d$ and $A_{\al\beta}, B_\al, C$ must satisfy the following
normalization conditions (see \cite{D4}):
\eqa
&&q_1=0,\ r_\al\ne 0 \quad \mbox{only if} \ q_\al=1,
\nn\\
&&A_{\al\beta}\ne 0\quad \mbox{only if} \ q_\al+q_\beta=d-1,
\nn\\
&&B_\al\ne 0\quad \mbox{only if} \ q_\al=d-2,
\nn\\
&&C\ne 0 \quad \mbox{only if} \ d=3,
\nn\\
&&A_{1\al}=\sum_{\ve}\eta_{\al\ve}\,r_{\ve},\ B_1=0.
\eeqa
We will usually normalize the coordinates $t^\al$ reducing $\eta_{\al\beta}$ 
to
the antidiagonal form
\beq
\eta_{\al\beta} =\delta_{\al + \beta , n+1}.
\eeq
This can always be done for $d\neq 0$. Then
\beq
q_\al + q_{n-\al +1} =d,\quad   q_n =d.
\eeq\par
Any solution of WDVV equations provides the space of parameters
$M^n\ni (t^1,\dots,t^n)$ with a structure of {\it Frobenius manifold}.
That means that there exists a unique structure of a Frobenius
algebra $\left( A_t,<\ ,\ >\right)$ on the tangent planes $T_t M^n$ 
such that
\beq
\left<\pal_\al\cdot\pal_\beta,\pal_\ga\right>=
\pal_\al\pal_\beta\pal_\ga F(t),\quad
\left<\pal_\al,\pal_\beta\right>=\eta_{\al\beta}.
\eeq
Explicitly
\beq
\pal_\al\cdot\pal_\beta=c^\ga_{\al\beta}(t)\,\pal_\ga 
\quad\mbox{where}\ c^\ga_{\al\beta}(t)=\eta^{\ga\ve}\,\pal_\ve\pal_\al
\pal_\beta F(t).
\eeq
The vector field 
\beq
e=\pal_1
\eeq
is the unity of the algebra. We introduce also the Euler vector field
on $M^n$
\beq\label{VF-Euler}
E(t)=E^\ve(t)\,\pal_\ve :=\sum_{\ve=1}^n \left[(1-q_\ve)\,t^\ve+
r_\ve\right]\,\pal_\ve .
\eeq
This is the generator of the scaling transformations (\ref{wdvv3}).
All the equations (\ref{wdvv1})--(\ref{wdvv3}) can be easily reformulated 
in a covariant way (see \cite{D3}).\par 

One of the main geometrical objects on a Frobenius manifold is a 
deformation of the Levi-Civita connection $\nabla$ for $<\ ,\ >$:
\beq\label{dfc}
\tn_u\,v=\nabla_u\,v+z\,u\cdot v.
\eeq
Here $u, v$ are two vector fields on $M^n,\ z$\ is the parameter of the
deformation. The connection (\ref{dfc}) is flat for any $z$. It can be 
extended to a flat connection on $ M^n\times {\bf C}^*$
\beq\label{mc}
\tn_{\frac{d}{d z}} v=\pal_z v+E\cdot v-\frac1{z}\,\mu\, v,
\eeq
where
\eqa
&&\mu :=-\nabla\,E+\frac12\,(2-d)=\mbox{diag}(\mu_1,\dots,\mu_n),
\quad \mu_\al=q_\al-\frac{d}2,
\\
&&\left<\mu\,a,b\right>=-\left<a,\mu\,b\right>.
\eeqa
(Comparing with \cite{D3} we change the normalization of the component
$\tn_{\frac{d}{d z}}$ doing an elementary gauge transform).
The connection on $M^n\times {\bf C}^*$ is still flat.\par

The Frobenius manifold is said to satisfy {\it the semisimplicity
condition}
(or, briefly, it is semisimple) if the algebras $A_t$ are semisimple for
generic $t$. On the open domain of the points of semisimplicity one can
introduce {\it canonical coordinates} $u_1, \dots, u_n$ such that
\beq
\frac{\partial}{\partial u_i} \cdot \frac{\partial}{\partial u_j}
=\de_{ij}\frac{\partial}{\partial u_i}, ~i,j=1,
\dots, n.
\eeq
(We will use all lower indices working with the canonical coordinates.
No summation over the repeated indices will be assumed in this case.)
In these coordinates WDVV can be reduced to a commuting family of
nonstationary Hamiltonian flows on the Lie algebra $so(n)$ with the
standard Poisson bracket
\beq
\frac{\partial V}{\partial u_i} = \{ V, H_i(V;u)\}, ~i=1, \dots, n
\eeq
(the definition of the matrix $V=(V_{ij})$, $V^T =-V \in so(n)$ see below
in Sect.4), the canonical coordinates $u_1, \dots, u_n$ play the role
of the times and the quadratic Hamiltonian has the form
\beq\label{ham-i}
H_i =\frac{1}{2} \sum_{j\neq i} \frac {V_{ij}^2}{u_i-u_j}.
\eeq
These are the equations of isomonodromy deformations of the operator
\beq
\frac{d}{dz} -U -\frac{1}{z} V, ~~U= diag(u_1, \dots, u_n)
\eeq
with rational coefficients \cite{D1}. The tau-function $\tau_I$ of a
solution in the theory of isomonodromy deformations is defined \cite{Sato}
by the quadrature
\beq\label{tau-i}
d\log \tau_I = \sum_{i=1}^n H_i du^i
\eeq
(We denote this function $\tau_I$ to avoid confusions with the
tau-function (\ref{tau-eps}) of the integrable hierarchy.)\par

Another geometric object is a deformation of the flat metric $<\ ,\ >$
on $M^n$ \cite{D1, D3}. We introduce the {\it intersection form}
\beq
(\omega_1,\omega_2)_t :=i_E\,(\omega_1\cdot\omega_2),\quad 
\omega_1, \omega_2\in T^*_t\,M^n.
\eeq
The metric 
\beq
(\ ,\ )_t-\lm\,<\ ,\ >_t
\eeq
on $T^*_t\,M^n$\ does not degenerate for almost all $(\lm,t)$.\
It is flat for these $(\lm,t)$. In the coordinates $t^\al$
\beq\label{def-g}
g^{\al\beta}(t) :=(d\,t^\al, d\,t^\beta)=E^\ve\,c^{\al\beta}_\ve=
(d+1-q_\al-
q_\beta)\,F^{\al\beta}(t)+A^{\al\beta},
\eeq
where
\beq\label{up-index}
F^{\al\beta}(t) :=\eta^{\al\al'}\eta^{\beta\beta'}\,\frac{\pal^2 F(t)}
{\pal t^{\al'}\pal t^{\beta'}},\quad
A^{\al\beta} :=\eta^{\al\al'}\eta^{\beta\beta'}\,A_{\al'\beta'}.
\eeq
We give also the formula for the Levi-Civita connection for the flat
(but not constant in the coordinates $t^\al$\,!) metric $(\ ,\ )$
\beq
\Gamma^{\al\beta}_\ga(t) :=-g^{\al\ve}(t)\,\Gamma^\beta_{\ve\ga}(t)
=\left(\frac{1+d}2-q_\beta\right)\,c^{\al\beta}_\ga(t),
\eeq
where
\beq
c^{\al\beta}_\ga(t)=\eta^{\al\al'}\eta^{\beta\beta'}\,\pal_{\al'}
\pal_{\beta'}\pal_\ga F(t).
\eeq
The flat metric (\ref{def-g}) is responsible not only for the second Poisson bracket
of the integrable hierarchy (see below), but also for the relation between
Frobenius manifolds and reflection groups \cite{D3}.

The genus zero approximation of the needed integrable hierarchy will be an
infinite family of dynamical systems on the loop space ${\cal L}(M^n)$.
We supply the loop space with a Poisson bracket
\beq\label{fpb}
\{v^{\al}(X),v^{\beta}(Y)\}_1^{(0)}=\eta^{\al\beta}\,\de'(X-Y),
\eeq
(to avoid confusions we redenote $\ t^\al\to v^\al$\ the coordinates
on $M^n$ when dealing with the hierarchy; comparing with
the above notations of Introduction we omit the label
$0$, i.e., $v^\al=\eta^{\al\ve}\,v^0_\ve$).
The second Poisson bracket on the same loop space has the form
\beq\label{spb}
\{v^{\al}(X),v^{\beta}(Y)\}_2^{(0)}=g^{\al\beta}(v(X))\,\de'(X-Y)+
        \Gamma^{\al\beta}_{\ga}(v(X))\, v^{\ga}_X\,\de(X-Y).
\eeq
Particularly, for $d\ne 1$ the Poisson bracket of 
\beq
T(X) :=\frac2{1-d}\,t^n(X)
\eeq
has the form
\beq\label{Vira-gz}
\{ T(X),T(Y)\}^{(0)}_2=\left[ T(X)+T(Y)\right]\,\de'(X-Y).
\eeq
This coincides with the Poisson bracket on the dual space to the Lie
algebra of one-dimensional vector fields. Therefore the full Poisson 
bracket (\ref{spb}) can be considered as a nonlinear extension of this
algebra (the classical W-algebra with zero central charge).
Observe
that
\beq
\{t^\alpha(X),T(Y)\}^{(0)}_2=\left(\frac{2\,(1-q_\alpha)}{1-d}\,t^\alpha(X)+
\frac{2\,r_\alpha}{1-d}\right)\,\de'(X-Y)+
\tx^\alpha\,\de(X-Y).
\eeq
So $T(X)$ plays the role of the stress-energy tensor, and 
the conformal dimensions of the fields $t^\alpha$ having $q_\alpha\neq
1$ are
\beq\label{conf-dim}
\Delta^\alpha= \frac{2(1-q_\alpha)}{1-d}.
\eeq
When $q_\alpha =1$ the variable $s^\alpha := \exp t^\alpha$ has the
Poisson bracket with the stress-energy tensor of the form

\beq
\{ s^\alpha (X), T(Y)\}^{(0)}_2 =\frac{2r_\alpha}{1-d} s^\al (X)\delta'(X-Y) 
+s^\alpha_X \delta(X-Y).
\eeq
So it is a primary field with the conformal dimension
\beq
\Delta_\alpha = \frac{2r_\alpha}{1-d}.
\eeq

The two Poisson brackets are compatible, i.e., any linear combination
\beq\label{pencil}
a_1\,\{\ ,\ \}^{(0)}_1+a_2\,\{\ ,\ \}^{(0)}_2
\eeq
with arbitrary constant coefficients $a_1, a_2$ gives a Poisson
bracket on ${\cal L}(M^n)$ \cite{D3}. This gives a possibility to 
construct a hierarchy of commuting flows on ${\cal L}(M^n)$ starting
from the Casimirs of the first Poisson bracket
\beq
H^{\al,-1}=\int v^\al(X)\,dX,\quad \al=1,\dots,n
\eeq
using the standard bihamiltonian recursion procedure \cite{Magri}
\beq
\{\cdot,H^{\al,p}\}^{(0)}_1=k_{\al,p}\,\{\cdot,H^{\al,p-1}\}^{(0)}_2
\eeq
for appropriate constants $k_{\al,p}$. These constants are to be chosen in a
clever way to make the hierarchy compatible with the genus zero recursion
relations for the topological correlators. For the genus zero
approximation the needed normalization of the Hamiltonians is given
by an alternative procedure \cite{D1} using the flat coordinates of the
deformed connection $\tn$.\par

The flat coordinates of $\tn$ are functions $\theta(t,z)$ such that
\beq
\tn\,d\theta=0.
\eeq
Let us forget for the moment about the last component (\ref{mc}) of the
connection $\tn$. Then the flat coordinates $\theta$ are specified
by the equation
\beq
\pal_\al\pal_\beta\theta=z\,c^\ga_{\al\beta}\,\pal_\ga\theta.
\eeq
A basis of the solutions $\theta_1(t,z),\dots,\theta_n(t,z)$ can be
obtained as power series 
\beq
\theta_\ga(t,z)=t_\ga+\sum_{p\ge 1}\theta_{\ga,p}(t)\,z^p,
\eeq
where the coefficients $\theta_{\ga,p}(t)$ are determined recursively
from the equations 
\beq\label{odefh}
\pal_{\al}\pal_\beta\,\theta_{\ga,p+1}(t)=c_{\al\beta}^\rho(t)\pal_\rho\theta_{\ga,p},\quad \theta_{\ga,0}(t)=t_\ga=\eta_{\ga\ve}\,t^\ve.
\eeq
One can normalize the deformed flat coordinates requiring
\beq
\left<\nabla\,\theta_\al(t,z),\nabla\,\theta_\beta(t,-z)\right>\equiv 
\eta_{\al\beta}.
\eeq
There still remains some freedom in the choice of the deformed flat
coordinates
\beq\label{free1}
\theta_\al(t,z)\mapsto \theta_\ve(t,z)\,G^\ve_\al(z)
\eeq
with an arbitrary matrix-valued series $G(z)=(G^\beta_\al(z))$
\eqa\label{free2}
&&G(z)=1+z\,G_1+z^2\,G_2+\dots,
\\
&& G(z)\,\eta\,G(-z)^T\equiv\eta.\label{free3}
\eeqa
Later we put also the equation (\ref{mc}) into the game. This 
will fix the deformed flat coordinates almost uniquely.\par

The Hamiltonians of the genus zero hierarchy have the form 
\beq\label{H-gz}
H_{\beta,p}=\int \theta_{\beta,p+1}(v(X))\,dX,\quad p=0,1,\dots.
\eeq
The hierarchy itself reads
\beq\label{h-eq}
\frac{\pal v}{\pal T^{\beta,p}}=K^{(0)}_{\beta,p}(v,v_X)=
\{v,H_{\beta,p}\}_1^{(0)}=\pal_X \nabla\,\theta_{\beta,p+1}(v)=
\nabla\,\theta_{\beta,p}(v)\cdot \pal_X v,
\eeq
(we treat $\pal_X v$ and $\pal_{T^{\beta,p}} v$ as tangent vectors to the
Frobenius manifold). Observe that the coefficients in front of $\pal_X v$ 
are functions well-defined everywhere on the Frobenius manifold.\par

The genus zero two-point functions 
\beq
v^0_\al(T)=\left<\phi_{\al,0}\,\phi_{1,0}\,e^{\sum T^{\beta,q}\,\phi_{\beta,q}}
\right>_0
\eeq
give a particular solution of the commutative hierarchy (\ref{h-eq})
specified by the following symmetry reduction
\beq
\left(\pal_{T^{1,1}}-\sum_{\al,p}T^{\al,p}\,\pal_{T^{\al,p}}\right)\,v^0=0.
\eeq
(I identify $T^{1,0}$ and $X$. So the variable $X$ is supressed in the 
formulae). The solution can be found in the form
\beq
v^0(T)=T_0+\sum_{q>0} T^{\beta,q}\,\nabla\,\theta_{\beta,q}(T_0)+
\sum_{p,q>0}T^{\beta,q}\,T^{\ga,p}\,\nabla \,\theta_{\beta,q-1}(T_0)\cdot
\nabla\,\theta_{\ga,p}(T_0)+\dots.
\eeq
This is a power series in $T^{\al,p>0}$ with the coefficients depending on 
$T_0 :=(T_{1,0},\dots,T_{n,0})$, $T_{\al ,0} :=\eta_{\al\beta} T^{\beta , 0}$. 
The series can be found as the fixed 
point $t=v^0$ of the gradient map $M^n\to M^n$
\beq
t=\nabla\,\Phi_T(t),
\eeq
where 
\beq
\Phi_T(t)=\sum T^{\al,p}\,\theta_{\al,p}(t).
\eeq
\par

Defining the functions $\Omega_{\al,p;\beta,q}(t)$ on the Frobenius
manifold by the following generating function
\beq
(z+w)^{-1}\,\left(\left<\nabla\,\theta_\al(t,z),\nabla\,\theta_\beta(t,w)
\right>-\eta_{\al\beta}\right)=
\sum_{p,q=0}^\infty \Omega_{\al,p;\beta,q}(t)\,z^p\,w^q,
\eeq
we complete the construction of the genus zero free energy of the
TFT coupled to gravity by setting
\beq\label{formula-F0}
\log\tau=
\F_0(T)=\frac12\,\sum \Omega_{\al,p;\beta,q}(v^0(T)){\widetilde T}^{\al,p}\,
{\widetilde T}^{\beta,q},
\eeq
where
\eqa
{\widetilde T}^{\al,p}&=&T^{\al,p}\quad \mbox {if}\quad (\al,p)\ne(1,1)
\nn\\
{\widetilde T}^{1,1}&=&T^{1,1}-1.\label{tildeT}
\eeqa
The resulting function $\F_0(T)$ satisfies the string equation
\beq
\frac{\pal\F_0(T)}{\pal T^{1,0}}=\sum T^{\al,p}\pal_{T^{\al,p-1}}\,
\F_0(T)+\frac12\,\eta_{\al\beta}\,T^{\al,0}\,T^{\beta,0}.
\eeq
On the small phase space $T^{\al,p>0}=0$ one has
\beq
\F_0(T)|_{\begin{array}{c} T^{\al,p>0}=0\\T^{\al,0}=t^\al\end{array}}
=F(t).
\eeq
Also the derivatives of the function $\F_0(T)$ satisfy the genus zero 
recursion relations of Dijkgraaf and Witten. Observe that
\beq
\frac{\pal^2\F_0(T)}{\pal T^{\al,p}\pal T^{\beta,q}}=
\Omega_{\al,p;\beta,q}(v^0(T)).
\eeq
The proofs of all these results can be found in \cite{D1}.

We now use the last component $\tn_{\frac{d}{d z}}$ of the deformed connection
to fix the densities of the commuting Hamiltonians $H_{\al,p}$. Let us
consider first the {\it non-resonant case} $\ \mu_\al-\mu_\beta\not\in 
{\bf Z}_{\ne 0}$ for $\al\ne \beta$. Then the system of deformed coordinates 
${\tilde t}_\al(t,z)$ of $\tn$ can be constructed in the form
\eqa
&&{\tilde t}_\al(t,z)=\theta_\al(t,z)\,z^{\mu_\al}=
\sum_{p=0}^\infty\theta_{\al,p}(t)\,z^{p+\mu_\al},
\\
&&\tn_{\frac{d}{d z}} d\tilde t_\al(t,z)=0.\label{dffc}
\eeqa
The coefficients $\theta_{\al,p}(t)$ 
are now defined uniquely by (\ref{odefh}) and
by the quasihomogeneity equation following from (\ref{dffc})
\beq
{\cal L}_E\theta_{\al,p}(t)=\left(p+\frac{2-d}2+\mu_\al\right)\,
\theta_{\al,p}(t).
\eeq
The functions $\Omega_{\al,p;\beta,q}(t)$ are also quasihomogeneous
of the degree $p+q+1+\mu_\al+\mu_\beta$.\ From this one easily derives the 
quasihomogeneity constraint for $\F_0$ (see \cite{D1}).\par

Let us now consider the non-generic case. We describe first the 
fundamental matrix solution of the linear system
\beq
\tn_{\frac{d}{d z}}\,d \tilde t=0.
\eeq
We rewrite  this system for the gradient
\beq
(\nabla\,\tilde t)^\al=\eta^{\al\beta}\,\pal_\beta \tilde t
\eeq
of the deformed flat coordinates. So the columns of the fundamental  
 matrix are the gradients of the deformed flat coordinates
${\tilde t}_1(t,z),\dots,{\tilde t}_n(t,z)$. The fundamental
matrix has the form
\beq\label{flat-g}
Y(t,z)=\left(\nabla{\tilde t}_1(t,z),\dots,\nabla{\tilde t}_n(t,z)\right)
=\left(\nabla{\theta}_1(t,z),\dots,\nabla{\theta}_n(t,z)\right)\,
z^\mu\,z^R,
\eeq
where the constant\footnote{
Constancy of the matrix $R$ is a manifestation of the general isomonodromicity
property proved in the theory of Frobenius manifolds \cite{D1,D3,D4}.}
 matrix $R=(R^\al_\beta)$ satisfies the following
requirements:
\begin{enumerate}
\item $R^\al_\beta\ne 0$ \ only if \ $\mu_\al-\mu_\beta$ is a 
positive integer,
\item Let $\ {R_k}^\al_{\beta}=\left\{\begin{array}{ll}
R^\al_\beta & \mbox{if \ $\mu_\al-\mu_\beta=k$}\\
0 &\mbox{otherwise} \end{array}\right.$. \newline
We have 
\beq\label{dec-R}
R=R_1+R_2+\dots
\eeq
(finite number of terms). Then we must have 
\beq
\left<R_k\,a,b\right>+(-1)^k\,\left<a,R_k\,b\right>=0,
\quad k=1,2,\dots.
\eeq
for any two vectors $a, b$.
\end{enumerate}
The matrix $R$ is determined uniquely up to the transformations
\beq\label{tranf-G}
R\mapsto G^{-1}\,R\,G,
\eeq
where the matrix $G=(G^\al_\beta)$ must satisfy the following 
conditions:
\begin{enumerate}
\item $G^\al_\beta\ne 0 $\ only if \ $\mu_\al-\mu_\beta
\ $ is a non-negative integer.
\item Define the decomposition
\beq
G=G_0+G_1+\dots
\eeq
similar to (\ref{dec-R}). We must have
\beq
G_0=1
\eeq
and the matrix $G$ must satisfy the following orthogonality condition
\beq
\left<G^+\,a,G\,b\right>=<a,b>,
\eeq
for any $a$, $b$ where
\beq
G^+=G_0-G^T_1+G^T_2-G^T_3+\dots.
\eeq
\end{enumerate}
Proof can be found in \cite{D4}. The class of equivalence of the matrix
$R$ modulo the transformations (\ref{tranf-G}) together with the 
matrix $\mu$ completely specifies the class of gauge equivalence of the 
operator $\ \tn_{\frac{d}{d z}}$ modulo gauge transformations of the
form (\ref{free1})--(\ref{free3}) near the singularity at $z=0$.
Particularly, the coefficients $A_{\alpha\beta}$ $B_\alpha$, $C$ in (2.3)
have the form
\beq
A_{\alpha\beta}= \eta_{\alpha\epsilon} (R_1)_\beta^\epsilon
\eeq
\beq
B_\alpha = \eta_{1\ve} (R_2)^\ve_\al,
\eeq
\beq
C=-\frac{1}{2}\eta_{1\ve} (R_3)^\ve_1
\eeq

Plugging the formula (\ref{flat-g}) into the equation 
$ \tn_{\frac{d}{d z}}=0$
we obtain the following quasihomogeneity constraint for the function
$\theta_{\al,p}(t)$
\beq\label{scale-theta}
{\cal L}_E\,\theta_{\al,p}(t)=\left( p+\frac{2-d}2 +\mu_\al\right)
\theta_{\al,p}(t)+\sum_{k=1}^p \theta_{\ve,p-k}(t)\,\left(R_k\right)^\ve_\al+
const.
\eeq
(Observe that the functions $\theta_{\al,p}(t)$ are defined up to
an additive constant). A more involved computation shows that
\eqa
&&{\cal L}_E\,\Omega_{\al,p;\beta,q}(t)=(p+q+1+\mu_\al+\mu_\beta)
\Omega_{\al,p;\beta,q}(t)+
\sum_{r=1}^p \left(R_r\right)^\ve_\al\,\Omega_{\ve,p-r;\beta,q}(t)
\nn\\
&&\quad+\sum_{r=1}^q \left(R_r\right)^\ve_\beta\,\Omega_{\al,p;\ve,q-r}(t)
+(-1)^q\left(R_{p+q+1}\right)^\ve_\al\,\eta_{\ve\beta}.\label{scale-omega}
\eeqa
Using this and the explicity formula (\ref{formula-F0}) we arrive at
\begin{prop}
The genus zero partition function $\tau$ satisfies the following 
constraint
\beq\label{Virasoro}
L_0\tau=0,
\eeq
where 
\eqa
L_0=&&\sum \left(\frac12+k+\mu_\lm\right){\widetilde T}^{\lm,k}\pal_{T^{\lm,k}}
+\sum{\widetilde T}^{\lm,k}\left(R_r\right)^\ve_\al\,\pal_{T^{\ve,k-r}} 
\nn\\
&&
+\frac12\,\sum(-1)^q {\widetilde T}^{\al,p}\,{\widetilde T}^{\beta,q}
\left(R_{p+q+1}\right)^\ve_\al\,\eta_{\ve\beta}.
\eeqa
Here ${\widetilde T}^{\al,p}$ are defined by (\ref{tildeT}).
\end{prop}

{\bf Example.} $\ $ For topological sigma-models $R$ coincides with the matrix
of multiplication by the first Chern class $c_1(X)$ in the classical 
cohomologies of the target  space $X$ \cite{D3}.   Since $\deg c_1(X)=1$ we
have
\beq
R=R_1.
\eeq
The recursion relation (\ref{scale-theta}) in this case coincide with the recursion
relation of Hori \cite{Hori}, and the particular case of (\ref{scale-omega}) 
was obtained in \cite{Egu4}.
We infer that the coefficients $\theta_{\al,p}(t)$
of the expansion of the deformed flat coordinates coincide\footnote{
Our normalization of the correlators differs from that of Hori}
with the two-point functions 
$\left<\phi_{\al,0}\phi_{1,0}\,e^{\sum_{\al=1}^n t^\al\,\phi_{\al,0}}
\right>_0$
defined in terms of  intersection theory on the moduli spaces of
instantons $S^2\to X$. The general identity (\ref{Virasoro}) in this
particular case coincides with the $L_0$ Virasoro constraint derived
in \cite{Hori}.

{\bf Remark.} $\ $ Applying an appropriate recursion procedure to the
operator $L_0$ we can derive a half-infinite sequence of the Virasoro
constraints 
\beq
L_k\tau=0,\quad k\ge -1
\eeq
generalizing the constraints of \cite{Egu3}. All the operators 
$L_k, \ k\ge -1$ are given in terms of the monodromy data $(\mu,R)$
at $z=0$.
We will present these results in a  separate publication.

We conclude this section with an explicit formula for the bihamiltonian
structure of genus zero hierarchy  (\ref{h-eq}).
\begin{prop}
Let $(\al,p)$ be a pair of indices such that 
\beq
p+\mu_\al+\frac12\ne 0.
\eeq
then the equation 
\beq
\frac{\pal v}{\pal T^{\al,p}}=\{v,H_{\al,p}\}^{(0)}_1
\eeq
of the hierarchy (\ref{h-eq}) is also a Hamiltonian flow w.r.t.
the second Poisson bracket (\ref{spb})
\beq\label{BH}
\{v,H_{\al,p}\}^{(0)}_1=\{v,{\hat H}_{\al,p}\}^{(0)}_2.
\eeq
The Hamiltonian ${\hat H}_{\al,p}$ has the form
\beq
{\hat H}_{\al,p}= \sum_{k, l} (-1)^{k} \left( R_{p-l, k}\right)_\al^\ve
\frac{H_{\ve , l-1}}{(p+\mu_\al +\frac{1}{2})^{k+1}}
\eeq
where the matrices $R_{k,l}$ are defined as follows
\beq\label{matR}
R_{0,0}=1, \quad R_{k,0}=0 \quad \mbox{for} \quad k>0,
R_{k,l}= \sum_{i_1 + ... + i_l=k} R_{i_1} ... R_{i_l} \quad \mbox{for} \quad l>0.
\eeq
\end{prop}
\pf We use the identity 
\beq
\{\,\cdot\,,\int {\tilde t}(v(X),z) dX\}^{(0)}_2=
\{\cdot,\int\left(\pal_z-\frac1{2\,z}\right){\tilde t}(v(X),z) dX\}^{(0)}_1
\eeq
valid for an arbitrary flat coordinate $\tilde t$ of $\tn$ (see Lemma H.3
in \cite{D3}). Inverting, we obtain
\beq
\{\,\cdot\,,\int\left(z^{\frac12}\int^z w^{-\frac12}\,{\tilde t}_\al(v(X),w)
dw\right) dX\}^{(0)}_2
=\{\,\cdot\,,\int {\tilde t}_\al(v(X),z) dX\}^{(0)}_1.
\eeq
Integrating the expansions in both sides of the equation and
using 
\beq
{\tilde t}_\al(t,z)=\sum \theta_{\ve,p}\,
z^{p+\mu_\ve}\,\left( z^R\right)^\ve_\al 
\eeq
we obtain the formula (\ref{BH}).
\epp
\vskip 0.8cm

\setcounter{equation}{0}
\section{Formulation of the main results}\par

We formulate now the main requirement to uniquely specify the
one-loop correction to the hierarchy (\ref{h-eq}). We want to find a hierarchy
of equations of the form
\beq\label{H-deformed}
\frac{\pal t}{\pal T^{\beta,p}}=K^{(0)}_{\beta,p}(t,\tx)+
\ve^2\,K^{(1)}_{\beta,p}(t,t_X,\dots)
\eeq
such that the following property holds true (cf. \cite{Dij1, Egu6}):

{\bf Main assumption.} $\ $ For {\it any} solution $v=v(T)$ of the hierarchy 
(\ref{h-eq}) the function $t(T)=(t_1(T),\dots,t_n(T))$
\beq\label{F-BT}
t(T) :=v(T)+\ve^2\,w(T)
\eeq
where
\beq\label{Def-w}
w_\al(T)=\frac{\pal^2}{\pal T^{\al,0}\,\pal T^{1,0}}\left\{
\left[\frac1{24}\,\log\det M_{\al\beta}(t,t_X)+G(t)\right]_{t=v(T)}
\right\}
\eeq
satisfies (\ref{H-deformed}) modulo terms of the order $\ve^4$. Here the matrix
$M_{\al\beta}(t,t_X)$ is defined by (\ref{Def-matrix}), 
and $G(t)$ is the G-function of the Frobenius manifold (see below).\par

We denote $t=(t^1,\dots,t^n)$ the dependent variables of the hierarchy to
emphasize that they live on the Frobenius manifold $M^n$. So (\ref{H-deformed})
is still a dynamical system on the loop space ${\cal L}(M^n)$.

It is clear that the corrections $K^{(1)}_{\beta,p}$ are determined uniquely.
Indeed, the deformed hierarchy  (\ref{H-deformed}) is obtained from the 
tree-level hierarchy (\ref{h-eq}) by the infinitesimal B\"{a}cklund transform
\beq\label{Back}
v_\al\mapsto v_\al+\ve^2\,w_\al(v,v_X,v_{XX},v_{XXX})=t_\al
\eeq
where the functions $w_\al$ are defined by the formula (\ref{Def-w}).
The functions $w_\al$ are polynomials in $v_{XX}, v_{XXX}$ but they 
are {\it rational} functions in $v_X$. Remarkably, all the 
denominators will {\it disappear} from the deformed hierarchy.\par

We will prove that the corrections are polynomials in $t_X, t_{XX},
t_{XXX}$ for the case of semisimple Frobenius manifold (see the 
definition in Sect. 2 above).
Observe that $M_{\al\beta}$ is the
matrix of multiplication by the vector $\pal_X t$. So the determinant
$\det M_{\al\beta}$ vanishes identically on the nilpotent part of the
algebra $A_t$.
\par

First we observe that  the correction $K^{(1)}_{\beta,p}$ can be subdivided
into two parts
\beq
K^{(1)}_{\beta,p}=K'_{\beta,p}+K''_{\beta,p}
\eeq
where  $K'_{\beta,p}$ is the contribution of the first term in the 
r.h.s. of (\ref{Def-w}), and $K''_{\beta,p}$ is the contribution 
of the second term respectively. The main difficulty is in the 
computation of $K'_{\beta,p}$.

\begin{theorem} There exists a unique hierarchy of the form
\eqa\label{eq-M}
&&\frac{\pal t}{\pal T^{\beta,p}}=K^{(0)}_{\beta,p}(t,\tx) 
+\ve^2\,\left[ K'_{\beta,p;\lm}(t)\,\txxx^\lm+
K'_{\beta,p;\lm\mu}(t)\,\txx^\lm\,\tx^\mu\right.
\nn\\
&&\quad+\left.
K'_{\beta,p;\lm\mu\nu}(t)\,\tx^\lm\,\tx^\mu\,\tx^\nu\right]
\eeqa
such that the function $t(T)=(t_1(T),\dots,t_n(T))$
satisfies (\ref{eq-M}) up to terms of order $\ve^4$ for an arbitrary
solution $v(T)$ of (\ref{h-eq})
\beq\label{Back-s}
t_\al(T)=v_\al(T)+\frac{\ve^2}{24}\,\frac{\pal^2}{\pal T^{\al,0}\,\pal T^{1,0}}
\left[\log\det M_{\al\beta}(t,t_X)\right]_{t=v(T)}.
\eeq
The coefficients $K'_{\beta,p;\lm\mu\nu}(t), K'_{\beta,p;\lm\mu}(t),
 K'_{\beta,p;\lm}(t)$ of the hierarchy are analytic functions on the
Frobenius manifold.

The hierarchy (\ref{eq-M}) admits a representation
\beq\label{eq-FHS}
\frac{\pal t}{\pal T^{\beta,p}}=\left\{t(X),H_{\beta,p}+\ve^2\,
\de H'_{\beta,p}\right\}'_1+{\cal O}(\ve^4)
\eeq
where the perturbation of the first Poisson bracket has the form
\eqa
\lefteqn{
\{t^\al(X),t^\beta(Y)\}'_1=}\nn \\
&&
 \{t^\al(X),t^\beta(Y)\}^{(0)}_1+
\frac{\varepsilon^2}{24}\,\left(
\eta^{\mu\nu}\,c^{\al\beta}_{\mu\nu}(t(X))+
 \eta^{\mu\nu}\,c^{\al\beta}_{\mu\nu}(t(Y))\right)\,\de'''(X-Y)
\nn \\
&&- \frac{\varepsilon^2}{24}\,
\left[\eta^{\mu\nu}\,\pal_X^2(c^{\al\beta}_{\mu\nu}(t(X)))+
\eta^{\mu\nu}\,\pal_Y^2(c^{\al\beta}_{\mu\nu}(t(Y)))\right]\,\delta'(X-Y)
+{\cal O}(\ve^4).\label{first-bracket}
\eeqa
The operation $\{\ ,\ \}'_1$ is skew-symmetric and it satisfies the
Jacobi identity modulo ${\cal O}(\ve^4)$. The perturbations of the 
Hamiltonians have the form 
\beq\label{H-go}
\de H'_{\beta,p}=\int \chi_{\beta,p+1;\mu\nu}(t(X))\,\tx^\mu\,\tx^\nu\,dX,
\eeq
where
$\chi_{\beta,p;\mu\nu}=\chi_{\beta,p;\nu\mu}$ are
given by
\eqa
&&\chi_{\al,0;\mu\nu}=0,
\nn\\
&&\chi_{\al,p+1;\mu\nu}=\frac1{24}\,w^\ga_{\mu\nu}\,
\frac{\pal\theta_{\al,
p}}{\pal t^\ga}-\frac1{24}\,c^{\ga}_{\xi\zeta}\,c^{\xi\sigma}_\nu\,
c^\zeta_{\sigma\mu}\,\frac{\pal \theta_{\al,p-1}}{\pal t^\ga},
\quad p\ge 0.\label{CHI}
\eeqa
Here $\theta_{\al,-1}=0$ and $w^\al_{\mu\nu}$ are defined by
\beq\label{w3}
w^\al_{\beta\ga}=c^{\mu\nu}_{\mu\beta}\,c^\al_{\nu\ga}-
c^{\mu\nu}_\ga\,c^\al_{\mu\nu\beta}
=c^{\mu\nu}_{\mu\beta}\,c^\al_{\nu\ga}+c^{\mu\nu}_{\mu\ga}\,c^\al_{\nu\beta}-
\pal_\mu\,(c^\nu_{\beta\ga}\,c^{\al\mu}_\nu).
\eeq
The Hamiltonians $H_{\beta,p}+\ve^2\,\de H'_{\beta,p}$ commute pairwise
 modulo ${\cal O}(\ve^4)$ w.r.t. the bracket (\ref{first-bracket}).\par
\end{theorem}

Here and below
$c^{\al}_{\ga\mu}, c^{\al}_{\ga\mu\nu}, c^{\al}_{\ga\mu\nu\sigma},
 c^{\al\beta}_{\ga\mu}, c^{\al\beta}_{\ga\mu\nu}, 
c^{\al\beta}_{\ga\mu\nu\sigma}
$ are obtained by taking derivatives of the function 
$F(t^1,\dots,t^N)$ with respect to the coordinates $t^1,\dots,t^N$ 
and by using $\eta^{\al\beta}$ to raise the indices,
for example,
\beq\label{Def-ofc}
c^{\al\beta}_{\ga\mu}=\eta^{\al{\al'}}\,\eta^{\beta{\beta'}}\,
\frac{\pal^4 F(t)}{\pal t^{\al'}\,\pal t^{\beta'}\,\pal t^\ga\,
\pal t^\mu}.
\eeq

{\bf Remark.} $\ $ The first theorem does not use the quasihomogeneity 
condition (\ref{wdvv3}). The next theorem does use it.

\begin{theorem} The following formulae give the perturbation of the
second Poisson bracket:
 \eqa
\lefteqn{
\{t^\al(X),t^\beta(Y)\}'_2=\{t^\al(X),t^\beta(Y)\}^{(0)}_2
}\nn\\
&&+\varepsilon^2\,\left[
h^{\al\beta}(t(X))\,\de'''(X-Y)+
r^{\al\beta}_\ga(t(X))\,t^\ga_X\,
\delta''(X-Y)\right.
\nn\\
&&+\left(f^{\al\beta}_\ga(t(X))\,t^\ga_{XX}
+q^{\al\beta}_{\ga\mu}(t(X))\,t^\ga_X\,t^\mu_X\right)\,
\delta'(X-Y)
\nn\\
&&\left.+\left(b^{\al\beta}_{\ga\mu}(t(X))\,t^\ga_X\,t^\mu_{XX}
+a^{\al\beta}_{\ga\mu\nu}(t(X))\,t^\ga_X\,t^\mu_X\,t^\nu_X
+p^{\al\beta}_\ga(t(X))\,t^\ga_{XXX}\right)\,\delta(X-Y)\right]
\nn\\
&&+{\cal O}(\ve^4).\label{second-bracket}
\eeqa
where
\eqa\label{EQh}
h^{\al\beta}&=&\frac1{12}\,\left(\pal_\nu
(g^{\mu\nu}\,c^{\al\beta}_\mu)+
\frac12\, c^{\mu\nu}_\nu\,c^{\al\beta}_\mu\right),\\
p^{\al\beta}_\ga&=&\frac1{12}\,(\frac{1}2-\mu_\beta)\,c^{\al\beta}_{\mu\nu}\,
c^{\mu\nu}_\ga,\label{ep}\\
a^{\al\beta}_{\ga\mu\nu}&=&\frac1{72}\,(\frac{1}2-\mu_\beta)\,
\left(\eta^{\al\sigma}\,(\pal_\sigma\,\pal_\nu\,w^\beta_{\ga\mu}+
\pal_\sigma\,\pal_\ga\,w^\beta_{\mu\nu}\right.
\nn \\ &&
+ \pal_\sigma\,\pal_\mu\,w^\beta_{\ga\nu}-2\, \pal_\mu\,\pal_\nu\,
w^\beta_{\ga\sigma}-2\, \pal_\mu\,\pal_\ga\,w^\beta_{\nu\sigma}-
2\, \pal_\nu\,\pal_\ga\,w^\beta_{\mu\sigma})
\nn \\ 
&&
+\eta^{\xi\zeta}\,(6\,c^{\al\sigma}_{\xi\zeta}\,
c^\beta_{\sigma\ga\mu\nu}+
 3\,c^{\al\sigma}_{\xi\zeta\ga}\,c^\beta_{\sigma\mu\nu}+
3\,c^{\al\sigma}_{\xi\zeta\mu}\,c^\beta_{\sigma\ga\nu}
\nn \\ &&\left.
+3\,c^{\al\sigma}_{\xi\zeta\nu}\,c^\beta_{\sigma\ga\mu}+
  c^{\al\sigma}_{\xi\zeta\ga\mu}\,c^\beta_{\sigma\nu}+
 c^{\al\sigma}_{\xi\zeta\ga\nu}\,c^\beta_{\sigma\mu}+
 c^{\al\sigma}_{\xi\zeta\mu\nu}\,c^\beta_{\sigma\ga})\right),\label{ea}\\
b^{\al\beta}_{\ga\mu}&=&\frac1{12}\,(\frac{1}2-\mu_\beta)\,
\left(\eta^{\al\sigma}\,(-2\,\pal_\ga\,w^\beta_{\mu\sigma}+
\pal_\sigma\,w^\beta_{\ga\mu}-\pal_\mu\,w^\beta_{\ga\sigma})\right.
\nn \\ &&\left.
+\eta^{\xi\zeta}\,(3\,c^{\al\sigma}_{\xi\zeta}\,c^\beta_{\sigma\ga\mu}+
\frac32\, 
c^{\al\sigma}_{\xi\zeta\ga}\,c^\beta_{\sigma\mu}+\frac12\, 
c^{\al\sigma}_{\xi\zeta\mu}\,c^\beta_{\sigma\ga})\right),\label{eb}
\\
r^{\al\beta}_\ga&=&\frac32\,\pal_\ga h^{\al\beta}+
\frac1{24} \left(\frac{3}2-\mu_\beta\right)
c^{\al\nu}_\ga\,c^{\beta\mu}_{\nu\mu}-
\frac1{24}\left(\frac{3}2-\mu_\al\right)
c^{\beta\nu}_\ga\,c^{\al\mu}_{\nu\mu},\label{expofr}
\\
f^{\al\beta}_\ga &=& r^{\al\beta}_\ga+ p^{\al\beta}_\ga+
p^{\beta\al}_\ga-\pal_\ga h^{\al\beta},\label{expoff}
\\
q^{\al\beta}_{\ga\mu}&=& 
\frac12\,(b^{\al\beta}_{\mu\ga}+b^{\beta\al}_{\mu\ga})+
\frac12\,\pal_\ga r^{\al\beta}_\mu+\frac12\,
\pal_\mu r^{\al\beta}_\ga-
\pal_\mu\pal_\ga h^{\al\beta}-\nn\\
&& -\frac12\,\pal_\mu (p^{\al\beta}_\ga+
p^{\beta\al}_\ga).\label{eq}
\eeqa
The Jacobi identity for a linear combination
\beq
a_1\,\{\ ,\ \}'_1+a_2\,\{\ ,\ \}_2'
\eeq
with arbitrary constant coefficients $a_1, a_2$ holds true modulo
${\cal O}(\ve^4)$. The equations of the perturbed hierarchy for
those $(\beta,p)$  for which
\beq
p+\mu_\beta+\frac12\ne 0
\eeq
are Hamiltonian flows also w.r.t. the second Poisson bracket 
(\ref{second-bracket}) with the Hamiltonian
\beq
{\hat H}'_{\beta,p}= \sum_{k,l} (-1)^k \left(R_{p-l,k}\right)^\epsilon_\beta
\frac{H_{\epsilon,l-1}+\ve^2\,\de H'_{\epsilon,l-1}}
{(p+\mu_\al+\frac12)^{k+1}}.
\eeq
Here $R_{l,k}$ are defined in (\ref{matR}).
\end{theorem}

The proofs will be given in Sect. 5. The deformations 
(\ref{first-bracket}) and (\ref{second-bracket})
of the Poisson brackets
are obtained by applying the same infinitesimal B\"{a}cklund
transform (\ref{Back-s}) to the Poisson brackets (\ref{fpb}) and (\ref{spb}) 
resp. 
We prove that after this transform each of
the deformed Poisson bracket is a combination of $\de(X-Y), \dots,
\de'''(X-Y)$ with the coefficients being polynomial in $\tx, \txx,\txxx$.
Coefficients of these polynomials are functions analytic on the Frobenius
manifold. 
(Assuming the Frobenius manifold to be analytic itself).
Applying similar procedure to the Hamiltonians (\ref{H-gz}) 
we obtained the
deformed Hamiltonians (\ref{H-go}). 
The structure (\ref{eq-M}) of the deformed hierarchy 
follows from the fromulae (\ref{eq-FHS})--(\ref{CHI}). 
Finally, the same infinitesimal 
B\"{a}cklund transform gives the deformation of the linear pencil 
(\ref{pencil}) of the 
Poisson brackets.

We describe now the effect of adding of the second term in the
formula (\ref{Def-w}). 
At the moment we consider $G(t)$ as an arbitrary function
on some domain in the Frobenius manifold. We will compute 
this function in Theorem 3 below.

\begin{prop}\label{prop4}
Inserting an arbitrary function $G(t)$ in (\ref{Def-w}) 
we preserve the structure 
of the hierarchy, 
of the Hamiltonian, and of the Poisson brackets. 
The Hamiltonians get a correction $\ve^2\,\de H''_{\al,p}$ with
\beq
\de H''_{\al,p}=
c^\ga_{\xi\nu}\,c^{\sigma\xi}_\mu\,\frac{\pal\theta_{\al,p}}
{\pal t^\ga}\,\frac{\pal G}{\pal t^\sigma}.
\eeq 
The deformations of the first and of the second Poisson brackets get the 
correction
$\ve^2\,\{\ ,\ \}''_1$ and $\ve^2\,\{\ ,\ \}''_1$ with
\eqa
\lefteqn{
\{t^\al(X),t^\beta(Y)\}''_1=}
\nn\\
&&{\tilde a}^{\al\beta}(t(X))\,
\de'''(X-Y)+{\tilde{b}}^{\al\beta}(t(X))\,\de''(X-Y)
\nn\\
&&+{\tilde e}^{\al\beta}(t(X))\,\de'(X-Y),
\eeqa
\eqa
\lefteqn{
\{t^\al(X),t^\beta(Y)\}''_2=}
\nn\\&&
a^{\al\beta}(t(X))\,
\de'''(X-Y)+b^{\al\beta}(t(X))\,\de''(X-Y)
\nn\\
&&+e^{\al\beta}(t(X))\,\de'(X-Y)+\pal_X(q^{\al\beta}(t(X)))\,\de(X-Y),
\eeqa
where
\eqa
{\tilde a}^{\al\beta}&=&2\,c^{\al\beta\mu}\,\frac{\pal G(t)}{\pal t^\mu},
\nn\\
{\tilde b}^{\al\beta}&=&\frac32\,\pal_X {\tilde a}^{\al\beta}+
\frac{\pal^2 G}{\pal t^\sigma\pal t^\rho}
\left(c^{\al\sigma}_\mu\,\eta^{\beta\rho}-c^{\beta\sigma}_\mu\,\eta^{\al\rho}
\right)\,\tx^\mu,
\nn\\
{\tilde e}^{\al\beta}&=&\pal_X {\tilde b}^{\al\beta}-
\pal_X^2 {\tilde a}^{\al\beta},
\nn\\
a^{\al\beta}&=&2\,c^{\al\mu}_\ga\,g^{\ga\beta}\,\frac{\pal G(t)}{\pal t^\mu},
\\
b^{\al\beta}&=&\frac32\,\pal_X a^{\al\beta}+
\left(c^{\al\rho}_\mu\,g^{\ga\beta}-
c^{\beta\rho}_\mu\,g^{\ga\al}\right)\,\frac{\pal^2 G(t)}
{\pal t^\ga\,\pal t^\rho}\,\tx^\mu
+(\mu_\al-\mu_{\beta})\,
c^{\al\rho}_\ga\,c^{\ga\beta}_\mu\,\frac{\pal G(t)}{\pal t^\rho}\,
\tx^\mu,
\nn\\
q^{\al\beta}&=&\left(\frac{1}2-\mu_\beta\right)\left(
\frac{\pal^2 G}{\pal t^\sigma \pal t^\rho} c^{\al\sigma}_\mu\,
c^{\beta\rho}_\nu-
\frac{\pal^2 G}{\pal t^\sigma \pal t^\nu} c^{\beta\sigma}_\rho\,
c^{\rho\al}_\mu+\frac{\pal G}{\pal t^\sigma}
c^{\al\beta}_{\rho\mu}\,c^{\rho\sigma}_\nu\right)\tx^\mu\,\tx^\nu,
\nn\\
e^{\al\beta}&=&q^{\al\beta}+q^{\beta\al}+\pal_X b^{\al\beta}-
\pal_X^2 a^{\al\beta}.\nn
\eeqa
\end{prop}

The full Poisson brackets of the one-loop deformed hierarchy are $\{\ ,\
\}_1=\{\ ,\ \}'_1+\{\ ,\ \}''_1$ and 
$\{\ ,\ \}_2=\{\ ,\ \}'_2+\{\ ,\ \}''_2$.
\par

For the case of quantum cohomology the function $G(t)$ must be the
generating function of the elliptic Gromov - Witten invariants of the
target space. The recursion relations for the elliptic Gromov - Witten
invariants were found by Getzler \cite{Getz}. He proved that the
generating function $G(t)$ must satisfy a complicated system of
differential equations (see (\ref{Getz}) below). This system makes sense on an
arbitrary Frobenius manifold. Our next result is the solution of this
system on an arbitrary semisimple Frobenius manifolds.
\begin{theorem}
For an arbitrary semisimple Frobenius manifold the system (\ref{Getz}) has a
unique, up to an additive constant, solution $G=G(t^2, \dots, t^n)$
satisfying the quasihomogeneity condition 
\beq
\label{anomaly}
{\cal L}_E\,G=\ga
\eeq
with a constant $\ga$. This solution is given by the formula
\beq
\label{egregia}
G=\log\frac{\tau_I}{J^{1/24}}
\eeq
where $\tau_I$ is the isomonodromic tau-function (\ref{tau-i}) and
\beq
J=\det \left(\frac{\partial t^\alpha}{\partial u^i} \right)
\eeq
is the Jacobian of the transform from the canonical coordinates to
the flat ones. The scaling anomaly $\ga$ in (\ref{anomaly}) is given by
the formula
\beq
\label{gamma}
\ga =-\frac{1}{4} \sum_{\alpha=1}^n \mu_\alpha^2 +\frac{n\, d}{48}
\eeq
where
\beq
\mu_\alpha = q_\alpha -\frac{d}{2}, ~\alpha=1, \dots, n.
\eeq
\end{theorem}

\begin{cor}
For $d\ne 1$ 
the variable $T(X)=\frac2{1-d}\,t^n(X)$ has
the Virasoro Poisson bracket
\beq
\{T(X),T(Y)\}_2=\left[ T(X)+T(Y)\right]\,\de'(X-Y)+
\ve^2\,\frac{c}{12}\,\de'''(X-Y)
\eeq
with the central charge
\beq\label{cen-ch}
c\, \varepsilon^2 =\frac{12 \varepsilon^2}{(1-d)^2} \left[ \frac{1}{2} n - 2
\sum_{\alpha=1}^n \mu_\alpha^2\right].
\eeq
\end{cor}
 
So, the bihamiltonian structure of the conjectured integrable hierarchy
at the one-loop approximation looks like a classical $W$-algebra with the
central charge (\ref{cen-ch}) and the conformal weights (\ref{conf-dim}).

\setcounter{equation}{0}
\section{Some formulas related to the canonical 
coordinates of Frobenius manifold}

The canonical
coordinates on a semisimple Frobenius manifold  $M^n$ are denoted 
by $(u_1,\dots,u_n) $. They satisfy the multiplication table 
\beq
\frac{\pal}{\pal u_i}\cdot\frac{\pal}{\pal u_j}=\de_{ij}\,\frac{\pal}
{\pal u_i}.
\eeq
The invariant metric becomes diagonal in the canonical coordinates,
i.e. $<\ ,\ >=\sum_{i=1}^n \eta_{ii}(u) d u_i^2$.  We assume that 
the unit vector field of the Frobenius manifold is 
$e=\frac{\pal}{\pal t^1}$, then the {\it rotation 
coefficients} $\gamma_{ij}(u)$ are defined by
\beq
\gamma_{ij}(u)=\frac{\pal_j\,\sqrt{\eta_{ii}(u)}}{\sqrt{\eta_{jj}(u)}}=
\frac12\,\frac{\pal_i\,\pal_j\,t_1(u)}{\sqrt{\pal_i\,t_1(u)\,\pal_j\,t_1(u)}},
\quad for  \ i\ne j,
\eeq
where $\pal_i=\frac{\pal}{\pal u_i}$. They are symmetric with respect
to their indices and satisfy the following equations:
\eqa
&& \frac{\pal \ga_{ij}}{\pal u_k}=\ga_{ik}\,\ga_{kj},\quad i, j, k \ distinct,
\\
&&\sum_{k=1}^N \frac{\pal \ga_{ij}}{\pal u_k}=0.
\eeqa
Define
\beq
\psi_{i\al}(u)=\frac{\pal_i t_\al(u)}{\sqrt{\eta_{ii}(u)}}.
\eeq
The matrix $(\psi_{i\al})$
satisfies the following identities:
\beq\label{eOrth}
\sum_{l=1}^n \psi_{l\al}\,\psi_{l\beta}=\eta_{\al\beta},\quad
\sum_{l=1}^n \psi_{l}^\al\,\psi_{l}^\beta=\eta^{\al\beta},
\eeq
where $\psi_j^\al=\psi_{j\ga}\,\eta^{\ga\beta}$. 
We list here the following useful identities: (see \cite{D3})
\eqa
c_{\al\beta\ga}&=&\sum_{i=1}^n \frac{\psi_{i\al}\,\psi_{i\beta}\,\psi_{i\ga}}
{\psi_{i1}},\label{C3}\\
\frac{\pal t^\al}{\pal u_i}&=&\psi_{i1}\,\psi^\al_i,\quad
\frac{\pal u_i}{\pal t^\al}=\frac{\psi_{i\al}}{\psi_{i1}},\\
\frac{\pal \psi_{i\al}}{\pal u_k}&=&\ga_{ik}\,\psi_{k\al},\quad i\ne k,
\quad \frac{\pal \psi_{i\al}}{\pal u_i}=-\sum_k \ga_{ik}\psi_{k\al}
\label{dpsi-du}\\
\frac{\pal \psi_{i\al}}{\pal t^\beta}&=&\sum_{l=1}^N \ga_{il}\,
\psi_{l\al}\,(\frac{\psi_{l\beta}}{\psi_{l1}}-
\frac{\psi_{i\beta}}{\psi_{i1}}),
\eeqa
Denote
\beq
\sigma_i=\psi_{i\al}\,\tx^\al.
\eeq
Then we have
\beq
\frac{\pal \sigma_i}{\pal t^\al}=\sum_{l=1}^n \ga_{il}\,\sigma_l\,
(\frac{\psi_{l\al}}{\psi_{l1}}-\frac{\psi_{i\al}}{\psi_{i1}}).
\eeq
Let's consider the matrix
\beq
A=(\psi_i^\al\,\psi_j^\beta\,c_{\al\beta\ga}\,\tx^\ga),
\eeq
by using the identity (\ref{eOrth}) we have
\beq
\det A=\det(\eta^{\al\beta})\,\det(c_{\al\beta\ga}\,\tx^\ga).
\eeq
We see that the matrix $c_{\alpha\beta\gamma}t^\gamma_X$ diagonalizes
in the canonical coordinates.
From  
(\ref{C3}) we see that the following expression for ${\cal F}^{(1)}(t,\tx)$ 
holds 
true:
\eqa
&&\F^{(1)}(t,\tx)=\frac1{24}\,\log\det(c_{\al\beta\ga}\,\tx^\ga)+G(t)\nn\\
&=&-\frac1{24}\,\log\det(\eta^{\al\beta})+\frac1{24}\,\log\det(A)+G(t)\nn\\
&=&-\frac1{24}\,\log\det(\eta^{\al\beta})+
\frac1{24}\,\log\det(\psi^\al_i\,\psi^\beta_j\sum_k\frac{\psi_{k\al}\,
\psi_{k\beta}\,\psi_{k\ga}}{\psi_{k1}}\,\tx^\ga)+G(t)\nn\\
&=&\frac1{24}\,\log(\prod_{l=1}^n \psi_{l\ga}\,\tx^\ga)-
\frac1{24}\,\log(\prod_{l=1}^n \psi_{l1})+G(t)-
\frac1{24}\,\log\det(\eta^{\al\beta}),\nn\\
&=&\frac1{24}\,\log\prod_{l=1}^n \sigma_l-
\frac1{24}\,\log(\prod_{l=1}^n \psi_{l1})+G(t)-
\frac1{24}\,\log\det(\eta^{\al\beta}) \label{F1s}
\eeqa
This expression of the function ${\cal F}^{(1)}$ 
is crucial in the proof of Theorems 1, 2.

Let's denote
\eqa\label{F-G0}
&&\f :=\frac1{24}\,\log\det(c_{\al\beta\ga}\,\tx^\ga)
\nn\\
&&=\frac1{24}\,\log\prod_{l=1}^n \sigma_l-
\frac1{24}\,\log(\prod_{l=1}^n \psi_{l1})-
\frac1{24}\,\log\det(\eta^{\al\beta}).
\eeqa
By a direct calculation we get also the following formulas:
\eqa\label{ftx}
\f_{\tx^\al}&=&\frac1{24}\,\sum_{i=1}^n \frac{\psi_{i\al}}{\sigma_i},\\
\f_{\tx^\al\tx^\beta}&=&-\frac1{24}\,\sum_{i=1}^n \frac{\psi_{i\al}\,
\psi_{i\beta}}{\sigma_i^2},\label{ftxtx}\\
\f_{\tx^\al t^\beta}&=&\frac1{24}\,\sum_{j,k=1}^n \frac{\ga_{jk}\,
\psi_{k\al}}{\sigma_j}\,\left(\frac{\psi_{k\beta}}{\psi_{k1}}-
\frac{\psi_{j\beta}}{\psi_{j1}}\right)\nn \\
&& 
-\frac1{24}\,\sum_{j,k,\ga=1}^N \frac{\ga_{jk}\,\sigma_k\,
\psi_{j\al}}{\sigma_j^2}\,\left(\frac{\psi_{k\beta}}{\psi_{k1}}-
\frac{\psi_{j\beta}}{\psi_{j1}}\right),\label{ftxt} \\
\f_{t^\al}&=&\frac1{24}\,\sum_{i,j=1}^N \ga_{ij}\,
\left(\frac{\sigma_j}{\sigma_i}\,\left(\frac{\psi_{j\al}}{\psi_{j1}}-
\frac{\psi_{i\al}}{\psi_{i1}}\right)-
\frac{\psi_{j\al}}{\psi_{i1}}+\frac{\psi_{i\al}\,\psi_{j1}}
{\psi_{i1}^2}\right),\label{ft}\\
c^{\al\ga}_{\beta\mu}&=&\sum_{i,j=1}^N \ga_{ij}\,\left(
\frac{\psi^\al_i\,\psi_{i\beta}\,\psi^\ga_j}{\psi_{i1}}+
\frac{\psi^\al_i\,\psi_{i}^\ga\,\psi_{j\beta}}{\psi_{i1}}+
\frac{\psi^\ga_i\,\psi_{i\beta}\,\psi^\al_j}{\psi_{i1}}-
\frac{\psi^\ga_i\,\psi_{i}^\al\,\psi_{i\beta}\,\psi_{j1}}{\psi_{i1}^2}
\right)\nn\\
&&\times\left(\frac{\psi_{j\mu}}{\psi_{j1}}-
\frac{\psi_{i\mu}}{\psi_{i1}}\right).\label{C4}
\eeqa

All these formulae do not use the quasihomogeneity (\ref{wdvv3}).
In the quasihomogeneous case the canonical coordinates $u_1(t),
\dots,u_n(t)$ are the roots of the characteristic equation
\beq
\det(g^{\al\beta}(t)-u\,\eta^{\al\beta})=0.
\eeq
Here $g^{\al\beta}$ is the intersection form. The matrix 
$\ga_{ij}$ in this case has the form
\beq
\ga_{ij}=-(u_i-u_j)^{-1}\, V_{ij}
\eeq
where
\beq
V_{ij} =\sum_{\al=1}^n \mu_\al\,\psi_{i\al}\,
\psi_j^\al.
\eeq
The columns of the matrix $\Psi =(\psi_{i\alpha})$ are the eigenvectors
of the matrix $V$ with the eigenvalues $\mu_\alpha$. Particularly,
$\psi_{i1}$ is the eigenvector of $V$ with the eigenvalue $\mu_1=-d/2$.
It follows that
\beq
\pal_k \ga_{ij}= \ga_{ik} \ga_{kj}, k\neq i, j, \quad \pal_i \ga_{ij}=
\frac{\sum_k (u_j-u_k) \ga_{ik}\ga_{kj}- \ga_{ij}}{u_i-u_j}
\eeq
\eqa
\frac{\pal \ga_{ij}}{\pal t^\al}&=&\sum_{k=1}^N 
(\ga_{ik}\,\ga_{kj}\,\frac{\psi_{k\al}}{\psi_{k1}}+
\frac{u_j-u_k}{u_i-u_j}\,
\ga_{ik}\,\ga_{kj}\,\frac{\psi_{i\al}}{\psi_{i1}}+
\frac{u_i-u_k}{u_j-u_i}\,
\ga_{ik}\,\ga_{kj}\,\frac{\psi_{j\al}}{\psi_{j1}}
\nn\\
&&-\frac{\ga_{ij}}{u_i-u_j}\,\frac{\psi_{i\al}}{\psi_{i1}}-
\frac{\ga_{ij}}{u_j-u_i}\,\frac{\psi_{j\al}}{\psi_{j1}}).
\eeqa
We also write down the following useful formulae:
\eqa
&&\psi_{i\ga}\,g^{\ga\beta}=u_i\,\psi^\beta_i\label{scaling},
\\
&&(u^j-u^i)\gamma_{ij}=\sum_\al (q_\al-\frac d2)\,\psi_{i\al}\,
\psi^\al_j.\label{eOrth2}
\eeqa
\vskip 0.6cm
\setcounter{equation}{0}
\section{Proofs of theorem 1 and 2}\par
We begin with the proof of  Theorem 1. 
So we assume here  that $G=0$ in the formulae (\ref{F-BT}), (\ref{Def-w}).
Doing the infinitesimal B\"acklund tranform (\ref{F-BT}) (with
$G(t)=0$) we obtain
\eqa
&&\{t^\al(X),t^\beta(Y)\}^{'}_1=\eta^{\al\beta}\,\de'(X-Y)\nn\\
&&+\ve^2\left[\left(\frac{\pal w^\beta(t(Y))}{\pal t^\ga}+
\frac{\pal w^\beta(t(Y))}{\pal t^\ga_Y}\,\pal_Y+
\frac{\pal w^\beta(t(Y))}{\pal t^\ga_{XX}}\,\pal^2_X
\right.\right.
\nn\\
&&+\left.\frac{\pal w^\beta(t(Y))}{\pal t^\ga_{YYY}}\,\pal^3_Y\right)
\,\eta^{\al\ga}\,\de'(X-Y)+
\left(\frac{\pal w^\al(t(X))}{\pal t^\ga}+
\frac{\pal w^\al(t(X))}{\pal t^\ga_X}\,\pal_X\right.
\nn\\
&&+\left.\left.\frac{\pal w^\al(t(X))}{\pal t^\ga_{XX}}\,\pal^2_X+
\frac{\pal w^\al(t(X))}{\pal t^\ga_{XXX}}\,\pal^3_X\right)
\, \eta^{\ga\beta}\,\de'(X-Y)\right]
+{\cal O}(\ve^4),\label{templ}
\eeqa
where $w^\al(t)=w^\al(t,\tx)$ is the function obtained from $w^\al(v,v_X)$ by
replacing $v^\mu$ and their $X$-derivatives by $t^\mu$ and by the 
correspondent $X$-derivatives of $t^\mu$. Recall that $w^\al(t)$ depends
not only on $t^\al$, but also on $\tx^\al, \txx^\al, \txxx^\al$.
More explicitly, we have
\eqa
w^\al(t)&=&\f_{\tx^\beta}\,c^{\al\beta}_\ga\,\txxx^\ga+
(\f_{\tx^\ga\,t^\beta}\,c^{\al\beta}_\mu+
\f_{\tx^\beta\,t^\mu}\,c^{\al\beta}_\ga+
3\,\f_{\tx^\beta}\,c^{\al\beta}_{\ga\mu})\,\txx^\ga\,\tx^\mu\nn \\
&&+\f_{\tx^\beta\tx^\mu}\,c^{\al\beta}_\ga\,\txx^\ga\,\txx^\mu
+\f_{\tx^\beta\tx^\nu}\,c^{\al\beta}_{\ga\mu}\,
\tx^\ga\,\tx^\mu\,\txx^\nu+\f_{t^\beta}\,c^{\al\beta}_\ga\,\txx^\ga \nn \\
&&+(\f_{\tx^\beta t^\nu}\,c^{\al\beta}_{\ga\mu}+\f_{\tx^\beta}\,
c^{\al\beta}_{\ga\mu\nu})\,\tx^\ga\,\tx^\mu\,\tx^\nu 
+(\f_{t^\beta t^\mu}\,c^{\al\beta}_{\ga}+\f_{t^\beta}\,
c^{\al\beta}_{\ga\mu})\,\tx^\ga\,\tx^\mu,\label{w}
\eeqa
where $\f=\f(t,t_X)$ is defined in (\ref{F-G0}). 
Whenever there is no risk of confusion
we will omit the
arguments of a function henceforth.

In the Poisson bracket $\{t^\al(X),t^\beta(Y)\}^{'}_1$, the coefficient of
$\ve^2\, \de^{(4)}(X-Y)$ is equal to zero, so it can be written as
\eqa
&&\{t^\al(X),t^\beta(Y)\}^{'}_1=\eta^{\al\beta}\,\de'(X-Y)+
\ve^2\,
\left(
{\hat{h}}^{\al\beta}\,\de^{(3)}(X-Y)
\right.
\nn\\
&&+\left. {\hat{r}}^{\al\beta}\,\de''(X-Y)+
{\hat{f}}^{\al\beta}\,\de'(X-Y)+
{\hat{p}}^{\al\beta}\,\de(X-Y) \right),\label{fpbg}
\eeqa
where ${\hat h}^{\al\beta}, 
{\hat r}^{\al\beta}, {\hat f}^{\al\beta}, {\hat p}^{\al\beta}$
are functions of $t^\mu(X)$ and their $X$-derivatives.

We have the following two lemmas on the coefficients $ {\hat h}^{\al\beta}$
and ${\hat r}^{\al\beta}$:
\begin{lemma} \label{lemma1}
The coefficients ${\hat h}^{\al\beta}$  
have the expression
\beq\label{fh}
{\hat h}^{\al\beta}
=\frac1{12}\,\eta^{\mu\nu}\,c^{\al\beta}_{\mu\nu}.
\eeq
\end{lemma}
\begin{lemma}\label{lemma2}
 The coefficients ${\hat r}^{\al\beta}$ are symmetric w.r.t.
$\al$ and $\beta$, i.e.,
\beq\label{anti-r}
{\hat r}^{\al\beta}={\hat r}^{\beta\al}.
\eeq
\end{lemma}
\vskip 0.4cm
The proofs of Lemma \ref{lemma1} and Lemma \ref{lemma2}  are
 similar to those of Lemma \ref{Lh} and Lemma \ref{symofr}
which will be given below, however, the computation is much more simple, 
so we omit the proofs here.
\begin{lemma}\label{Lham}
For the Hamiltonians $H_{\al,p}$ defined in
(\ref{H-gz}) the following identity holds true:
\beq\label{f-HP}
H_{\al,p}=\int \theta_{\al,p+1}(v(X)) dX=
\int \theta_{\al,p+1}(t(X)) dX+\ve^2\,\de H'_{\al,p}+{\cal O}(\ve^4),
\eeq
where $\de H'_{\al,p}$ are defined by (\ref{H-go}) and (\ref{CHI}).
\end{lemma}

\pf
By using (\ref{odefh}) we have
\eqa
\lefteqn{
H_{\al,p}=\int \theta_{\al,p+1}(v(X))\,dX}
\nn\\
&=&\int \theta_{\al,p+1}(t(X))\,dX-\ve^2\,\int \frac{\pal \theta_{\al,p+1}
(t(X))}{\pal t^\mu}\,w^\mu\,dX+{\cal O}(\ve^4)
\nn\\
&=&\int \theta_{\al,p+1}(t(X))\,dX+\ve^2\,\int \frac{\pal^2 \theta_{\al,p+1}
(t(X))}{\pal t^\mu\,\pal t^\nu}\,\tx^\nu\,
\frac{\pal \f(t(X),\tx(X))}{\pal T_{\mu,0}}\,dX+{\cal O}(\ve^4)
\nn\\
&=&\int \theta_{\al,p+1}(t(X))\,dX+
\ve^2\,\int c^\gamma_{\mu\nu}\,\frac{\pal\theta_{\al,p}}{\pal t^\gamma}\,
\tx^\nu\,
\left( \f_{t^\sigma}\,c^{\sigma\mu}_\rho\,\tx^\rho+
\f_{\tx^\sigma}\,c^{\sigma\mu}_{\rho\xi}\,\tx^\rho\,\tx^\xi
\right.
\nn\\
&&+\left.\f_{\tx^\sigma}\,c^{\sigma\mu}_{\rho}\,\txx^\rho\right)dX
+{\cal O}(\ve^4).
\eeqa
Now formulas (\ref{C3}), (\ref{ftx}), (\ref{ft}) and (\ref{C4}) 
amount to
\eqa
\lefteqn{
c^\gamma_{\mu\nu}\,\frac{\pal\theta_{\al,p}}{\pal t^\gamma}\,
\tx^\nu\,
\left( \f_{t^\sigma}\,c^{\sigma\mu}_\rho\,\tx^\rho+
\f_{\tx^\sigma}\,c^{\sigma\mu}_{\rho\xi}\,\tx^\rho\,\tx^\xi\right)}
\nn\\
&=&
\frac1{24}\,\gamma_{ij}\,\frac{\pal\theta_{\al,p}}{\pal t^\gamma}
\left(2\,\sigma_i^2\,\frac{\psi_i^\gamma\,\psi_{j1}}{\psi_{i1}^4}+
\sigma_j^2\,\frac{\psi_j^\gamma}{\psi_{i1}\,\psi_{j1}^2}
+\sigma_j^2\,\frac{\psi_i^\gamma}{\psi_{i1}^2\,\psi_{j1}}\right.
\nn\\
&&-\left.3\,\sigma_i\,\sigma_j\,\frac{\psi_i^\gamma}{\psi_{i1}^3}-
\sigma_i\,\sigma_j\,\frac{\psi_j^\gamma}{\psi_{i1}^2\,\psi_{j1}}\right),\nn
\eeqa
and
\eqa
\lefteqn{
\int c^\gamma_{\mu\nu}\,\frac{\pal\theta_{\al,p}}{\pal t^\gamma}\,
\tx^\nu\,
\f_{\tx^\sigma}\,c^{\sigma\mu}_{\rho}\,\txx^\rho\,dX}
\nn\\
&=&\frac1{24}\int\frac{\pal\theta_{\al,p}}{\pal t^\gamma}\,
\frac{\psi^\ga_j\,\psi_{j\mu}\,\psi_{j\nu}}{\psi_{j1}}\,t^\nu_X\,
\frac{\psi_{i\sigma}}{\sigma_i}\,\frac{\psi_k^\sigma\,\psi^\mu_k\,\psi_{k\rho}}
{\psi_{k1}}\,\txx^\rho
\nn\\
&=&\frac1{24}\int\frac{\pal\theta_{\al,p}}{\pal t^\gamma}\,
\frac{\psi^\ga_j\,\psi_{j\mu}\,\sigma_{j}}{\psi_{j1}}\,
\frac{\psi_{i\sigma}}{\sigma_i}\,\frac{\psi_k^\sigma\,\psi^\mu_k\,\psi_{k\rho}}
{\psi_{k1}}\,\txx^\rho
\nn\\
&=&
\frac1{24}\,\int \frac{\pal\theta_{\al,p}}{\pal t^\gamma}\,
\frac{\psi_j^\gamma\,\psi_{j\rho}}{\psi_{j1}^2}\,\txx^\rho\,dX
=- \frac1{24}\,\int \frac{\pal}{\pal t^\nu}
\left(\frac{\pal\theta_{\al,p}}{\pal t^\gamma}\,
\frac{\psi_j^\gamma\,\psi_{j\rho}}{\psi_{j1}^2}\right)\,\tx^\rho\,
\tx^\nu\,dX.\nn
\eeqa
So the Hamiltonians $H^{\al,p}$ can be expressed in the form
(\ref{f-HP}), (\ref{H-go}) with
\eqa
\lefteqn{
\chi_{\al,p+1;\mu\nu}=\frac{\ga_{ij}}{24}\,\frac{\pal\theta_{\al,p}}
{\pal t^\ga}\left(\frac{2\,\psi_{i\mu}\,\psi_{i\nu}\,\psi^\ga_i\,\psi_{j1}}
{\psi_{i1}^4}+
\frac{\psi_{j\mu}\,\psi_{j\nu}\,\psi^\ga_j}
{\psi_{i1}\,\psi_{j1}^2}
+\frac{\psi_{j\mu}\,\psi_{j\nu}\,\psi^\ga_i}
{\psi_{i1}^2\,\psi_{j1}}\right.}
\nn\\
&&-
\frac{3\,\psi_{i\mu}\,\psi_{j\nu}\,\psi^\ga_i}
{2\,\psi_{i1}^3}-
\frac{3\,\psi_{i\nu}\,\psi_{j\mu}\,\psi^\ga_i}
{2\,\psi_{i1}^3}-
\frac{\psi_{i\mu}\,\psi_{j\nu}\,\psi^\ga_j}
{2\,\psi_{i1}^2\,\psi_{j1}}
\nn\\
&&\left.-
\frac{\psi_{i\nu}\,\psi_{j\mu}\,\psi^\ga_j}
{2\,\psi_{i1}^2\,\psi_{j1}}\right)-
\frac1{48}\,\frac{\pal}{\pal t^\nu}\left(
\frac{\pal \theta_{\al,p}}{\pal t^\ga}\,\frac{\psi^\ga_j\,
\psi_{j\mu}}{\psi_{j1}^2}\right)
-\frac1{48}\,\frac{\pal}{\pal t^\mu}\left(
\frac{\pal \theta_{\al,p}}{\pal t^\ga}\,\frac{\psi^\ga_j\,
\psi_{j\nu}}{\psi_{j1}^2}\right)
\nn\\
&=&
\frac1{24}\,w^\ga_{\mu\nu}\,
\frac{\pal\theta^{\al,
p-1}}{\pal t^\ga}-\frac1{24}\,c^{\ga}_{\xi\zeta}\,c^{\xi\sigma}_\nu\,
c^\zeta_{\sigma\mu}\,\frac{\pal \theta_{\al,p-1}}{\pal t^\ga},
\quad p\ge -1,\nn
\eeqa
where $\theta_{\al,-2}=\theta_{\al,-1}=0$ 
and $w^\al_{\mu\nu}$ are defined in (\ref{w3}).
\epl

\vskip 0.5cm
\noindent{\bf Proof of Theorem 1 \ }
We first prove the formula (\ref{first-bracket}) for the
first Poisson bracket. From Lemma \ref{lemma1}
and Lemma \ref{lemma2} we already know the expression of the
coefficients ${\hat h}^{\al\beta}$
and the anti-symmetric part of the coefficients ${\hat r}^{\al\beta}$
in the formula (\ref{fpbg}). 
Now from the fact that the Casimirs of the first Poisson bracket 
$\int v^{\ga}(X)\,dX$ have the expression 
\beq\label{casimir}
\int v^{\ga}(X)\,dX
=\int t^{\ga}(X)\,dX +{\cal O}(\ve^4)
\eeq
we see that
\beq\label{equation0}
{\hat p}^{\al\beta}=0
\eeq
in the formula (\ref{fpbg}). 
So the anti-symmetry condition of the first Poisson bracket gives
us the following relations:
\eqa\label{equation1}
&&{\hat r}^{\al\beta}+{\hat r}^{\beta\al}=3\,\pal_X {\hat h}^{\al\beta},
\\
&&
{\hat f}^{\beta\al}-{\hat f}^{\al\beta}+2\,\pal_X {\hat r}^{\al\beta}=
3\,\pal_X^2 {\hat h}^{\al\beta},\label{equation2}
\\
&&
\pal_X {\hat f}^{\al\beta}+\pal_X^3\,{\hat h}^{\al\beta}-
\pal_X^2 {\hat r}^{\al\beta}=0.\label{lasteq}
\eeqa
Identity (\ref{equation1}) together with 
(\ref{anti-r}) gives us the expression for ${\hat r}^{\al\beta}$, 
while from the identity (\ref{lasteq}) it follows that
\beq
{\hat f}^{\al\beta}=\pal_X {\hat r}^{\al\beta}
-\pal_X^2 {\hat h}^{\al\beta},\label{equation3}
\eeq
there is no integration constant because ${\hat f}^{\al\beta}$
must depend on $\tx^\ga$ or $\txx^\ga$. So we get the expression for
the coefficients ${\hat f}^{\al\beta}$ 
and complete the proof of formula (\ref{first-bracket}).
The remaining part of the Theorem follows from 
Lemma \ref{Lham}. 
\ept 
\vskip 0.5cm
We now proceed to  prove Theorem 2. 
So we still assume here  that $G=0$ in the formulae (\ref{F-BT}), 
(\ref{Def-w}).
Doing the same infinitesimal B\"acklund tranform (\ref{F-BT}) (with
$G(t)=0$) we obtain
\eqa
&&\{t^\al(X),t^\beta(Y)\}^{'}_2=g^{\al\beta}(t(X))\,\de'(X-Y)+
        \Gamma^{\al\beta}_{\ga}(t(X))\, t^{\ga}_X\,\de(X-Y)\nn \\ 
&& -\ve^2\,\left[\frac{\pal g^{\al\beta}(t(X))}{\pal t^{\ga}}\, 
w^{\ga}(t(X))\, 
\de'(X-Y)+ \frac{\pal \Gamma^{\al\beta}_\ga (t(X))}{\pal 
t^\mu}\,w^\mu(t(X))\,t^{\ga}_X\,\de(X-Y)\right. \nn \\ 
&&+\Gamma^{\al\beta}_\ga (t(X))\,(\pal_X w^\ga(t(X))\,\de(X-Y))\nn \\ 
&&+(\frac{\pal w^\beta(t(Y))}{\pal t^\ga}+
\frac{\pal w^\beta(t(Y))}{\pal t^\ga_Y}\,\pal_Y+
\frac{\pal w^\beta(t(Y))}{\pal t^\ga_{YY}}\,\pal^2_Y+
\frac{\pal w^\beta(t(Y))}{\pal t^\ga_{YYY}}\,\pal^3_Y)
\nn \\ 
&& \ \ \times (g^{\al\ga}(t(X))\,\de'(X-Y)+\Gamma^{\al\ga}_{\mu}(t(X))\,
t^\mu_X\,\de(X-Y))\nn \\ 
&&+ (\frac{\pal w^\al(t(X))}{\pal t^\ga}+
\frac{\pal w^\al(t(X))}{\pal t^\ga_X}\,\pal_X+
\frac{\pal w^\al(t(X))}{\pal t^\ga_{XX}}\,\pal^2_X+
\frac{\pal w^\al(t(X))}{\pal t^\ga_{XXX}}\,\pal^3_X)
\nn \\ 
&& \left. \times (g^{\ga\beta}(t(X))\,\de'(X-Y)+\Gamma^{\ga\beta}_{\mu}(t(X))\,
t^\mu_X\,\de(X-Y))+{\cal O}(\ve^4)\right],\label{spbg}
\eeqa
In the Poisson bracket $\{t^\al(X),t^\beta(Y)\}_2$, the coefficient of
$\ve^2\, \de^{(4)}(X-Y)$ is equal to
\beq\label{ord4}
-\frac{\pal w^\beta(t(X))}{\pal \txxx^\ga}\,g^{\ga\al}+
\frac{\pal w^\al(t(X))}{\pal \txxx^\ga}\,g^{\ga\beta}=
-\f_{\tx^\mu}\,c^{\mu\beta}_\ga\,g^{\ga\al}+
\f_{\tx^\mu}\,c^{\mu\al}_\ga\,g^{\ga\beta}=0,
\eeq
the last equality above is due to the associativity equation
\beq\label{asso}
c^{\mu\beta}_\ga\,c^{\ga\al}_{\nu}=c^{\mu\al}_\ga\,c^{\ga\beta}_{\nu},
\eeq
and the definition (\ref{def-g}) of the intersection form.

The coefficient of $\ve^2\,\de^{(3)}(X-Y)$ is equal to
\eqa
h^{\al\beta}&=&
2\,\f_{t^\mu}\,c^{\al\mu}_\ga\,g^{\ga\beta}+3\,\f_{\tx^\mu}\,
c^{\mu\al}_{\ga\xi}\,g^{\beta\ga}\,\tx^\xi-\f_{\tx^\mu}\,
c^{\beta\mu}_{\ga\xi}\,g^{\al\ga}\,\tx^\xi-
2\,\f_{\tx^\mu t^\xi}\,c^{\al\mu}_\ga\,g^{\ga\beta}\,\tx^\xi\nn\\
&&+\f_{\tx^\ga t^\mu}\,(c^{\al\mu}_\xi\,g^{\ga\beta}+
c^{\beta\mu}_\xi\,g^{\ga\al})\,\tx^\xi+
\f_{\tx^\mu \tx^\ga}\,(c^{\mu\al}_{\nu\xi}\,g^{\ga\beta}+
c^{\mu\beta}_{\nu\xi}\,g^{\ga\al})\,\tx^\nu\,\tx^\xi \nn \\
&&+3\,\f_{\tx^\mu}\,c^{\al\mu}_\ga\,\frac{\pal g^{\ga\beta}}{\pal t^\xi}\,
\tx^\xi-\f_{\tx^\mu}\,c^{\beta\mu}_\ga\,\Gamma^{\al\ga}_\xi\,\tx^\xi+
\f_{\tx^\mu}\,c^{\al\mu}_\ga\,\Gamma^{\ga\beta}_\xi\,\tx^\xi+S^{\al\beta},
\label{defh}
\eeqa
where
\eqa
S^{\al\beta}&=&(\f_{\tx^\mu\tx^\xi}\,c^{\beta\mu}_\ga\,g^{\al\ga}+
\f_{\tx^\mu\tx^\xi}\,c^{\al\mu}_\ga\,g^{\beta\ga}+
\f_{\tx^\mu\tx^\ga}\,c^{\beta\mu}_\xi\,g^{\al\ga}+
\f_{\tx^\mu\tx^\ga}\,c^{\al\mu}_\xi\,g^{\beta\ga}\nn \\
&&-4\,\f_{\tx^\mu\tx^\xi}\,c^{\beta\mu}_\ga\,g^{\al\ga})\,\txx^\xi\nn\\
&=&(\f_{\tx^\mu\tx^\ga}\,c^{\beta\mu}_\xi\,g^{\al\ga}+
\f_{\tx^\mu\tx^\ga}\,c^{\al\mu}_\xi\,g^{\beta\ga}-
\f_{\tx^\mu\tx^\xi}\,c^{\beta\mu}_\ga\,g^{\al\ga}-
\f_{\tx^\mu\tx^\xi}\,c^{\al\mu}_\ga\,g^{\beta\ga})\,\txx^\xi.
\eeqa

\begin{lemma} 
$\quad  S^{\al\beta}=0$.
\end{lemma}
\pf
By using (\ref{def-g}), (\ref{C3}) and (\ref{ftxtx}) we have the identity
\beq
\f_{\tx^\mu\tx^\ga}\,c^{\beta\mu}_\nu\,g^{\al\ga}=
\f_{\tx^\mu\tx^\nu}\,c^{\al\mu}_\ga\,g^{\beta\ga},
\eeq
since both sides of the above identity are equal to
$$
-\frac1{24}\,\sum_i \frac{E^\xi\,\psi^\al_i\,\psi^\beta_i\,\psi_{i\nu}\,
\psi_{i\xi}}{\sigma_i^2\,\psi_{i1}^2}.
$$
The lemma follows from the above identity immediately.
\epl

\begin{lemma}\label{Lh}
The coefficients $ h^{\al\beta}$ defined in (\ref{defh}) have 
the expression
\beq\label{h}
h^{\al\beta}=\frac1{12}\,\left( \frac{\pal}{\pal t^\nu}\,(g^{\mu\nu}\,
c^{\al\beta}_{\mu})+\frac12\,c^{\mu\nu}_\nu\,c^{\al\beta}_\mu\right).
\eeq
\end{lemma}
\pf
Let's rewrite $24\,h^{\al\beta}$ as the sum of $A^{\al\beta}$ and 
$B^{\al\beta}$, where
\eqa
A^{\al\beta}&=&24\left(
2\,\f_{t^\mu}\,c^{\al\mu}_\ga\,g^{\ga\beta}+3\,\f_{\tx^\mu}\,
c^{\mu\al}_{\ga\xi}\,g^{\beta\ga}\,\tx^\xi-\f_{\tx^\mu}\,
c^{\beta\mu}_{\ga\xi}\,g^{\al\ga}\,\tx^\xi-
2\,\f_{\tx^\mu t^\xi}\,c^{\al\mu}_\ga\,g^{\ga\beta}\,\tx^\xi \right.
\nn\\
&&
\left.+\f_{\tx^\ga t^\mu}\,(c^{\al\mu}_\xi\,g^{\ga\beta}+
c^{\beta\mu}_\xi\,g^{\ga\al})\,\tx^\xi+
\f_{\tx^\mu \tx^\ga}\,(c^{\mu\al}_{\nu\xi}\,g^{\ga\beta}+
c^{\mu\beta}_{\nu\xi}\,g^{\ga\al})\,\tx^\nu\,\tx^\xi\right),\nn
\eeqa
and
$$
B^{\al\beta}=24\,\left(
3\,\f_{\tx^\mu}\,c^{\al\mu}_\ga\,\frac{\pal g^{\ga\beta}}{\pal t^\xi}\,
\tx^\xi-\f_{\tx^\mu}\,c^{\beta\mu}_\ga\,\Gamma^{\al\ga}_\xi\,\tx^\xi+
\f_{\tx^\mu}\,c^{\al\mu}_\ga\,\Gamma^{\ga\beta}_\xi\,\tx^\xi\right).
$$
By using the formulas given in Section 4 we have
\eqa
&&
A^{\al\beta}=
\frac{\ga_{ij}\,\sigma_j}{\sigma_i\,\psi_{i1}\,\psi_{j1}}\left(
3\,\psi_{j\ga}\,\psi_i^\al\,g^{\ga\beta}-
3\,\psi_{i\ga}\,\psi_j^\beta\,g^{\ga\al}+
\psi_{i\ga}\,\psi_j^\al\,g^{\ga\beta}-
\psi_{j\ga}\,\psi_i^\beta\,g^{\ga\al}\right)
\nn\\
&&
+2\,\ga_{ij}\left(
\frac{\psi_i^\al\,\psi_{i\ga}\,g^{\ga\beta}}{\psi_{i1}\,\psi_{j1}}-
\frac{\psi_j^\al\,\psi_{i\ga}\,g^{\ga\beta}}{\psi_{i1}^2}-
2\,\frac{\psi_i^\al\,\psi_{j\ga}\,g^{\ga\beta}}{\psi_{i1}^2}+
\frac{\psi_j^\beta\,\psi_{i\ga}\,g^{\ga\al}}{\psi_{i1}^2}\right.
\nn\\
&&\left.+\frac{\psi_i^\al\,\psi_{i\ga}\,\psi_{j1}\,
g^{\ga\beta}}{\psi_{i1}^3}\right).
\eeqa
For $B^{\al\beta}$, by using the formulas given in Section 4 and
the following formulas
\beq
\Gamma^{\al\beta}_\ga=\left(\frac{1+d}2-q_\beta\right)c^{\al\beta}_\ga,
\quad
\frac{\pal g^{\al\beta}}{\pal t^\ga}=\Gamma^{\al\beta}_\ga+
\Gamma^{\beta\al}_\ga,
\eeq
we obtain
\eqa
B^{\al\beta}&=&24\,\f_{\tx^\mu}\,\left(
(3+3\,d-4\,q_\beta)\,c^{\al\mu}_\ga\,c^{\ga\beta}_\xi
-3\,q_\ga\,c^{\al\mu}_\ga\,c^{\ga\beta}_\xi+
q_\ga\,c^{\beta\mu}_\ga\,c^{\ga\al}_\xi\right)\,\tx^{\xi}
\nn\\
&=&
(3+3\,d-4\,q_\beta)\,\frac{\psi_i^\al\,\psi_i^\beta}{\psi_{i1}^2}+
\frac{\sigma_j}{\sigma_i}\,\frac{q_\ga\,\psi_{i\ga}\,\psi_j^\ga}{\psi_{i1}\,
\psi_{j1}}\left(\psi_j^\al\,\psi_i^\beta-3\,
\psi_j^\beta\,\psi_i^\al\right)
\nn\\
&=&
(3+3\,d-4\,q_\beta)\,\frac{\psi_i^\al\,\psi_i^\beta}{\psi_{i1}^2}+
\frac{\sigma_j}{\sigma_i}\left(
(u_j-u_i)\,\ga_{ij}+\frac d2\,\delta_{ij}\right)
\frac1{\psi_{i1}\,
\psi_{j1}}\left(\psi_j^\al\,\psi_i^\beta-3\,
\psi_j^\beta\,\psi_i^\al\right)
\nn\\
&=&
(3+2\,d-4\,q_\beta)\,\frac{\psi_i^\al\,\psi_i^\beta}{\psi_{i1}^2}+
\frac{\sigma_j}{\sigma_i}\,\frac{(u_j-u_i)\,\ga_{ij}}{\psi_{i1}\,\psi_{j1}}
\left(\psi_j^\al\,\psi_i^\beta-3\,
\psi_j^\beta\,\psi_i^\al\right),
\eeqa
above we have used formula (\ref{eOrth2}).
From formula (\ref{scaling}) it follows that
\eqa
&&24\,h^{\al\beta}=
A^{\al\beta}+B^{\al\beta}
\nn\\
=&&2\,\ga_{ij}\left(
\frac{\psi_i^\al\,\psi_{i\ga}\,g^{\ga\beta}}{\psi_{i1}\,\psi_{j1}}-
\frac{\psi_j^\al\,\psi_{i\ga}\,g^{\ga\beta}}{\psi_{i1}^2}-
2\,\frac{\psi_i^\al\,\psi_{j\ga}\,g^{\ga\beta}}{\psi_{i1}^2}+
\frac{\psi_j^\beta\,\psi_{i\ga}\,g^{\ga\al}}{\psi_{i1}^2}+
\frac{\psi_i^\al\,\psi_{i\ga}\,\psi_{j1}\,
g^{\ga\beta}}{\psi_{i1}^3}\right)
\nn\\
&&+(3+2\,d-4\,q_\beta)\,\frac{\psi_i^\al\,\psi_i^\beta}{\psi_{i1}^2}.
\eeqa
On the other hand, for the right-hand side of (\ref{h}) we have
\eqa
&&2\,\frac{\pal}{\pal t^\nu}\,(g^{\mu\nu}\,
c^{\al\beta}_{\mu})+c^{\mu\nu}_\nu\,c^{\al\beta}_\mu
\nn\\
=&&-2\,\frac{\pal}{\pal t^\nu}\,(g^{\mu\nu}\,
c^{\al\beta}_{\mu})+4\,\frac{\pal}{\pal t^\nu}\,(g^{\mu\beta}\,
c^{\al\nu}_{\mu})+c^{\mu\nu}_\nu\,c^{\al\beta}_\mu
\nn\\
=&&
-2\,(1+d-q_\mu-q_\nu)\,c^{\mu\nu}_\nu\,c^{\al\beta}_\mu+
4\,(1+d-q_\mu-q_\beta)\,c^{\mu\beta}_\nu\,c^{\al\nu}_\mu
\nn\\
&&
+c^{\mu\nu}_\nu\,c^{\al\beta}_\mu-2\,g^{\mu\nu}\,c^{\al\beta}_{\mu\nu}+
4\,g^{\mu\beta}\,c^{\al\nu}_{\mu\nu}
\nn\\
=&&(3+2\,d-4\,q_\beta)\,\frac{\psi_i^\al\,\psi_i^\beta}{\psi_{i1}^2}-
2\,\ga_{ij}\,(u_i-u_j)\,\frac{\psi_i^\al\,\psi_i^\beta}
{\psi_{i1}\,\psi_{j1}}
\nn\\
&&-2\,\ga_{ij}\,\left(
\frac{u_j\,\psi_i^\al\,\psi_i^\beta}{\psi_{i1}\,\psi_{j1}}-
\frac{u_i\,\psi_i^\beta\,\psi_j^\al}{\psi_{i1}^2}-
\frac{u_i\,\psi_i^\al\,\psi_j^\beta}{\psi_{i1}^2}+
\frac{u_i\,\psi_i^\al\,\psi_i^\beta\,\psi_{j1}}{\psi_{i1}^3}\right)
\nn\\
&&
+4\,\ga_{ij}\,\left(\frac{\psi_i^\al\,\psi_{i\ga}}{\psi_{i1}\,\psi_{j1}}-
\frac{\psi_j^\al\,\psi_{i\ga}}{\psi_{i1}^2}-
\frac{\psi_i^\al\,\psi_{j\ga}}{\psi_{i1}^2}+
\frac{\psi_i^\al\,\psi_{i\ga}\,\psi_{j1}}{\psi_{i1}^3}\right)\,g^{\ga\beta}
\nn\\
=&&
2\,\ga_{ij}\left(
\frac{\psi_i^\al\,\psi_{i\ga}\,g^{\ga\beta}}{\psi_{i1}\,\psi_{j1}}-
\frac{\psi_j^\al\,\psi_{i\ga}\,g^{\ga\beta}}{\psi_{i1}^2}-
2\,\frac{\psi_i^\al\,\psi_{j\ga}\,g^{\ga\beta}}{\psi_{i1}^2}+
\frac{\psi_j^\beta\,\psi_{i\ga}\,g^{\ga\al}}{\psi_{i1}^2}+
\frac{\psi_i^\al\,\psi_{i\ga}\,\psi_{j1}\,
g^{\ga\beta}}{\psi_{i1}^3}\right)
\nn\\
&&+(3+2\,d-4\,q_\beta)\,\frac{\psi_i^\al\,\psi_i^\beta}{\psi_{i1}^2}
\nn\\
=&&24\,h^{\al\beta}.
\eeqa
\epl

Knowing $h^{\al\beta}(t)$ we can compute the symmetrized 
coefficient in front of $\de''(X-Y)$ using the skew-symmetry condition
\beq
r^{\al\beta}_\ga(t)+r^{\beta\al}_\ga(t)=3\,\pal_\ga h^{\al\beta}.
\eeq
The antisymmetrization of the same coefficients is given by the following:
\begin{lemma}\label{symofr}
Let's denote ${\tilde r}^{\al\beta}$  the coefficients
before $\ve^2\,\de''(X-Y)$ in the second Poisson bracket
$\{t^\al(X),t^\beta(Y)\}^{'}_2$ of (\ref{spbg}), 
then the following identity holds true:
\beq
{\tilde r}^{\al\beta}-{\tilde r}^{\beta\al}=
\frac1{24} (d+3-2\,q_\beta)\,c^{\al\nu}_\ga\,c^{\beta\mu}_{\nu\mu}\,\tx^\ga-
\frac1{24} (d+3-2\,q_\al)\,c^{\beta\nu}_\ga\,c^{\al\mu}_{\nu\mu}\,\tx^\ga.
\eeq
\end{lemma}
\pf
From  the expression (\ref{spbg}) we have
\eqa
\lefteqn{
{\tilde r}^{\al\beta}=
g^{\ga\beta}\,\frac{\pal w^\al}{\pal \tx^\ga}-g^{\ga\al}\,
\frac{\pal w^\beta}{\pal \tx^\ga}+3\,g^{\ga\al}\,
\pal_X\left(\frac{\pal w^\beta}{\pal \txx^\ga}\right)
-6\,g^{\ga\al}\,\pal_X^2\left(\frac{\pal w^\beta}{\pal
\txxx^\ga}\right)}
\nn\\
&&+3\,\Gamma^{\ga\beta}_\mu\,\frac{\pal w^\al}{\pal \txx^\ga}\,
\tx^\mu+2\,\Gamma^{\beta\ga}_\mu\,\frac{\pal w^\al}{\pal \txx^\ga}\,
\tx^\mu+\Gamma^{\al\ga}_\mu\,\frac{\pal w^\beta}{\pal \txx^\ga}\,
\tx^\mu
\nn\\
&&-3\,\Gamma^{\al\ga}_\mu\,\pal_X\left(\frac{\pal w^\beta}{\pal
\txxx^\ga}\right)\,\tx^\mu+
3\,\frac{\pal_X^2 g^{\ga\beta}}{\pal X^2}\frac{\pal w^\al}{\pal
\txxx^\ga}+3\,\frac{\pal\left(\Gamma^{\ga\beta}_\mu\,\tx^\mu\right)}
{\pal X}\,\frac{\pal w^\al}{\pal \txxx^\ga}.
\eeqa
By using the formulas given in Section 2  we get, through a long
calculation, the following: 
\eqa
\lefteqn{
{\tilde r}^{\al\beta}-{\tilde r}^{\beta\al}=
2\,g^{\ga\beta}\,\frac{\pal w^\al}{\pal \tx^\ga}-
3\,g^{\ga\beta}\,
\pal_X \left(\frac{\pal w^\al}{\pal \txx^\ga}\right)
+6\,g^{\ga\beta}\,\pal_X^2 \left(\frac{\pal w^\al}{\pal
\txxx^\ga}\right)}
\nn\\
&&+3\,\Gamma^{\ga\beta}_\mu\,\frac{\pal w^\al}{\pal \txx^\ga}\,
\tx^\mu+\Gamma^{\beta\ga}_\mu\,\frac{\pal w^\al}{\pal \txx^\ga}\,
\tx^\mu
+3\,\Gamma^{\beta\ga}_\mu\,\pal_X\left(\frac{\pal w^\al}{\pal
\txxx^\ga}\right)\,\tx^\mu
\nn\\
&&+3\,\frac{\pal^2 g^{\ga\beta}}{\pal X^2}\,\frac{\pal w^\al}{\pal
\txxx^\ga}+3\,\frac{\pal\left(\Gamma^{\ga\beta}_\mu\,\tx^\mu\right)}
{\pal X}\,\frac{\pal w^\al}{\pal \txxx^\ga}
\nn\\
&&-\left(\mbox{the precedent sum with $\al$ and $\beta$ changed}\right)
\nn\\
&=&
2\,g^{\ga\beta}\,\pal_X \left(\frac{\pal}{\pal \tx^\ga}
\,\frac{\pal \f}{\pal T_{\al,0}}\right)+
2\,g^{\ga\beta}\,\left(\frac{\pal}{\pal t^\ga}
\,\frac{\pal \f}{\pal T_{\al,0}}\right)
-3\,g^{\ga\beta}\,
\pal_X \left(\frac{\pal w^\al}{\pal \txx^\ga}\right)
\nn\\
&&+6\,g^{\ga\beta}\,\pal_X^2 \left(\frac{\pal w^\al}{\pal
\txxx^\ga}\right)
+3\,\Gamma^{\ga\beta}_\mu\,\frac{\pal w^\al}{\pal \txx^\ga}\,
\tx^\mu+\Gamma^{\beta\ga}_\mu\,\frac{\pal w^\al}{\pal \txx^\ga}\,
\tx^\mu
\nn\\
&&+3\,\Gamma^{\beta\ga}_\mu\,\pal_X\left(\frac{\pal w^\al}{\pal
\txxx^\ga}\right)\,\tx^\mu
+3\,\frac{\pal^2 g^{\ga\beta}}{\pal X^2}\,\frac{\pal w^\al}{\pal
\txxx^\ga}+3\,\frac{\pal\left(\Gamma^{\ga\beta}_\mu\,\tx^\mu\right)}
{\pal X}\,\frac{\pal w^\al}{\pal \txxx^\ga}
\nn\\
&&-\left(\mbox{the precedent sum with $\al$ and $\beta$ exchanged}\right)
\nn\\
&=&
2\,g^{\ga\beta}\,\left(
\f_{t^\mu t^\ga}\,c^{\al\mu}_\nu\,\tx^\nu+\f_{\tx^\mu\,t^\ga}\,
c^{\al\mu}_{\nu\rho}\,\tx^\nu\,\tx^\rho+
\f_{\tx^\mu\,t^\ga}\,c^{\al\mu}_\nu\,\txx^\nu\right)
\nn\\
&&-g^{\ga\beta}\,\pal_X \left(
\f_{t^\mu\,\tx^\ga}\,c^{\al\mu}_\nu\,\tx^\nu+\f_{\tx^\mu\,\tx^\ga}\,
c^{\al\mu}_{\nu\sigma}\,\tx^\nu\,\tx^\sigma\right)
\nn\\
&&+g^{\ga\beta}\,\pal_X \left(
\f_{\tx^\mu}\,c^{\al\mu}_{\ga\nu}\,\tx^\nu+3\,\f_{\tx^\mu\,t^\nu}\,
c^{\al\mu}_\ga\,\tx^\nu+
3\,\f_{\tx^\mu\,\tx^\nu}\,c^{\al\mu}_\ga\,\txx^\nu
-\f_{\tx^\mu\,\tx^\ga}\,c^{\al\mu}_\nu\,\txx^\nu
\right.
\nn\\
&&\left.-\f_{t^\mu}\,c^{\al\mu}_\ga\right)
+2\,g^{\ga\beta}\,\left(
\f_{t^\mu}\,c^{\al\mu}_{\ga\nu}\,\tx^\nu+\f_{\tx^\mu}\,
c^{\al\mu}_{\nu\rho\ga}\,\tx^\nu\,\tx^\rho+
\f_{\tx^\mu}\,c^{\al\mu}_{\nu\ga}\,\txx^\nu\right)
\nn\\
&&+3\,\Gamma^{\ga\beta}_\mu\,\frac{\pal w^\al}{\pal \txx^\ga}\,
\tx^\mu+\Gamma^{\beta\ga}_\mu\,\frac{\pal w^\al}{\pal \txx^\ga}\,
\tx^\mu
+3\,\Gamma^{\beta\ga}_\mu\,\pal_X\left(\frac{\pal w^\al}{\pal
\txxx^\ga}\right)\,\tx^\mu
\nn\\
&&+3\,\frac{\pal^2 g^{\ga\beta}}{\pal X^2}\,\frac{\pal w^\al}{\pal
\txxx^\ga}+3\,\frac{\pal\left(\Gamma^{\ga\beta}_\mu\,\tx^\mu\right)}
{\pal X}\,\frac{\pal w^\al}{\pal \txxx^\ga}
\nn\\
&&-\left(\mbox{the precedent sum with $\al$ and $\beta$ exchanged}\right)
\nn\\
&=&
2\,g^{\ga\beta}\,\left(
\f_{t^\mu\,t^\ga}\,c^{\al\mu}_\nu\,\tx^\nu+\f_{\tx^\mu\,t^\ga}\,
c^{\al\mu}_{\nu\sigma}\,\tx^\nu\,\tx^\sigma\right)
-\frac1{12}\,u_j\,\ga_{ij}\,\left(\frac1{\sigma_j}+
\frac{\sigma_j}{\sigma_i^2}\right)
\,\frac{\psi_i^\al\,\psi_j^\beta\,\psi_{i\nu}}
{\psi_{i1}\,\psi_{j1}}\,\txx^\nu
\nn\\
&&+\frac1{24}\,\pal_X \left(2\,u_j\,\ga_{ij}\,
\frac{\sigma_i}{\sigma_j}\,\frac{\psi_i^\al\,
\psi^\beta_j}{\psi_{i1}\,\psi_{j1}}+(u_i+u_j)\,\ga_{ij}\,
\frac{\psi_i^\al\,\psi_j^\beta}{\psi_{i1}^2}\right)
\nn\\
&&+\left(\pal_X g^{\ga\beta}\right) \left(
\f_{t^\mu\,\tx^\ga}\,c^{\al\mu}_\nu\,\tx^\nu+\f_{\tx^\mu\,\tx^\ga}\,
c^{\al\mu}_{\nu\sigma}\,\tx^\nu\,\tx^\sigma\right)
\nn\\
&&+g^{\ga\beta}\,\pal_X \left(
\f_{\tx^\mu}\,c^{\al\mu}_{\ga\nu}\,\tx^\nu+3\,\f_{\tx^\mu\,t^\nu}\,
c^{\al\mu}_\ga\,\tx^\nu+
3\,\f_{\tx^\mu\,\tx^\nu}\,c^{\al\mu}_\ga\,\txx^\nu
-\f_{\tx^\mu\,\tx^\ga}\,c^{\al\mu}_\nu\,\txx^\nu\right.
\nn\\
&&\left.-\f_{t^\mu}\,c^{\al\mu}_\ga\right)
+2\,g^{\ga\beta}\,\left(
\f_{t^\mu}\,c^{\al\mu}_{\ga\nu}\,\tx^\nu+\f_{\tx^\mu}\,
c^{\al\mu}_{\nu\rho\ga}\,\tx^\nu\,\tx^\rho+
\f_{\tx^\mu}\,c^{\al\mu}_{\nu\ga}\,\txx^\nu\right)
\nn\\
&&+3\,\Gamma^{\ga\beta}_\mu\,\frac{\pal w^\al}{\pal \txx^\ga}\,
\tx^\mu+\Gamma^{\beta\ga}_\mu\,\frac{\pal w^\al}{\pal \txx^\ga}\,
\tx^\mu
+3\,\Gamma^{\beta\ga}_\mu\,\pal_X\left(\frac{\pal w^\al}{\pal
\txxx^\ga}\right)\,\tx^\mu
\nn\\
&&+3\,\frac{\pal^2 g^{\ga\beta}}{\pal X^2}\,\frac{\pal w^\al}{\pal
\txxx^\ga}+3\,\frac{\pal\left(\Gamma^{\ga\beta}_\mu\,\tx^\mu\right)}
{\pal X}\,\frac{\pal w^\al}{\pal \txxx^\ga}
\nn\\
&&-\left(\mbox{the precedent sum with $\al$ and $\beta$ exchanged}\right)
\nn\\
&=&
2\,g^{\ga\beta}\,\left(
\f_{t^\mu\,t^\ga}\,c^{\al\mu}_\nu\,\tx^\nu+\f_{\tx^\mu\,t^\ga}\,
c^{\al\mu}_{\nu\sigma}\,\tx^\nu\,\tx^\sigma\right)
\nn\\
&&
+\frac1{24}\,\tx^\sigma \frac{\pal}{\pal t^\sigma}\,
\left(2\,u_j\,\ga_{ij}\,\frac{\sigma_i}{\sigma_j}\,\frac{\psi_i^\al\,
\psi^\beta_j}{\psi_{i1}\,\psi_{j1}}+(u_i+u_j)\,\ga_{ij}\,
\frac{\psi_i^\al\,\psi_j^\beta}{\psi_{i1}^2}\right)
\nn\\
&&
-2\,q_\ga\,\left(\f_{\tx^\mu\,\tx^\ga}\,c^{\al\mu}_{\nu\sigma}\,
c^{\beta\ga}_\rho\,\tx^\rho\,\tx^\nu\,\tx^\sigma+
\f_{t^\mu\,\tx^\ga}\,c^{\al\mu}_{\sigma}\,
c^{\beta\ga}_\rho\,\tx^\rho\,\tx^\sigma\right)
\nn\\
&&
+\left\{\left(3\,(1+d)-4\,q_\beta\right)
\left(\f_{\tx^\mu\,\tx^\ga}\,c^{\al\mu}_{\nu\sigma}\,
c^{\beta\ga}_\rho\,\tx^\rho\,\tx^\nu\,\tx^\sigma+
\f_{t^\mu\,\tx^\ga}\,c^{\al\mu}_{\sigma}\,
c^{\beta\ga}_\rho\,\tx^\rho\,\tx^\sigma\right)\right.
\nn\\
&&\left.+q_\beta\,\f_{\tx^\mu\,t^\nu}\,c^{\al\mu}_\ga\,c^{\beta\ga}_\sigma
\,\tx^\sigma\,\tx^\nu
-3\,q_{\beta}\,\f_{\tx^\mu}\,c^{\al\mu}_{\ga\nu}\,c^{\beta\ga}_\sigma\,
\tx^\sigma\,\tx^\nu-
3\,q_{\beta}\,\f_{\tx^\mu}\,c^{\al\mu}_{\ga}\,c^{\beta\ga}_{\sigma\nu}\,
\tx^\sigma\,\tx^\nu\right.
\nn\\
&&\left.-2\,q_{\beta}\,\f_{t^\mu}\,c^{\al\mu}_{\ga}\,c^{\ga\beta}_\sigma\,
\tx^\sigma\right\}
-\left(\mbox{the precedent sum with $\al$ and $\beta$ exchanged}\right)
\nn\\
&=&\frac1{24}\left\{
(u_j-u_k)\,\ga_{ik}\,\ga_{kj}\,\sigma_k\,\frac{\psi_i^\al\,\psi_j^\beta}
{\psi_{i1}^2\,\psi_{k1}}
+3\,(u_l-u_j)\,\ga_{ij}\,\ga_{jl}\,\sigma_i\,\frac{\psi_i^\al\,\psi_l^\beta}
{\psi_{i1}\,\psi_{j1}^2}\right.
\nn\\
&&
+2\,(u_i-u_j)\,\ga_{ij}\,\ga_{il}\,\sigma_i\,\frac{\psi_i^\al\,\psi_j^\beta\,
\psi_{l1}}{\psi_{i1}^4}
-(u_i+u_j)\,\ga_{ij}\,\ga_{il}\,\sigma_i\,\frac{\psi_j^\beta\,\psi_l^\al}
{\psi_{i1}^3}
\nn\\
&&
+(u_j-u_k)\,\ga_{ik}\,\ga_{kj}\,\sigma_i\,\frac{\psi_i^\al\,\psi_j^\beta}
{\psi_{i1}^3}
+2\,(u_j-u_l)\,\ga_{ij}\,\ga_{jl}\,\frac{\sigma_j^2}{\sigma_i}\,
\frac{\psi_j^\al\,\psi_l^\beta}
{\psi_{j1}^3}
\nn\\
&&\left.+d\,\ga_{ik}\,\sigma_k\,\frac{\psi_i^\al\,\psi_k^\beta}{\psi_{i1}^2\,
\psi_{k1}}
+2\,d\,\ga_{ij}\,\sigma_i\,\frac{\psi_i^\al\,\psi_j^\beta}
{\psi_{i1}\,\psi_{j1}^2}
-d\,\ga_{ij}\,\sigma_i\,\frac{\psi_j^\al\,\psi_i^\beta}{\psi_{i1}^3}\right\}
\nn\\
&&
+\frac{\ga_{ij}}{24}
\left\{3\,q_\beta\,\sigma_i\,\frac{\psi_i^\al\,\psi_j^\beta}
{\psi_{i1}^3}+
\left(q_\beta-3\,(1+d)\right) \left(\sigma_j\,\frac{\psi_i^\al\,
\psi_j^\beta}{\psi_{j1}\,\psi_{i1}^2}+
\sigma_i\,\frac{\psi_i^\al\,\psi_j^\beta}{\psi_{i1}^3}
\right)\right.
\nn\\
&&
+\left(5\,q_\beta-6\,(1+d)\right)\,
\sigma_j\,\frac{\psi_j^\al\,\psi_i^\beta}{\psi_{j1}\,\psi_{i1}^2}
-4\,q_\beta\,
\sigma_i\,\frac{\psi_i^\al\,\psi_i^\beta\,\psi_{j1}}
{\psi_{i1}^4}
-2\,q_\beta\,
\sigma_j\,\frac{\psi_j^\al\,\psi_j^\beta}{\psi_{i1}\,\psi_{j1}^2}
\nn\\
&&\left.
-2\,q_\beta\,\frac{\sigma_j^2}{\sigma_i}\,
\frac{\psi_j^\al\,\psi_j^\beta}{\psi_{j1}^3}
\right\}
-\left(\mbox{the precedent sum with $\al$ and $\beta$ exchanged}\right)
\nn\\
&=&
\frac1{24} \left(\frac{d}2-q_\beta\right)\,
\left(
\ga_{ij}\,\sigma_j\,\frac{\psi_i^\al\,\psi_j^\beta}{\psi_{i1}^2\,
\psi_{j1}}+3\,
\ga_{ij}\,\sigma_i\,\frac{\psi_i^\al\,\psi_j^\beta}{\psi_{i1}\,
\psi_{j1}^2}
-2\,\ga_{ij}\,\sigma_i\,\frac{\psi_i^\al\,\psi_i^\beta\,\psi_{j1}}
{\psi_{i1}^4}\right.
\nn\\
&&
\left.+\ga_{ij}\,\sigma_i\,\frac{\psi_i^\al\,\psi_j^\beta}{\psi_{i1}^3}
+\ga_{ij}\,\sigma_i\,\frac{\psi_i^\al\,\psi_j^\beta}{\psi_{i1}^3}
-2\,\ga_{ij}\,\frac{\sigma_j^2}{\sigma_i}\,
\frac{\psi_j^\al\,\psi_j^\beta}{\psi_{j1}^3}\right)
\nn\\
&&+\frac{d}{24} \left\{
\ga_{ik}\,\sigma_k\,\frac{\psi_i^\al\,\psi_k^\beta}{\psi_{i1}^2\,
\psi_{k1}}
+2\,\ga_{ij}\,\sigma_i\,
\frac{\psi_i^\al\,\psi_j^\beta}{\psi_{i1}\,\psi_{j1}^2}
-\ga_{ij}\,\sigma_i\,\frac{\psi_j^\al\,\psi_i^\beta}{\psi_{i1}^3}\right\}
\nn\\
&&+\frac{\ga_{ij}}{24}\,\left\{3\,q_\beta\,\sigma_i\,\frac{\psi_i^\al\,\psi_j^\beta}
{\psi_{i1}^3}+
\left(q_\beta-3\,(1+d)\right) \left(\sigma_j\,\frac{\psi_i^\al\,
\psi_j^\beta}{\psi_{j1}\,\psi_{i1}^2}+
\sigma_i\,\frac{\psi_i^\al\,\psi_j^\beta}{\psi_{i1}^3}
\right)\right.
\nn\\
&&
+\left(5\,q_\beta-6\,(1+d)\right)\,
\sigma_j\,\frac{\psi_j^\al\,\psi_i^\beta}{\psi_{j1}\,\psi_{i1}^2}
-4\,q_\beta\,
\sigma_i\,\frac{\psi_i^\al\,\psi_i^\beta\,\psi_{j1}}
{\psi_{i1}^4}
-2\,q_\beta\,
\sigma_j\,\frac{\psi_j^\al\,\psi_j^\beta}{\psi_{i1}\,\psi_{j1}^2}
\nn\\
&&\left.-2\,q_\beta\,\frac{\sigma_j^2}{\sigma_i}\,
\frac{\psi_j^\al\,\psi_j^\beta}{\psi_{j1}^3}
\right\}
-\left(\mbox{the precedent sum with $\al$ and $\beta$ exchanged}\right)
\nn\\
&=&\frac1{24}
\left(2\,q_\beta-d-3\right)\,\ga_{ij}\,\sigma_i\,
\left(\frac{\psi_i^\al\,\psi_j^\beta}{\psi_{i1}\,\psi_{j1}^2}+
\frac{\psi_i^\al\,\psi_j^\beta}{\psi_{i1}^3}
-\frac{\psi_i^\al\,\psi_i^\beta\,\psi_{j1}}{\psi_{i1}^4}
-\frac{\psi_i^\al\,\psi_i^\beta}{\psi_{i1}^2\,\psi_{j1}}
\right)
\nn\\
&&-\left(\mbox{the precedent sum with $\al$ and $\beta$ exchanged}\right)
\nn\\
&=&
\frac1{24} (d+3-2\,q_\beta)\,c^{\al\nu}_\ga\,c^{\beta\mu}_{\nu\mu}\,\tx^\ga-
\frac1{24} (d+3-2\,q_\al)\,c^{\beta\nu}_\ga\,c^{\al\mu}_{\nu\mu}\,\tx^\ga.
\eeqa
\epl

\vskip 0.5cm

\vskip 0.5cm
\noindent {\bf Proof of Theorem 2 \ }
Let's denote 
$$
{\widetilde H}^{\beta,0}=\frac1{24} 
\int w^\beta_{\mu\nu}\,\tx^\mu\,\tx^\nu\,dX,\quad
F^\beta_\ga=\eta^{\beta{\beta}'}\,\frac{\pal^2 F}{\pal t^{\beta'}\,\pal 
t^\ga},
$$
where $ w^\beta_{\mu\nu}$ are defined in (\ref{w3}), then
from Lemma \ref{Lham} we see that the equations
in (\ref{eq-FHS}) with $p=0$  can be written as
\eqa
\lefteqn{
\frac{\pal t^\al}{\pal T_{\beta,0}}=
c^{\al\beta}_\ga(t)\,\tx^\ga+
\ve^2\,\pal_X \left(
\eta^{\al\ga}\,\frac{\de {\widetilde H}^{\beta,0}}{\de t^\ga}+
{\hat h}^{\al\ga}\,\pal_X^2 F^\beta_\ga+
\frac12 \pal_X {\hat h}^{\al\ga}\,
\pal_X F^\beta_\ga\right)}
\nn\\
&=&c^{\al\beta}_\ga(t)\,\tx^\ga+\ve^2\,
\left(
{\hat b}^{\al\beta}_{\ga\mu}(t)\,\tx^\ga\,\txx^\mu+
{\hat a}^{\al\beta}_{\ga\mu\nu}(t)\,\tx^\ga\,\tx^\mu\,\tx^\nu+
\right. \nn\\
&&+\left. {\hat p}^{\al\beta}_\ga(t)\,\txxx^\ga\right)+{\cal O}(\ve^4),
\label{dxp}
\eeqa
where the coefficients ${\hat b}^{\al\beta}_{\ga\mu}(t), 
{\hat a}^{\al\beta}_{\ga\mu\nu}(t), {\hat p}^{\al\beta}_\ga(t)$ 
have the expression
\eqa
{\hat p}^{\al\beta}_\ga&=&\frac1{12}\,c^{\al\beta}_{\mu\nu}\,
c^{\mu\nu}_\ga,\nn\\
{\hat a}^{\al\beta}_{\ga\mu\nu}&=&\frac1{72}\,
\left(\eta^{\al\sigma}\,(\pal_\sigma\,\pal_\nu\,w^\beta_{\ga\mu}+
\pal_\sigma\,\pal_\ga\,w^\beta_{\mu\nu}\right.
\nn \\ &&
+ \pal_\sigma\,\pal_\mu\,w^\beta_{\ga\nu}-2\, \pal_\mu\,\pal_\nu\,
w^\beta_{\ga\sigma}-2\, \pal_\mu\,\pal_\ga\,w^\beta_{\nu\sigma}-
2\, \pal_\nu\,\pal_\ga\,w^\beta_{\mu\sigma})
\nn \\ 
&&
+\eta^{\xi\zeta}\,(6\,c^{\al\sigma}_{\xi\zeta}\,
c^\beta_{\sigma\ga\mu\nu}+
 3\,c^{\al\sigma}_{\xi\zeta\ga}\,c^\beta_{\sigma\mu\nu}+
3\,c^{\al\sigma}_{\xi\zeta\mu}\,c^\beta_{\sigma\ga\nu}
\nn \\ &&\left.
+3\,c^{\al\sigma}_{\xi\zeta\nu}\,c^\beta_{\sigma\ga\mu}+
  c^{\al\sigma}_{\xi\zeta\ga\mu}\,c^\beta_{\sigma\nu}+
 c^{\al\sigma}_{\xi\zeta\ga\nu}\,c^\beta_{\sigma\mu}+
 c^{\al\sigma}_{\xi\zeta\mu\nu}\,c^\beta_{\sigma\ga})\right),\\
{\hat b}^{\al\beta}_{\ga\mu}&=&\frac1{12}\,
\left(\eta^{\al\sigma}\,(-2\,\pal_\ga\,w^\beta_{\mu\sigma}+
\pal_\sigma\,w^\beta_{\ga\mu}-\pal_\mu\,w^\beta_{\ga\sigma})\right.
\nn \\ &&\left.
+\eta^{\xi\zeta}\,(3\,c^{\al\sigma}_{\xi\zeta}\,c^\beta_{\sigma\ga\mu}+
\frac32\, 
c^{\al\sigma}_{\xi\zeta\ga}\,c^\beta_{\sigma\mu}+\frac12\, 
c^{\al\sigma}_{\xi\zeta\mu}\,c^\beta_{\sigma\ga})\right),
\eeqa
with $w^\al_{\beta\ga}$ defined by (\ref{w3}),
and ${\hat h}^{\al\beta}$ are defined
in (\ref{fh}).     
On the other hand, from the bihamiltinian relation
(\ref{BH}) we have 
\beq
(\frac{1+d}2-q_\beta)\,
\frac{\pal t^\al}{\pal T_{\beta,0}}=\{t^\al,
\int t^\beta(X)\,dX\}^{'}_2+{\cal O}(\ve^4),
\eeq
which together with (\ref{dxp}) leads to the expression for the
coefficients $ p^{\al\beta}_\ga(t), a^{\al\beta}_{\ga\mu\nu}(t),
b^{\al\beta}_{\ga\mu}(t)$  in the formula (\ref{second-bracket})
\eqa
&&p^{\al\beta}_\ga(t)=\left(\frac12-\mu_\beta\right)\,
{\hat p}^{\al\beta}_\ga(t),\quad
a^{\al\beta}_{\ga\mu\nu}(t)=\left(\frac12-\mu_\beta\right)\,
{\hat a}^{\al\beta}_{\ga\mu\nu},\nn\\
&&
b^{\al\beta}_{\ga\mu}(t)=\left(\frac12-\mu_\beta\right)\,
{\hat b}^{\al\beta}_{\ga\mu}(t).
\eeqa
The expression (\ref{EQh}) of the coefficients $h^{\al\beta}$ 
follows from Lemma \ref{Lh}, formulas (\ref{expofr})--(\ref{eq}) 
are obtained by using Lemma \ref{symofr} and  the anti-symmetry condition 
of the second Poisson bracket. In fact, if we denote ${\tilde r}^{\al\beta},
{\tilde f}^{\al\beta}$ and $ {\tilde p}^{\al\beta}$ the coefficients 
before $\ve^2\,\de''(X-Y), \ve^2\,\de'(X-Y)$ and $\ve^2\,\de(X-Y)$
in the second Poisson bracket $\{t^\al(X),t^\beta(Y)\}_2$
respectively, then the antisymmetry condition of the second Poisson
bracket gives us
\eqa
&&
{\tilde r}^{\al\beta}+{\tilde r}^{\beta\al}=3\,\pal_X h^{\al\beta},
\label{ant1}
\\
&&
\pal_X {\tilde f}^{\al\beta}+\pal_X^3 h^{\al\beta}-\pal_X^2 
{\tilde r}^{\al\beta}={\tilde p}^{\al\beta}+{\tilde p}^{\beta\al}.\label{ant2}
\eeqa
Formula (\ref{expofr}) follows immediately from  (\ref{ant1}) 
and Lemma \ref{symofr}. From (\ref{dxp}) it follows that
\beq
{\tilde p}^{\al\beta}=
\left(\frac{1+d}2-q_\beta\right) \pal_X \left(\eta^{\al\ga}\
\frac{\de {\widetilde H}^{\beta,0}}{\de t^\ga}+
h^{\al\ga}\,\pal_X^2 F^\beta_\ga+\frac12 \pal_X h^{\al\ga}\,
\pal_X F^\beta_\ga\right).\label{ddxp}
\eeq
So from (\ref{ant2}) and the above expression of $ {\tilde p}^{\al\beta}$
we obtain
\eqa
\lefteqn{
{\tilde f}^{\al\beta}=-\pal_X^2 h^{\al\beta}+\pal_X {\tilde r}^{\al\beta}}
\nn\\
&&+\left(\frac{1+d}2-q_\beta\right) \left(\eta^{\al\ga}\
\frac{\de {\widetilde H}^{\beta,0}}{\de t^\ga}+
h^{\al\ga}\,\pal_X^2 F^\beta_\ga+\frac12 \pal_X h^{\al\ga}\,
\pal_X F^\beta_\ga\right).
\eeqa
which leads to formula (\ref{expoff}) and (\ref{eq}). We have thus 
verified the formula (\ref{second-bracket}). 
The remaining part of the theorem 
follows from (\ref{BH}).
\ept
\vskip 0.5cm

\noindent{\bf Proof of Proposition \ref{prop4} \ } For the correction
of the expression of the first Possion bracket, let's
replace the functions $w^\al$
by $\frac{\pal^2 G}{\pal X\,\pal T_{\al,0}}$ in the identity 
(\ref{templ}), 
then a direct calculation aided by the anti-symmetry condition
of the Poisson bracket gives 
the expression for ${\tilde a}^{\al\beta},
{\tilde b}^{\al\beta},{\tilde e}^{\al\beta}
$. For the
correction of the expression of the second  Poisson bracket,
from the identity (\ref{spbg}) with $ w^\al$
replaced by $\frac{\pal^2 G}{\pal X\,\pal T_{\al,0}}$ we can 
easily get the expression for the coefficients $a^{\al\beta}$
and $b^{\al\beta}$, however, it's not easy to get the simplified
expression for the coefficients $q^{\al\beta}$ and $e^{\al\beta}$
in this way. We use instead the relation
\beq
\left(\frac{1-d}2+q_\beta\right) \{v^\al(X),H_{\beta,0}\}_1
=\{v^\al(X),H_{\beta,-1}\}_2
\eeq
with $H_{\beta,-1}=\int v_\beta(X) dX,\ 
H_{\beta,0}=\int \frac{\pal F(v(X))}{\pal v^\beta(X)} dX$
and the infinitesimal B\"acklund transform 
\beq
t^\al=v^\al+\ve^2\,\frac{\pal^2 G}{\pal X\,\pal T_{\al,0}},
\eeq
to get the expression for the coefficient $q^{\al\beta}$, then
by using the anti-symmetry condition of the Poisson 
bracket we get the expression for the coefficient $e^{\al\beta}$.
Proposition is proved. $\quad \Box$

\vskip 0.6cm
\setcounter{equation}{0}

\section {Genus one Gromov-Witten invariants and G-function 
of a Frobenius manifold}\par

In the paper \cite{Getz} Getzler studied recursion relations for the 
genus one Gromov-Witten invariants of smooth projective varieties.
He derived a remarkable system of linear differential 
equations for a generating function $G=G(t^2,\dots,t^n)$ 
of these invariants. The system can be written in the following form:
\eqa
&&\sum_{1\le \al_1,\al_2,\al_3,\al_4\le n} z_{\al_1} z_{\al_2}
z_{\al_3} z_{\al_4} \left(3\,c^{\mu}_{\al_1\al_2}\,c^{\nu}_{\al_3\al_4}
\,\frac{\pal^2 G}{\pal t^\mu\pal t^{\nu}}-4\,
c^{\mu}_{\al_1\al_2}\,c^{\nu}_{\al_3\mu}
\,\frac{\pal^2 G}{\pal t^{\al_4}\pal t^{\nu}}\right.
\nn\\
&&
-c^{\mu}_{\al_1\al_2}\,c^{\nu}_{\al_3\al_4\mu}
\,\frac{\pal G}{\pal t^{\nu}}+2\,
c^{\mu}_{\al_1\al_2\al_3}\,c^{\nu}_{\al_4\mu}
\,\frac{\pal G}{\pal t^{\nu}}+\frac16 
c^{\mu}_{\al_1\al_2\al_3}\,c^{\nu}_{\al_4\mu\nu}
\nn\\
&&\left.
+\frac1{24} c^{\mu}_{\al_1\al_2\al_3\al_4}\,c^{\nu}_{\mu\nu}-
\frac14 c^{\mu}_{\al_1\al_2\nu}\,c^{\nu}_{\al_3\al_4\mu}\right)=0
\label{Getz}
\eeqa
The l.h.s. must be equal to zero identically in $z_1,\dots,z_n$.
The notations for the coefficients $c^\al_{\beta\de}, c^{\al\beta\ga\de}$
are defined in (\ref{Def-ofc}). Now we solve this system for an arbitrary
semisimple Frobenius manifold.
\vskip 0.5cm
\noindent {\bf Proof of Theorem 3.} Let us rewrite the system (6.1) in the canonical
coordinates. At this end we first do a linear change of the indeterminates
\beq
z_\alpha\mapsto w_i := \sum_{\alpha=1}^n \frac{\psi_{i\alpha}}{\psi_{i1}}z_\al .
\eeq
Instead of the partial derivatives of $G(t)$ and of $F(t)$ we substitute
in (6.1) the corresponding covariant derivatives. For example,
$$
\frac{\partial^2 G}{\partial t^\lambda \partial t^\mu} \to \nabla_i
\nabla_j G
$$
$$c_{\alpha\beta\gamma}^\delta \to \nabla_i \nabla_j \nabla_k \nabla^l F
$$
etc. Here $\nabla$ is the Levi-Civita flat connection for the metric $<\ ,
\ >$
written in the
curvilinear coordinates $u_i$. Recall that the metric becomes diagonal
in the canonical coordinates
\beq
<\ , \ > = \sum_{i=1}^n \psi_{i1}^2 d\, {u_i}^2.
\eeq
The only nontrivial Christoffel coefficients of the connection are
\beq
\Gamma_{ij}^i = \gamma_{ij} \frac{\psi_{j1}}{\psi_{i1}}, \
\Gamma_{ii}^j = - \gamma_{ij} \frac{\psi_{i1}}{\psi_{j1}}, i\neq j
\eeq
\beq
\Gamma_{ii}^i =-\sum_{k\neq i} \gamma_{ik} \frac{\psi_{k1}}{\psi_{i1}}.
\eeq
From the definition of the canonical coordinates we have
\beq
\nabla_i \nabla_j \nabla^k F = \delta_i^k \delta_j^k.
\eeq
This simplifies the computation. Finally we obtain for the polynomial
(6.1) in $w_1, \dots, w_n$ the following structure\par

\noindent 1).\ The coefficient in front of $w_i^4$ is equal to
$$
-\frac{\partial^2 G}{{\partial u^i}^2} + P_{ii}.
$$
2). \ The coefficient in front of $w_i^3 w_j$ for $i<j$ is equal to
$$
-4 \frac{\partial ^2 G}{\partial u^i \partial u^j} + 4 P_{ij}.
$$
3).\ The coefficient in front of $w_i^2 w_j^2$ for $i<j$ is equal to
$$
6  \frac{\partial ^2 G}{\partial u^i \partial u^j} - 6 P_{ij}.
$$
4). \ All other coefficients of the polynomial (6.1) vanish.
Here $P_{ij}=P_{ji}$ is a complicated expression in $u_1$, \dots, $u_n$, 
$\psi_{11}, \dots, \psi_{n1},\ \ga_{12}, \dots, \ga_{n-1\, n}$.

From the above structure of the coefficients we immediately derive
the uniqueness part of Theorem 4. Indeed, the general solution of the
corresponding linear homogeneous system
$$
 \frac{\partial ^2 G}{\partial u^i \partial u^j}=0
$$
is 
$$G=\sum_i c_i u_i + c_0
$$
for arbitrary constant coefficients. The quasihomogeneity equation 
(\ref{anomaly})
in the canonical coordinates reads
\beq
\sum_{j=1}^n u^j \frac{\partial G}{\partial u^j} = \ga .
\label{quasi}
\eeq
Hence $c_1 = ... = c_n=0$ and $G=const$.

To find the first derivatives of $G$ we differentiate (\ref{quasi})
w.r.t. $u_i$. This gives
$$
 \frac{\partial G}{\partial u^i}
=- \sum_j u^j  \frac{\partial^2 G}{\partial u^i \partial u^j}, \
i=1, ..., n.
$$
So
$$ \frac{\partial G}{\partial u^i}
=-\sum_j u^j P_{ij}.
$$
After tedious calculations we obtain the following formula
\eqa
24 \,  \frac{\partial G}{\partial u^i} =&&
\sum_j \frac{\ga_{ij}^2 (u_j -u_i) \left[ \psi_{i1}^4 - 10 \psi_{i1}^2
\psi_{j1}^2 + \psi_{j1}^4 \right]}{\psi_{i1}^2 \psi_{j1}^2} 
\nn\\
&&+ \sum_j \gamma_{ij} \frac{(\psi_{i1}^2 +\psi_{j1}^2
)}{\psi_{i1}\psi_{j1}}
\left[ \frac{1}{\psi_{i1}}\sum_{k\neq j} V_{ik} \psi_{k1}
-\frac{1}{\psi_{j1}}\sum_{k\neq i} V_{jk}\psi_{k1}\right] 
\nn\\
&&+ \sum_j \ga_{ij} \left( \frac{\psi_{j1}}{\psi_{i1}} 
- \frac{\psi_{i1}}{\psi_{j1}}\right)
\eeqa
where, we recall, $V_{ij} =(u_j-u_i)\, \ga_{ij}$.
Using that $\psi_{i1}$ is an eigenvector of $V$ we rewrite the formula in
the following way
\beq
 \frac{\partial G}{\partial u^i} =\frac{1}{2} \sum_{j\neq i} \frac{V_{ij}^2}
{u_i-u_j} -\frac{1}{24} \sum_{k\neq i} \ga_{ik} \left(
\frac{\psi_{i1}}{\psi_{k1}} -\frac{\psi_{k1}}{\psi_{i1}}\right).
\eeq
Using (\ref{tau-i}) and (\ref{dpsi-du}) we recognize in the r.h.s. the 
derivative
$$
\frac{\partial}{\partial u^i} 
\left[ \log \tau_I -\frac{1}{24}\log (\psi_{11} ... \psi_{n1})\right] .
$$
It remains to observe that
$$
\det \frac{\partial t^\alpha}{\partial u^i} = \psi_{11} ... \psi_{n1}
$$
up to an inessential constant. One can easily check that
$$
\sum_i \frac{\partial G}{\partial u^i} =0,
$$
so
$$\frac{\partial G}{\partial t^1}=0.
$$
The formula (\ref{egregia}) is proved.

Let us derive the formula (\ref{gamma}) for the constant $\ga$. We have
$$
\sum_{i=1}^n u^i \partial_i \log \tau_I =
\frac12 \sum_{j\neq i} \frac{u_i\,V_{ij}^2}{u_i-u_j} = \frac12 \sum_{i<j}
V_{ij}^2 = -\frac14 \mbox{Trace}\ V^2 
=- \frac14 \sum_{\alpha=1}^n \mu_\alpha^2.
$$
The second term in the formula for $G$ gives
$$
\sum_{i=1}^n u_i \partial_i \log (  \psi_{11} ... \psi_{n1}
) = \sum_{j=1}^n \psi_{j1}^{-1} \sum_{i=1}^n  u_i \partial_i \psi_{j1}.
$$
But
$$
\psi_{j1}=\sqrt{\frac{\partial t^n}{\partial u^j}}
$$
and
$$
\sum_{i=1}^n u_i \partial_i t^n = (1-d) t^n \ \mbox{for} \ d\neq 1
$$
$$\sum_{i=1}^n u_i \partial_i t^n =r_n \ \mbox{for} \ d=1
$$
(the Euler identity). So
$$
\sum_{i=1}^n u^i \partial_i \psi_{j1}=-\frac{d}{2} \psi_{j1}.
$$
This proves the formula (\ref{gamma}). Theorem is proved.
\par
\noindent 
{\bf Definition.} The function (\ref{egregia}) is called {\it $G$-function
of the Frobenius manifold}.

\vskip 0.5cm 

We begin our examples with the case $n=2$.
In the 2-dimensional case, we write the free energy $F$ in the form 
$$
F=\frac12 (t^1)^2\,t^2+f(t^2).
$$
The Getzler's equations (\ref{Getz}) are reduced to
\beq
48 f^{(3)}\,\frac{\pal^2 G}{\pal t^2\pal t^2}-24 f^{(4)}\,
\frac{\pal G}{\pal t^2}-f^{(5)}=0
\eeq
(cf. \cite{Kimura}).
For the free energy 
$$
F=\frac12 (t^1)^2\,t^2+c\,(t^2)^{h+1},
$$
where $c$ is an arbitrary non-zero constant,
the $G$-function is
$$
G=-\frac1{24} \frac{(2-h) (3-h)}h\, \log(t^2).
$$
Particularly, the $G$-function vanishes for the $A_2$ topological
minimal model (the case $h=3$). The constant $\ga$ equals
$$
\ga=\frac{d\,(1-3\,d)}{24}
$$
since $d=1-\frac2{h}$.
For the free energy of the $CP^1$ model
$$
F=\frac12 (t^1)^2\,t^2+e^{t^2}
$$
the $G$-function reads
$$
G=-\frac1{24} t^2.
$$
The constant is
$$
\ga=-\frac1{12}.
$$
Observe that the $G$-function is analytic everywhere on the 
Frobenius manifold only 
for $d=\frac13$ (the $A_2$ topological minimal model) and
for $d=1$, i.e., for the $CP^1$ topological sigma model.
\vskip 0.5cm
 
Let us consider now examples with $n=3$. We will take the list of
examples of Frobenius manifolds with good analytic properties from \cite{D3}.

\noindent 1)\  For the polynomial free energy related to $A_3$, 
\beq\label{FM-A3}
F=\frac12 (t^1)^2\,t^3+\frac12 t^1\,(t^2)^2+ 
\frac14 (t^2)^2\, (t^3)^2+\frac1{60} (t^3)^5,
\eeq
we have $G=0, \ \ga=0$;

\noindent 2)\  For the polynomial free energy related to $B_3$, 
$$
F=\frac12 (t^1)^2\,t^3+\frac12 t^1\,(t^2)^2+\frac16 (t^2)^3\,t^3+
\frac16 (t^2)^2\,(t^3)^3+\frac1{210} (t^3)^7,
$$
we have
$$
G=-\frac1{48}\,\log(2\,t^2-3\,(t^3)^2),\quad \ga=-\frac{1}{72}.
$$

\noindent 3)\ For the polynomial free energy related to the 
symmetry group of icosahedron, 
$$
F=\frac12 (t^1)^2\,t^3+\frac12 t^1\,(t^2)^2+\frac16 (t^2)^3\,(t^3)^2+
\frac1{20} (t^2)^2\,(t^3)^5+\frac1{3960} (t^3)^{11},
$$
we have
$$
G=-\frac1{20}\,\log(t^2-(t^3)^3),\quad \ga=-\frac3{100}.
$$

\noindent 4)\ For the free energy of the $CP^2$ model, 
\beq\label{FE-cp2}
F=\frac12 (t^1)^2\,t^3+\frac12 t^1\,(t^2)^2+
\sum_{k\ge 1} N^{(0)}_k\,\frac{(t^3)^{3\,k-1}}{(3\,k-1)!}\,e^{k t^2},
\eeq
where $N^{(0)}_k$ are the number of rational  curves of degree $k$ on
$CP^2$ 
which meet $3\,k-1$ generic points, for example, 
$N^{(0)}_1=N^{(0)}_2=1, \ N^{(0)}_3=12,\ N^{(0)}_4=620$. 
The $G$-function has the form
\beq\label{CP2-G}
G=-\frac{t^2}8+\sum_{k\ge 1} N^{(1)}_k\,\frac{(t^3)^{3\,k}}{(3\,k)!}\,
e^{k t^2}, \quad
\left. \frac{\pal G}{\pal t^2}\right|_{t^2=z, t^3=1}
=\frac{\phi'''-27}{8(27+2\,\phi'-3\,\phi'')},\quad \ga=-\frac38,
\eeq
where $\phi$ is defined by (\ref{Defofphi}), and 
$N^{(1)}_k$ are the number of elliptic plane curves of
degree $k$ which meet $3\,k$ generic points, for example, 
$N^{(1)}_1=N^{(1)}_2=0, \ N^{(1)}_3=1,\ N^{(1)}_4=225$. 

\vskip 0.5cm
\noindent 5) For the free energy
\beq
F=\frac12\,(t^1)^2\,t^3+\frac12\,t^1\,(t^2)^2+(t^2)^4\,\phi(t^3-2\,r\,
\log(t^2))
\eeq
(here $d=1$, $r>0$) we obtain a solution of WDVV with good analytic 
properties only for $r=\frac32, 1$ or $\frac12$ \cite{D3}. These solutions
correspond to extended affine Weyl groups of type ${\tilde A}_2,
{\tilde C}_2, {\tilde G}_2$ respectively \cite{D5}. For all of them 
$\gamma=-1/16$. Particularly, 
for ${\tilde A}_2$
\beq\label{FM-At2}
\phi(z)=-\frac1{24}+e^z,
\eeq
then $G=-\frac1{24}\,t^3, $.\par
 For ${\tilde C}_2$
\beq
\phi(z)=-\frac1{48}+a\,e^z+\frac{a^2}2\,e^{2\,z},
\eeq
where $a$ is an arbitrary non-zero constant,
the $G$-function is
$$
G=-\frac1{24}\,t^3-\frac1{48}\,\log(16\,a\,e^{t^3}-(t^2)^2).
$$

Finally, for ${\tilde G}_2$
\beq
\phi(z)=-\frac1{72}+\frac23\,e^z+\frac32\,e^{2\,z}+\frac9{16}\,e^{4\,z},
\eeq
the $G$-function is 
$$
G=-\frac1{24}\,t^3-\frac1{12}\,\log(12\,e^{t^3}-t^2).
$$
\vskip 0.5cm

\noindent 6) We now take the free energy
\beq
F=\frac12\,(t^1)^2\,t^3+\frac12\,t^1\,(t^2)^2
-\frac1{16}\,(t^2)^4\,\phi(t^3),
\eeq
where $\phi(z)$ satisfies the Chazy equation
\beq
\phi'''=6\,\phi\,\phi''-9\,\,(\phi')^2,
\eeq
(here $d=1$, $r=r_3=0$). Then the $G$
function can be obtained from the equations
\beq
\frac{\pal G}{\pal t^2}=-\frac1{8\,t^2},\quad
\frac{\pal G}{\pal t^3}=-\frac14\,\phi(t^3),\quad \ga=-\frac1{16}.
\eeq
Particularly, for the case
\beq
\phi(t^3)=8\,\pi\,i\,E_2(t^3)=4\,\frac{d}{d t^3}\log\eta(t^3)
\eeq
where $\eta(\tau)$ is the Dedekind function, $E_2(\tau)$ is the second
Eisenstein series (see \cite{D3}) we obtain
\beq
G=-\log\left[\eta(t^3)\,(t^2)^{\frac18}\right].
\eeq

We see that, for $n=3$, only on the Frobenius manifold (\ref{FM-A3})
(the free energy of the $A_3$ topological minimal model), and on the Frobenius
manifold (\ref{FM-At2}) related to the extended affine Weyl group 
${\tilde A}_2$ the $G$-function are manifestly analytic 
everywhere. For the $CP^2$ sigma model the 
$G$-function is regular on the open subset where
\beq
27+2\,\phi'-3\,\phi''\ne 0.
\eeq
From equations of associativity for the function (\ref{FE-cp2}) it can
be seen (see \cite{DI}) that in the points $x_0$ where
\beq
3\,\phi''(x_0)-2\,\phi'(x_0)-27=0
\eeq
the series (\ref{CP2-G}) diverges. Analytic
properties of the $G$-function (\ref{CP2-G}) deserve a
separate investigation.

{\bf Remark.} In  the  cases $B_3$, $H_3$, $\tilde B_2$, $\tilde G_2$ the
$G$-function has logarithmic branching on the part of the nilpotent
locus of the Frobenius manifold, where $u_i=u_j$ for some $i\neq j$. The
coefficients of our hierarchy will have singularities in these points.
Probably, appearance of these singularities suggests not to select these
Frobenius manifolds for a construction of a physically consistent model
of 2D TFT. 

\vskip 0.6cm
\setcounter{equation}{0}

\section {Some examples}
We now compare the dispersion expansions of  
some well known examples of bi-Hamiltonian integrable
systems with
those given in Theorem 1 and Theorem 2.\par

\vskip 0.5cm

{\bf Example 1} $\ $ Let's start with the detailed consideration 
of the simplest example
of KdV hierarchy. We take the Lax operator in the form
\beq
L=\frac12\,(\ve\,\pal_X)^2+u(X).
\eeq
Then the two compatible Poisson brackets related to this operator
is given by 
\eqa
&&\{u(X),u(Y)\}_1=\de'(X-Y),
\\
&& \{u(X),u(Y)\}_2=u(X)\,\de'(X-Y)+\frac12\,u_X(X)\,\de(X-Y)+\frac{\ve^2} 8
\,\de'''(X-Y).
\eeqa
(They are derived from the formulae (\ref{HA}) and (\ref{BR}) in Example 2).
Starting from the Casimir
\beq
H_{-1}=\int u(X) dX
\eeq
of the first Poisson bracket we can construct a hierarchy of 
commuting Hamiltonians $H_p$ by using the following recursion
relation
\beq
\{u(X), H_{p-1}\}_2=\left(\frac12+p\right)\{u(X),H_p\}_1,
\eeq
i.e.,
\beq
\left(u(X)\pal_X+\frac12\,u_X(X)+
\frac{\ve^2}8\,\pal_X^3\right)\frac{\de H_{p-1}}
{\de u}=\left(\frac12+p\right)\,\pal_X\frac{\de H_{p}}{\de u}.
\eeq

Note that the factor $\left(\frac12+p\right)$ does not appear 
in the usual recursion relation for the  KdV hierarchy, we use 
this factor here to meet the topological recursion relation 
of the $A_1$ topological minimal model.  Let's list the first four 
Hamiltonians
\eqa
&&H_{-1}=\int u(X) dX,\quad H_{0}=\int \frac12 u(X)^2 dX,
\nn\\
&& H_1=\int\left(\frac16\,u(X)^3-\frac1{24}\,\ve^2\,u_X(X)^2\right) dX,
\nn\\
&&H_2=\int\left(\frac1{24} u(X)^4-\frac1{24}\,\ve^2\,u(X)\,u_X(X)^2+
\frac1{480}\,\ve^4\,u_{XX}(X)^2\right) dX.
\eeqa
The KdV hierarchy is then given recursively as
\eqa
&&\frac{\pal u}{\pal T^0}=u_X,
\nn\\
&& \frac{\pal u}{\pal T^1}=u\,u_X+\frac1{12}\,\ve^2\,u_{XXX},
\nn\\
&& \frac{\pal u}{\pal T^p}=(\frac12+p)^{-1}\left(
\frac12 u_X\pal_X^{-1}+u+
\frac18\ve^2\pal_X^2\right)\frac{\pal u}{\pal T^{p-1}}.
\eeqa
Let's note that each flow of the KdV hierarchy can be written
as a polynomial in $\ve^2$. The parameter $\ve$ can be introduced 
to the usual KdV hierarchy through the rescaling $X\mapsto\ve\,X,\
T^p\mapsto\ve\,T^p$.
 We write down explicitly the 
$\ve^0$ and $\ve^2$ terms in the
hierarchy 
\eqa
&&\frac{\pal u}{\pal T^0}=u_X,
\nn\\
&& \frac{\pal u}{\pal T^1}_1=u\,u_X+\frac1{12}\,\ve^2\,u_{XXX},
\nn\\
&&\frac{\pal u}{\pal T^p}=\{u(X),H^{(0)}_p\}_1+
\ve^2\,\{u(X),H^{(1)}_p\}_1+
{\cal O}(\ve^4).
\eeqa
where
\beq
H^{(0)}_p=\int \frac{u(X)^{p+2}}{(p+2)!} dX
\eeq
and
\eqa
&&H^{(1)}_{-1}=H^{(1)}_0=0,
\nn\\
&&H^{(1)}_p=\int\left(-\frac1{24}\right)\frac{u(X)^{p-1}}{(p-1)!}
\,u_X(X)^2 dX.
\eeqa  

We now take the free energy to be
\beq
F=\frac1{6}\,(t^1)^3,
\eeq
in this case the $G$-function $G=0$. Plugging this free energy into
Theorem 1 and Theorem 2, and identify $t^1, T^{p,1}, H_{1,p}, \de H'_{1,p}$
with $u, T^p, H^{(0)}_{p}, H^{(1)}_p$ respectively, we obtain,
modulo ${\cal O}(\ve^4)$, the  above described KdV hierarchy and
it's bihamiltonian structure.
\vskip 0.5cm

{\bf Example 2} $\ $ 
More generally, let's  
consider the differential operators
\begin{equation}\label{L}
L=(\ve\,\pal)^{N+1}+u_{N}(X) (\ve\,\pal)^{N-1}
+\dots+u_1(X),
\end{equation}
where $\pal=\frac{\pal}{\pal X}$.
For any pseudo-differential operator $Z$ of the form
\begin{equation}
Z=(\ve\,\pal)^{-1} Z_1+(\ve\,\pal)^{-2} Z_2+
\dots+(\ve\,\pal)^{-N} Z_{N},
\end{equation}
define the following two Hamiltonian mappings \cite{Dickey}:
\begin{equation}\label{HA}
H_1: \ Z\mapsto [Z,L]_+,\quad
H_2:\ Z\mapsto L (Z L)_+-(L Z)_+L+\frac1{N+1}[L, \int^X Res[Z,L]\,dX],
\end{equation}
and the corresponding Poisson brackets 
\begin{equation}\label{BR}
\{ \tilde f,\tilde g\}_i=\int Res (H_i(\frac {\delta f}{\delta L})
\frac{\delta g}{\delta L}) dx,\quad i=1,2,
\end{equation}
for the functionals
$$
\tilde f=\int f dx,\quad \tilde g=\int g dx,
$$
where
\beq\label{VF}
\frac {\delta f}{\delta L}=\sum_{i=1}^{N} (\ve\,\pal)^{-i}
\frac {\delta f}{\delta u_i},\quad 
\frac {\delta g}{\delta L}=\sum_{i=1}^{N} 
(\ve\,\pal)^{-i} \frac {\delta 
g}{\delta u_i}.
\eeq

In the case of $N=2$, if we define
\beq
t^1=u_1-\frac12\,\ve\,u_2',\quad t^2=\frac13 \,u_2,
\eeq
then  the first Poisson bracket is given as follows:
\eqa
&\{t^1(X),t^1(Y)\}_1=0,\ \ &\{t^1(X),t^2(Y)\}_1=\de'(X-Y),
\\
&\{t^2(X),t^2(Y)\}_1=0,\ \ &
\eeqa
and the second Poisson bracket is given as 
\eqa
\{t^1(X),t^1(Y)\}_2&=&
-6\,(t^2)(X)^2\,\de'(X-Y)-6\,t^2(X)\,(t^2)'(X)\,\de(X-Y)
\nn\\
&&-\ve^2\left(\frac{15}4 (t^2)'(X)\,\de''(X-Y)+
\frac94\,(t^2)''(X)\,\de'(X-Y)
\right.
\nn\\ 
&&+\left.\frac52\,t^2(X)\,\de^{(3)}(X-Y)
+\frac12\,(t^1)^{(3)}(X)\,\de(X-Y)\right)-
\frac{\ve^4}6\,\de^{(5)}(X-Y),
\nn\\
\{t^1(X),t^2(Y)\}_2&=&t^1(X)\,\de'(X-Y)+\frac13\,(t^1)'(X)\,\de(X-Y),
\nn\\
\{t^2(X),t^2(Y)\}_2&=&\frac23\,t^2(X)\,\de'(X-Y)+
\frac13\,(t^2)'(X)\,\de(X-Y)
\nn\\
&&+\frac{2\ve^2}9\,\de^{(3)}(X-Y).
\eeqa
The integrable hierarchy has the form
\beq\label{addinte}
\frac{\pal t^\al}{\pal T^{\beta,p}}=\{t^\al(X),H_{\beta,p}\}_1,
\eeq
where the Hamiltonians $H_{\beta,p}$ are recursively defined by
\beq\label{addinteb}
\{t^\al(X),H_{\beta,p-1}\}_2=\left(\frac{1-d}2 +p+q_\beta\right)\,
\{t^\al(X),H_{\beta,p}\}_1
\eeq
with $H_{\beta,-1}=\int t_\beta(X) dX$.\par
Up to the $\ve^2$ terms, the above Poisson brackets and the
integrable hierarchy 
coincide with the Poisson brackets and the integrable 
hierarchy given in Theorem 1 and Theorem 2
with the free energy defined by
\beq
F=\frac12\,(t^1)^2\,t^2-\frac38\,(t^2)^4,
\eeq 
and with the $G$-function $G=0$.
This is the primary free energy of the
$A_2$ topological minimal model of \cite{Dij2}.\par

In the case of $N=3$, if we define
\beq\label{WD}
t^1=u_1-\frac18 u_3^2-\frac{\ve}2 u_2'+\frac{\ve^2}{12}\, u_3'',\quad
t^2=u_2-\ve\,u_3',\quad t^3=u_3,
\eeq
then the Poisson brackets defined by (\ref{HA}), (\ref{BR}) 
and the integrable system given in the form of (\ref{addinte}) and
(\ref{addinteb}) 
coincide, modulo $\ve^4$, with the Poisson
brackets and the integrable system
given in Theorem 1 and Theorem 2 with the free energy defined
by
\beq
F=\frac18\,t^1\,(t^2)^2 + \frac18\,(t^1)^2\,t^3 -
\frac1{64}\,(t^2)^2\,(t^3)^2 +
\frac1{3840} \,(t^3)^5,
\eeq
and with the $G$-function $G=0$.
This is the primary free energy of the
$A_3$ topological minimal model [{\it ibid}]. Formulae in 
(\ref{WD}) coincide with formulae in (4.38) of \cite{Dij1}.
\vskip 0.5cm

{\bf Example 3} $\ $
The explicit bihamiltonian structure related to the Lie 
algebra of type $B_2$ is given, for example, in
\cite{Casati} in the following form:
\eqa
\{u_1(X),u_1(Y)\}_1&=&
2\, u_2(X)\,\de'(X-Y)+u_2'(X)\,\de(X-Y)- \ve^2\,\de^{(3)}(X-Y),
\nn\\
\{u_1(X),u_2(Y)\}_1&=&
2\,\de'(X-Y),
\nn\\
\{u_2(X),u_1(Y)\}_1&=&2\,\de'(X-Y),
\nn\\
\{u_2(X),u_2(Y)\}_1&=&0;\label{pbb1}
\eeqa
\eqa
\{u_1(X),u_1(Y)\}_2&=&
2 u_2(X)\, u_1(X)\, \de'(X-Y)+u_2'(X)\, u_1(X)\, \de(X-Y)
\nn\\
&&
+u_2(X)\,u_1'(X)\, \de(X-Y)
-\ve^2\,[
\frac32\, u_2'(X)^2\, \de'(X-Y)
\nn\\
&&+6\, u_2(X)\, u_2'(X)\, \de''(X-Y)+\frac32\, u_1'(X)\, \de''(X-Y)
\nn\\
&&+4\, u_2(X)\, u_2''(X)\, \de'(X-Y)+\frac12\, u_2'(X)\, u_2''(X)\,
 \de(X-Y)
\nn\\
&&+\frac32\, u_1''(X)\, \de'(X-Y)+2\, u_2(X)^2 \,\de^{(3)}(X-Y)
\nn\\
&&+u_1(X)\, \de^{(3)}(X-Y)
+u_2(X)\, u_2^{(3)}(X)\, \de(X-Y)
\nn\\
&&+\frac12\, u_1^{(3)}(X)\,\de(X-Y)]
+\ve^4 [8 u_2''(X)\, \de^{(3)}(X-Y)
\nn\\
&&+7\, u_2^{(3)}(X)\, 
\de''(X-Y)+5 u_2'(X)\, \de^{(4)}(X-Y)
\nn\\
&&+3 u_2^{(4)}(X) \de'(X-Y)+2 u_2(X) \de^{(5)}(X-Y)
\nn\\
&&+\frac12\, u_2^{(5)}(X)\, \de(X-Y)]
-\frac12\, \ve^6 \,\de^{(7)}(X-Y),
\nn\\
\{u_1(X),u_2(Y)\}_2&=&
2 u_1(X)\, \de'(X-Y)+\frac12\, u_1'(X)\, \de(X-Y)
\nn\\
&&-\ve^2\left(u_2'(X)\, \de''(X-Y)+2\,  u_2(X)\, \de^{(3)}(X-Y)\right)
+\ve^4\, \de^{(5)}(X-Y),
\nn\\
\{u_2(X),u_1(Y)\}_2&=&
2 u_1(X) \de'(X-Y)+\frac32\, u_1'(X) \de(X-Y)-\ve^2\,[
u_2^{(3)}(X) \de(X-Y)
\nn\\
&&+5 u_2'(X)\, \de''(X-Y)+2\, u_2(X)\, \de^{(3)}(X-Y)+
        +4\, u_2''(X)\, \de'(X-Y)]
\nn\\
&&+\ve^4\, \de^{(5)}(X-Y),
\nn\\
\{u_2(X),u_2(Y)\}_2&=&
u_2(X)\, \de'(X-Y)+\frac12\, u_2'(X)\, \de(X-Y)-\frac52\, 
\ve^2\, \de^{(3)}(X-Y),\label{pbb2}
\eeqa
here we note that the above  coordinates $u_1, u_2$ should be the coordinates
$u_1, u_0$ respectively in \cite{Casati}, and there is a
sign difference between the above  first Poisson bracket and that of
\cite{Casati}.\par 
We now compare the above Poisson brackets with the
Poisson brackets given by Theorem 1 and Theorem 2 with the
free energy related to $B_2$. For this, let
\beq\label{Fb2}
F=\frac12\, (t^1)^2\, t^2+\frac1{15}\,(t^2)^5,
\eeq
then the $G$-function is given by $G=-\frac1{48}\,\log(t^2)$,
and the first and second Poisson brackets of Theorem 1 and
Theorem 2  are given by

\eqa
&&\{t^1(X),t^1(Y)\}_1=\{t^2(X),t^2(Y)\}_1={\cal O}(\ve^4),
\nn\\
&&\{t^1(X),t^2(Y)\}_1=\de'(X-Y)+{\cal O}(\ve^4);
\eeqa
\eqa
&&\{t^1(X),t^1(Y)\}_2=
2\,{{t^2(X)}^3}\,\de'(X-Y) + 
  3\,{{t^2(X)}^2}\,(t^2)'(X)\,\de(X-Y)
\nn\\
&&\quad + 
  {\ve^2}\,\left( {{{{(t^1)'(X)}^2 \,\de'(X-Y)}}\over 
       {32\,{{t^2(X)}^2}}} 
- {{{{(t^1)'(X)}^2}\,
         (t^2)'(X) \,\de(X-Y)}\over {32\,{{t^2(X)}^3}}}\right. 
\nn\\
&&\quad+ 
     {{29\,{{(t^2)'(X)}^2\,\de'(X-Y)  }}\over {24}} + 
     {{13\,t^2(X)\,(t^2)'(X)\,\de''(X-Y)}\over 4} 
\nn\\
&&\quad+ 
     {{(t^1)'(X)\,(t^1)''(X)\,\de(X-Y) }\over {32\,{{t^2(X)}^2}}} 
+ 
     {{25\,t^2(X)\,(t^2)''(X)\,\de'(X-Y) }\over {12}} 
\nn\\
&&\quad+ 
    {{5\,(t^2)'(X)\,(t^2)''(X)\,\de(X-Y)  }\over 8} + 
     {{13\,{{t^2(X)}^2}\,\de^{(3)}(X-Y)}\over {12}} 
\nn\\
&&\quad\left.+ 
     {{\de(X-Y)\,t^2(X)\,(t^2)^{(3)}(X)\,\de(X-Y)  }\over 2} \right), 
\nn\\
&&
\{t^1(X),t^2(Y)\}_2
=t^1(X)\,\de'(X-Y) + {{(t^1)'(X)\,\de(X-Y) }\over 4} 
\nn\\
&&\quad+ 
  {\ve^2}\,\left( {{-\left( (t^1)'(X)\,\de''(X-Y) \right) }\over 
       {24\,t^2(X)}}
+{{t^1(X)\,(t^2)'(X)\,\de''(X-Y)}\over 
       {24\,{{t^2(X)}^2}}} 
\right.
\nn\\
&&\quad\left.- {{t^1(X)\,\de^{(3)}(X-Y)}\over 
       {24\,t^2(X)}} \right), 
\nn\\
&&
\{t^2(X),t^2(Y)\}_2
=
{{t^2(X)\,\de'(X-Y)}\over 2} + 
  {{(t^2)'(X)\,\de(X-Y) }\over 4}
\nn\\
&&\quad + 
  {{3\,{\ve^2}\,\de^{(3)}(X-Y)}\over {16}}.
\eeqa
Now if we relate the variables $u_1, \ u_2$ to the variables $
t^1,\ t^2$ \ by the following relation:
\beq
t^1=u_1-\frac14\,u_2^2+\frac{\ve^2}4\,u_2'',\quad
t^2=\frac12\,u_2,
\eeq
then the above first Poisson brackets coincide, modulo $\ve^4$,
with the Poisson brackets given in (\ref{pbb1}).
While for the second Poisson brackets, they coincide with the
Poisson brackets given in (\ref{pbb2}) only up to $\ve^0$, 
and starting from the $\ve^2$ terms, the two second Poisson brackets
no longer coincide. This result is in accordance with the
result of \cite{Egu6}, where it was shown, by imposing the
commutativity of the flows, that the integrable system (the tree-level one)
related to the free energy (\ref{Fb2}) can not be
extended beyond $\ve^2$ terms.
\par
{\bf Remark.}\ In this section, 
the free energies corresponding to the Lie algebras of the types
$A_2, A_3, B_2$
are different from those given in Section 6, they are related
by a rescaling.
\vskip 0.5cm

{\bf Example 4} $\ $
Consider the Toda lattice equation with open boundary
\eqa
\frac{\pal u_n}{\pal t}&=&v_n-v_{n-1},\nn\\
\frac{\pal v_n}{\pal t}&=&e^{u_{n+1}}-e^{u_n},\quad n\in {\bf Z}.
\eeqa
If we introduce the slow variables $\ T=t\,\ve,\ X=n\,\ve$,\ and the
new dependent variables $\ {\tilde u}(X)=u_n,\
{\tilde v}(X)=v_n$,\ then the Toda lattice equations lead  to
\eqa
\frac{\pal {\tilde u}}{\pal T}&=&\frac1{\ve} \left(
{\tilde v}(X)-{\tilde v}(X-\ve)\right),\nn\\
\frac{\pal {\tilde v}}{\pal T}&=&\frac1{\ve} \left(
e^{{\tilde u}(X+ \ve)}-e^{{\tilde u}(X)}\right).
\eeqa
This system has the bi-Hamiltonian structure
\eqa
\frac{\pal {\tilde u}}{\pal T}&=&\{{\tilde u}(X), H_0\}_1=
\{{\tilde u}(X), H_{-1}\}_2,\nn\\
\frac{\pal {\tilde v}}{\pal T}&=&\{{\tilde v}(X), H_0\}_1=
\{{\tilde v}(X), H_{-1}\}_2,
\eeqa
where the Poisson brackets are defined by
\eqa
&&\{{\tilde u}(X), {\tilde u}(Y)\}_1=\{{\tilde v}(X), {\tilde v}(Y)\}_1=0,
\nn\\
&&\{{\tilde u}(X), {\tilde v}(Y)\}_1=\frac1\ve \left(\del(X-Y)-
\del(X-Y-\ve)\right);\\
&&\{{\tilde u}(X), {\tilde u}(Y)\}_2=\frac1\ve \left(\del(X-Y+\ve)-
\del(X-Y-\ve)\right),\nn\\
&&\{{\tilde v}(X), {\tilde u}(Y)\}_2=\frac1\ve \left(\del(X-Y+\ve)-
\del(X-Y)\right)\, {\tilde v}(X),\nn\\
&&\{{\tilde v}(X), {\tilde v}(Y)\}_2=\frac1\ve \left(
e^{{\tilde u}(X+\ve)}\del(X-Y+\ve)-
e^{{\tilde u}(X)}\del(X-Y-\ve)\right);
\eeqa
and the Hamiltonians are given by
\beq
H_{-1}=\int {\tilde v}(X) dX,\quad
H_0=\int \left(\frac12 {\tilde v}(X)^2+e^{{\tilde u}(X)}\right) dX.
\eeq
We construct the hierarchy of integrable systems
\beq\label{tdsys}
\frac{\pal {\tilde u}}{\pal T^p}=\{{\tilde u}(X), H_p\}_1,
\quad
\frac{\pal {\tilde v}}{\pal T^p}=\{{\tilde v}(X), H_p\}_1
\eeq
with the Hamiltonians $H_{p}$ recursively defined by
\beq\label{tdham}
\{{\tilde u}(X), H_{p-1}\}_2=(p+1)\{{\tilde u}(X), H_{p}\}_1,
\quad
\{{\tilde v}(X), H_{p-1}\}_2=(p+1)\{{\tilde v}(X), H_{p}\}_1.
\eeq
We identify $T^0$ with $T$.

Let's define again the following new variables:
\eqa
t^1(X)&=&{\tilde v}(X)-\frac{\ve^2}{24}\, {\tilde v}''(X)+{\cal O}(\ve^4),
\\
t^2(X)&=&{\tilde u}(X)+\frac{\ve}2\,{\tilde u}'(X)
+\frac{\ve^2}{24} {\tilde u}''(X)-\frac{\ve^3}{48}\,{\tilde u}'''(X)+
{\cal O}(\ve^4),
\eeqa
and expand the above Poisson brackets in Taylor series in $\ve$,
we obtain
\eqa
&&\{t^1(X),t^1(Y)\}_1=\{t^2(X),t^2(Y)\}_1=0,\nn\\
&&\{t^1(X),t^2(Y)\}_1=\de'(X-Y)-\frac{\ve^2}{12}\,\de^{(3)}(X-Y)+
{\cal O}(\ve^4);\\
&&\{t^1(X),t^1(Y)\}_2=2\,e^{t^2(X)}\,\de'(X-Y)+e^{t^2(X)}\, (t^2(X))'\,
\de(X-Y)\nn\\
&&\quad +\ve^2\,\left(\frac16\, \del^{(3)}(X-Y)+\frac14 \,
(t^2(X))'\,\de''(X-Y)
+\frac1{12}\,{((t^2(X))')^2}\,\de'(X-Y)\right.
\nn\\
&&\quad
+\frac14\, (t^2(X))''\,\de'(X-Y)+
\frac1{12}\, (t^2(X))'\,(t^2(X))''\,\de(X-Y)
\nn\\
&&\quad\left.
+\frac1{12}\, (t^2(X))^{(3)}\,
\de(X-Y)\right)\,e^{t^2(X)}
+{\cal O}(\ve^4),
\nn\\
&&\{t^1(X),t^2(Y)\}_2=t^1(X)\,\de'(X-Y)-\ve^2\,\left(\frac1{12}\, t^1(X)\,
\de^{(3)}(X-Y)\right.
\nn\\
&&\quad\left.+\frac1{12}\,(t^1(X))'\,\de''(X-Y)\right)+
{\cal O}(\ve^4),
\nn\\
&&\{t^2(X),t^2(Y)\}_2=2\,\de'(X-Y)+{\cal O}(\ve^4).\label{tdpbk} 
\eeqa
We also expand the Hamiltonians (\ref{tdham})
and the integrable system (\ref{tdsys}) in Taylor series in $\ve$.
The Hamiltonians $H_{-1}$ and $H_0$  have the form
\eqa
H_{-1}&=&\int t^1(X) dX+{\cal O}(\ve^4),
\nn\\
H_0&=&\int \left(\frac12 (t^1(X))^2+e^{t^2(X)}\right) dX
\nn\\
&&\quad -
\frac{\ve^2}{12}\int \left(\frac12 \left(t^1_X(X)\right)^2+
e^{t^2(X)}\,\left(t^2_X(X)\right)^2\right) dX
+{\cal O}(\ve^4).
\eeqa
Now if we put the $CP^1$ free energy 
\beq
F=\frac12 (t^1)^2\,t^2+e^{t^2}
\eeq
into Theorem 1--Theorem 3, and with $G$-function 
$G=-\frac1{24} t^2$,
we get the Poisson brackets which coincide
 with those given in (\ref{tdpbk}) modulo $\ve^4$, 
and the Hamiltonians $H_{2,p}$ we get also 
coincide with $H_{p}$ modulo $\ve^4$.
This suggests that the Toda lattice hierarchy
is the appropriate hierarchy of integrable systems behind the
$CP^1$ model, as it was suggested in \cite{Egu6} from the point of view of
commuting flows.

\setcounter{equation}{0}
\section{Discussion}\par

We formulate here the conjectural shape of the 
integrable hierarchy to be considered starting from a 
Frobenius manifold and of its bihamiltonian 
structure in the form of genus expansion. The hierarchy must have the form
\beq\label{F-Hier}
\frac{\pal t}{\pal T^{\al,p}}=K^{(0)}_{\al,p}(t,\tx)+
\sum_{k\ge 1}\ve^{2\,k}\,K^{(k)}_{\al,p}(t,\tx,\txx,\dots)
=\{t(X),H_{\al,p}\}_1
\eeq
where the Hamiltonians and the first Poisson bracket must have the 
expansions
\eqa
&&H_{\al,p}=H^{(0)}_{\al,p}+\sum_{k\ge 1}\ve^{2\,k}\, H^{(k)}_{\al,p},
\\
&&\{t^\al(X),t^\beta(Y)\}_1=\{t^\al(X),t^\beta(Y)\}^{(0)}_1+
\sum_{k\ge 1}\ve^{2\,k}\,\{t^\al(X),t^\beta(Y)\}^{(k)}_1\label{F-PB}
\eeqa
where
\eqa
&&H^{(k)}_{\al,p}=\int P^{(k)}_{\al,p}(t;\tx,\txx,\dots) dX,
\\
&&\{t^\al(X),t^\beta(Y)\}^{(k)}_1=\sum_{s=0}^{2\,k+1} A_{k,s}^{\al,\beta}(
t;\tx,\txx,\dots)|_{t=t(X)}\,\de^{(s)}(X-Y).\label{F-PBS}
\eeqa
The densities $P^{(k)}_{\al,p}(t;\tx,\txx,\dots)$ and the coefficients 
$A_{k,s}^{\al,\beta}(t;\tx,\txx,\dots)$ of the Poisson bracket are 
quasihomogeneous polynomials in $\tx, \txx,\dots$ of the degrees
\eqa
&&\deg P^{(k)}_{\al,p}(t;\tx,\txx,\dots)=2\,k,
\\
&&\deg A_{k,s}^{\al,\beta}(t;\tx,\txx,\dots)=2\,k+1-s
\eeqa
where we assign the degrees
\beq\label{Def-deg}
\deg \pal_X^m t=m
\eeq
for any $m=1,2,\dots$. The coefficients $K^{(k)}_{\al,p}(t;\tx,\txx,\dots)$
of the hierarchy are also polynomials in the same variables of the degree
\beq
\deg K^{(k)}_{\al,p}(t;\tx,\txx,\dots)=2\,k+1,\quad k=0,1,\dots.
\eeq
All the Hamiltonians must commute.

{\bf Remark.} $\ $ The dispersion expansions of the known integrable
hierarchies obtained by simultaneous rescaling $x\mapsto \ve\,x,\ t\mapsto
\ve\,t$ for any time variable $t$ contain also odd powers of $\ve$.
However, doing an appropriate $\ve$-dependent change of dependent variables
we can reduce the hierarchy and their Poisson brackets to the 
form postulated in this section. (See examples above in Section 7).
\vskip 0.4cm 

We expect that the quasihomogeneity (\ref{wdvv3}) will not be involved in the 
construction of the hierarchy. If, however, it takes place then the 
coefficients of the first Poisson bracket must satisfy another 
quasihomogeneity condition. Let us introduce the extended Euler vector 
field
\beq
{\cal E} :=E+\sum_{m\ge 1}\sum_{\al}(1-m-q_\al)\,\pal_X^m t^\al\,\frac
{\pal}{\pal(\pal_X^m t^\al)},
\eeq
where the Euler vector field $E$ has the form (\ref{VF-Euler}). Then the
coefficients of the first Poisson bracket (\ref{F-PB}) must 
satisfy the quasihomogeneity conditions
\beq\label{hom-A}
{\cal L}_{{\cal E}} A^{\al\beta}_{k,s}(t;\tx,\txx,\dots)=
(k(d-3)+d+s-1-q_\al-q_\beta)\,A^{\al\beta}_{k,s}(t;\tx,\txx,\dots).
\eeq
Moreover, there exists another Poisson bracket with the structure 
similar to (\ref{F-PB}), (\ref{F-PBS})
\eqa
&&\{t^\al(X),t^\beta(Y)\}_2=\{t^\al(X),t^\beta(Y)\}^{(0)}_2+
\sum_{k\ge 1}\ve^{2\,k}\,\{t^\al(X),t^\beta(Y)\}^{(k)}_2\label{S-PB}
\\
&&\{t^\al(X),t^\beta(Y)\}^{(k)}_2=\sum_{s=0}^{2\,k+1} B_{k,s}^{\al,\beta}(
t;\tx,\txx,\dots)|_{t=t(X)}\,\de^{(s)}(X-Y),\label{S-PBS}
\eeqa
where $B_{k,s}^{\al,\beta}(
t;\tx,\txx,\dots)$ are polynomials in $\tx,\txx,\dots$ of the same degree 
$2\,k+1-s$ in the sense of (\ref{Def-deg}). The quasihomogeneity conditions
for the coefficients of the second Poisson bracket have the form
\beq\label{hom-B}
{\cal L}_{{\cal E}} B^{\al\beta}_{k,s}(t;\tx,\txx,\dots)=
(k(d-3)+d+s-q_\al-q_\beta)\,B^{\al\beta}_{k,s}(t;\tx,\txx,\dots).
\eeq 
The Poisson brackets $\{\ ,\ \}_1$ and $\{\ ,\ \}_2$ must be compatible,
i.e., any linear combination of them with arbitrary constant coefficients
must be again a Poisson bracket. Besides
\beq
\frac{\pal}{\pal t^1} B^{\al\beta}_{k,s}=A^{\al\beta}_{k,s},
\quad
 \frac{\pal}{\pal t^1} A^{\al\beta}_{k,s}=0.
\eeq
All the equations of the hierarchy (\ref{F-Hier}) with the numbers 
$(\al,p)$ such that
\beq
\frac12+\mu_\al+p\ne 0
\eeq
are Hamiltonian flows also w.r.t. the second Poisson bracket.

Additional conjecture about the bihamiltonian structure 
(\ref{F-PB}), (\ref{S-PB}) is that, for $d\neq 1$
\eqa
&&\{t^n(X), t^n(Y)\}^{(k)}_1=0\quad\mbox{for}\ k>0,
\nn\\
&&\{t^n(X), t^n(Y)\}^{(k)}_2=0\quad\mbox{for}\ k>1.
\eeqa
Here the invariant definition of the coordinate $t^n$ 
is $t^n :=\eta_{1\ve}\,t^\ve$. This conjecture means that
the Virasoro algebra with the central charge (\ref{cen-ch}) 
found for $d\ne 1$ 
in Corollary 1 above does not get deformations coming from the 
genera $\ge 2$.
In other words, our bihamiltonian structure is a classical $W$-algebra
with the conformal dimensions (\ref{conf-dim}) 
and the central charge (\ref{cen-ch}).

We recall that a Frobenius manifold $M^n$ is said to have  good
analytic properties if the primary free energy $F(t)$ has the form
\beq
F(t)=\mbox{cubic terms}+\mbox {analytic perturbation}
\eeq
near some point  $t_0\in M^n$. (See \cite{D3}). For example, the point $t_0$
is the origin in the topological minimal models and it is the point of
classical limit in the topological sigma-models.  For Frobenius manifolds
with good analytic properties we expect that all the coefficients
of the polynomials $A^{\al\beta}_{k,s}(t;\tx,\txx,\dots),\quad 
B^{\al\beta}_{k,s}(t;\tx,\txx,\dots)$ are analytic in $t$ near the 
point $t_0$. For the case $t_0=0,\ d<1$, i.e., the charges satisfy
\beq
0\le q_\al\le d<1,
\eeq
the analyticity implies finiteness of all of the expansions of the 
Poisson bracket.  Indeed, from 
(\ref{F-PBS}) and  (\ref{hom-A})
we obtain that
\beq
k(d-3)+d+s-1-q_\al-q_\beta\le k(d-1)+d.
\eeq
This number is nonnegative only if
\beq
k\le \frac{d}{1-d}.
\eeq
But all the degrees of the variables $t^\al$ are $1-q_\al>0$.
So all the terms $\{\ ,\ \}_1^{(k)}$ must vanish for $k>\frac{d}{1-d}$.
Similarly, the terms in the expansion of the second Poisson bracket must
vanish for $k> \frac{1+d}{1-d}$. All the examples of $1+1$ integrable 
hierarchies labeled by A-D-E Dynkin graphs are of this type. All the
coefficients of the genus expansions are polynomials.

Recall, that a polynomial Frobenius manifold can be constructed for an
arbitrary finite Coxeter group \cite{D3}. For this case 
$$
d=1-\frac{2}{h}, \quad q_\alpha= 1-\frac{m_\alpha +1}{h}
$$
where $h$ is the Coxeter number and $m_\alpha$ are the exponents of the
Coxeter group. However, the bihamiltonian hierarchy (\ref{F-Hier}) 
can be constructed
for
only simply-laced Dynkin graphs. Indeed, our formula (\ref{cen-ch})
for the central charge coincides with the formula \cite{FatLuk}
\beq
c \ve^2= 12 \varepsilon^2 \rho^2
\eeq
of the central charge of the classical $W$-algebras with the same Dynkin
diagram {\it exactly} for the simply-laced case! Here $\rho$ is one half
of the sum of positive roots. Our $\varepsilon$ is equal to $i\alpha$ of
\cite{FatLuk}. Recall, that for the simply-laced Coxeter groups our polynomial
Frobenius manifolds correspond to the topological minimal models \cite{Dij2}.
The constant $\ga$ in (\ref{anomaly}) equals 0. So the $G$-function
is identically equal to 0 for the A - D - E polynomial Frobenius manifolds.

For $d\ge 1$ the expansions probably are infinite. 
The Jacobi identity for the Poisson brackets, commutativity of the
Hamiltonians etc. are understood as identities for the formal power
series in $\ve^2$.
In the paper we have
constructed the first terms of the expansions and showed that they are in 
agreement with the assumptions we formulate in this section. 

To proceed to the next order ${\cal O}(\ve^4)$ we are to compute 
the Poisson brackets
\beq
\{H^{(0)}_{\al,p},H^{(1)}_{\beta,q}\}^{(1)}+
\{H^{(1)}_{\al,p},H^{(0)}_{\beta,q}\}^{(1)}+
\{H^{(1)}_{\al,p},H^{(1)}_{\beta,q}\}^{(0)} :=Q_{\al,p;\beta,q}.
\eeq
Then the corrections to the Hamiltonians and to the Poisson brackets
are to be determined from the linear equations
\beq\label{Lin-Sys}
\{H^{(0)}_{\al,p},H^{(2)}_{\beta,q}\}^{(0)}+
 \{H^{(2)}_{\al,p},H^{(0)}_{\beta,q}\}^{(0)}+
\{H^{(0)}_{\al,p},H^{(0)}_{\beta,q}\}^{(2)}=-Q_{\al,p;\beta,q}.
\eeq
We {\it do not} expect that the deformed hierarchy and the Poisson 
brackets can be constructed for an arbitrary Frobenius manifold 
(cf. \cite{Egu6}). However, solvability of the linear system
(\ref{Lin-Sys}) together with the bihamiltonian property 
could give a clue to the problem of selection of ``physical''
solutions of WDVV equations of associativity. We plan to investigate
this solvability in subsequent publications.

We do not discuss in this paper the relations between 
the one-loop deformations of the hierarchy and the 
Virasoro algebra of \cite{Egu3, Egu2}. This is to be done in a 
subsequent publication. Another interesting problem is a relation
between the hierarchy we contsruct and the recursion 
relations of \cite{KontsMan}.   


\vskip 0.6cm
\begin{thebibliography}{99}

\bibitem{Casati} P. Casati and M. Pedroni, Drinfeld-Sokolov reduction
on a simple Lie algebra from the bihamiltonian 
point of view. Lett. Math. Phys. 25(1992),
89--101. 
\bibitem{Dickey} L. Dickey, Soliton equations and Hamiltonian systems.
World Scientic, 1991.
\bibitem{DIZ} P.Di Francesco, C. Itzykson, J.-B. Zuber, 
Classical $W$-algebras,
Comm. Math. Phys. {\bf 140} (1991) 543--567.
\bibitem{DI} P. Di Francesco, C. Itzykson, 
Quantum intersection rings, hep-th/9412175.
\bibitem{Dij3} R. Dijkgraaf, E. Verlinde, H. Verlinde, Notes on topological
string theory and 2D quantum gravity, IASSNS-HEP-90/80.
\bibitem{Dij1} R. Dijkgraaf, E. Witten, Mean field theory, 
topological field theory, and multi-matrix models, Nucl. 
Phys. {\bf B342} (1990), 486--522.
\bibitem{Dij2} R. Dijkgraaf, E. Verlinde, H. Verlinde, Topological
strings in $d<1$. 
Nucl. Phys. {\bf B352} (1991), 59.
\bibitem{Dij5} R. Dijkgraaf, 
Intersection Theory, Integrable Hierarchies and Topological Field Theory,
Lectures given at the Cargese Summer School on
    `New Symmetry Principles in Quantum Field Theory,' July 16-27, 1991,
hep-th/9201003.
\bibitem{D1} B. Dubrovin, Integrable systems in topological 
field theory, Nucl. Phys. {\bf B379} (1992), 627--689.
\bibitem{D2} B. Dubrovin, Topological conformal 
field theory from the point of view of integrable systems, 
In: Integrable Quantum Field Theories, Edited
by L.Bonora, G.Mussardo, A.Schwimmer, L.Girardello, and M.Martellini,
Plenum Press, NATO ASI series {\bf B310} (1993), 283 - 302.
\bibitem{D3} B. Dubrovin, Geometry of 2D topological field theories,
in: Integrable Systems and Quantum Groups, Montecalini, Terme, 1993.
Editor: M.Francaviglia, S. Greco. Springer Lecture Notes in
Math. {\bf 1620} (1996), 120--348.
\bibitem{D4} B. Dubrovin, Painlev\'e equations in 2D topological field 
theories. In: Painlev\'e Property, One Century Later, Carg\`ese, 1996,
math.AG/9803107.
\bibitem{D5} B. Dubrovin, Y. Zhang, Extended affine Weyl groups and
Frobenius manifolds, Compositio Math. {\bf 111} (1998), 167--219.
\bibitem{Egu1}T. Eguchi,  H. Kanno,
Toda Lattice Hierarchy and the Topological 
Description of the c=1 String Theory,
Phys. Lett. {\bf B331} (1994), 330--334.
\bibitem{Egu5}T. Eguchi and S.-K. Yang,
The Topological $CP^1$ Model and the Large-N Matrix Integral,
Mod. Phys. Lett. {\bf A9} (1994), 2893--2902.
\bibitem{Egu6} T. Eguchi, Y. Yamada and S.-K. Yang,
On the Genus Expansion in the Topological String Theory, 
Rev. Math. Phys. {\bf 7} (1995) 279.
\bibitem{Egu4}T. Eguchi, K. Hori, C.S. Xiong,
Gravitational Quantum Cohomology,
Int. J. Mod. Phys. {\bf A12} (1997), 1743--1782.
\bibitem{Egu3}T. Eguchi, K. Hori, C.S. Xiong,
 Quantum Cohomology and Virasoro Algebra,
 Phys.Lett. {\bf B402} (1997), 71--80.
\bibitem{Egu2}T. Eguchi, M. Jinzenji, C.S. Xiong,
Quantum Cohomology and Free Field Representation,
hep-th/9709152.
\bibitem{FatLuk} V.A. Fateev, S.L. Lukyanov, 
Additional symmetries 
and exactly-solvable models in two-dimensional 
conformal field theory, parts I, II, and III,
Sov. Sci. Rev. {\bf A15} (1990) 1.
\bibitem{Getz} E. Getzler, Intersection theory on ${\bar M}_{1,4}$
and elliptic Gromov-Witten invariants, alg-geom/9612004, to appear
in J. Amer. Math. Soc..
\bibitem{Giv} A. Givental, Elliptic Gromov-Witten invariants and
the generalized mirror conjecture, math.AG/9803053. 
\bibitem{Hori} K. Hori, Constraints for topological strings
in $d\ge 1$, Nucl. Phys. {\bf B439} (1995), 395. 
\bibitem{Sato} M. Jimbo, T. Miwa, Y. Mori, M. Sato, 
Physica {\bf 1D} (1980) 80;
\noindent M. Jimbo, T. Miwa, Physica {\bf 2D} (1981) 407--448. 
\bibitem{Kimura} A. Kabanov, T. Kimura, 
Intersection numbers and rank one cohomological 
field theories in genus one, alg-geom/9706003.
\bibitem{Kon1} M. Kontsevich, Intersection theory on the moduli space
of curves and the matrix Airy function. Comm. Math.
Phys. {\bf 147} (1992), 1--23.
\bibitem{Kon2}  M. Kontsevich, Yu. Manin, Gromov-Witten classes, 
quantum cohomology and enumerative geometry, Comm. Math. Phys. {\bf 164}
(1994), 525--562.
\bibitem{KontsMan} M. Kontsevich, Yu. I. Manin, 
Relations between the correlators of the topological 
sigma-model coupled to gravity, alg-geom/9708024.
\bibitem{Magri} F. Magri, A simple model of the integrable Hamiltonian systems,
J. Math. Phys. {\bf 19}(1978), 1156-1162.
\bibitem{Miwa} T. Miwa, Painlev\'e property of monodromy preserving 
deformation equations and the analyticity of $\tau$-function, 
Publ. RIMS {\bf 17} (1981), 703--721.
\bibitem{Witten1} E. Witten, On the structure of the topological phase 
of two-dimensional gravity, Nucl. Phys. {\bf B340} (1990), 281--332.
\bibitem{Witten2} E. Witten, Two-dimensional gravity and intersection theory 
on moduli space, Surv. in Diff. Geom. {\bf 1} (1991), 243--310.
\bibitem{Witten3} E. Witten, On the Kontsevich model and other models of
two-dimensional gravity, preprint IASSNS-HEP-91/24.
\end{thebibliography}
\end{document}